\documentclass[11pt]{article}
\pdfoutput = 1

\usepackage[utf8]{inputenc}
\usepackage{color,graphicx}
\usepackage{epsfig}
\usepackage{amsmath}
\usepackage{verbatim}
\usepackage{amssymb}
\usepackage{mathabx}
\usepackage[mathcal]{eucal}
\usepackage{stmaryrd}
\usepackage{esint}
\usepackage{braket}
\usepackage{physics}
\usepackage{amsfonts}
\usepackage{cite}
\usepackage{array}
\usepackage{setspace}
\usepackage{braket}
\usepackage{float}
\usepackage{url}
\usepackage{mathtools}
\usepackage{bbold}
\usepackage{slashed}
\usepackage{tikz}
\usepackage{graphicx}
\usepackage{epstopdf}
\usepackage{subfigure}
\usepackage[labelfont=bf,font={sf}]{caption}
\usepackage{pgfplots}
\usepackage{mathrsfs}
\usepackage{fancybox}
\usepackage{eurosym}
\usepackage{tcolorbox}
\usepackage{tensor}
\allowdisplaybreaks

\usepackage[margin = 2.2cm]{geometry}
\usepackage[ragged]{footmisc}
\usepackage[bookmarks=true,bookmarksnumbered=false,
hyperindex=true,bookmarksopen=true,hyperfigures=true,
colorlinks=true,linkcolor=webblue,citecolor=webgreen,urlcolor=webblue,breaklinks]{hyperref}
\definecolor{webgreen}{rgb}{0, 0.5, 0}
\definecolor{webblue}{rgb}{0, 0, 0.5}
\definecolor{webred}{rgb}{0.5, 0, 0}
\definecolor{darkgreen}{rgb}{0,0.5,0}

\renewcommand{\d}{\mathrm{d}}
\renewcommand{\i}{\mathrm{i}}
\newcommand{\Ss}{\textsf{S}_0}

\newcommand{\average}[1]{\left\langle #1 \right\rangle}

\newcommand{\Vol}{\text{Vol}}

\def\ben{\begin{equation}}
\def\een{\end{equation}}

\let\a=\alpha \let\b=\beta \let\g=\gamma \let\d=\delta \let\e=\varepsilon
   
\let\l=\lambda     \let\r=v
\let\s=\sigma \let\t=\tau

\let\w=\omega \let\G=\Gamma \let\D=\Delta  \let\L=\Lambda

\def\nn{\nonumber}

\def\be{\begin{equation}}
\def\ee{\end{equation}}
\def\ba{\begin{array}}
\def\ea{\end{array}}

\def\mo{\mathcal{O}}

\def\dalemb#1#2{{\vbox{\hrule height .#2pt
       \hbox{\vrule width.#2pt height#1pt \kern#1pt
               \vrule width.#2pt}
       \hrule height.#2pt}}}

\newcommand{\bea}{\begin{eqnarray}}
\newcommand{\eea}{\end{eqnarray}}

\let\tilde=\widetilde

\def\md{\mathcal{D}}

\renewcommand{\d}{\mathrm{d}}
\renewcommand{\i}{\mathrm{i}}
\renewcommand{\S}{\textsf{S}_0}

\numberwithin{equation}{section}

\begin{document}

\thispagestyle{empty}
\begin{center}
    ~\vspace{5mm}

     {\LARGE \bf 
   The power of Lorentzian wormholes
   }
    
   \vspace{0.4in}
    
    {\bf Andreas Blommaert$^1$, Jorrit Kruthoff$\,^2$ and Shunyu Yao$\,^3$}

    \vspace{0.4in}
    {$^1$SISSA and INFN, Via Bonomea 265, 34127 Trieste, Italy\\
    $^2$ School of Natural Sciences, Institute for Advanced Study, Princeton, NJ 08540\\
    $^3$Department of Physics, Stanford University, Stanford, CA 94305, USA}
    \vspace{0.1in}
    
    {\tt ablommae@sissa.it, kruthoff@ias.edu, shunyu.yao.physics@gmail.com}
\end{center}

\vspace{0.4in}

\begin{abstract}
\noindent As shown by Louko and Sorkin in 1995, topology change in Lorentzian signature involves spacetimes with singular points, which they called crotches. We modify their construction to obtain Lorentzian semiclassical wormholes in asymptotically AdS. These solutions are obtained by inserting crotches on known saddles, like the double-cone or multiple copies of the Lorentzian black hole. The crotches implement swap-identifications, and are classically located at an extremal surface. The resulting Lorentzian wormholes have an instanton action equal to their area, which is responsible for topological suppression in any number of dimensions. 

We conjecture that including these Lorentzian wormhole spacetimes is gauge-equivalent to path integrating over all mostly Euclidean smooth spacetimes. We present evidence for this by reproducing semiclassical features of the genus expansion of the spectral form factor, and of a late-time two point function, by summing over the moduli space of Lorentzian wormholes. As a final piece of evidence, we discuss the Lorentzian version of West-Coast replica wormholes.

\end{abstract}

\pagebreak
\setcounter{page}{1}
\tableofcontents

\newpage
\section{Introduction and summary}

Path integrals in Lorentzian signature are subtle. The difficulties arise from two aspects. First, due to their oscillatory behaviour their convergence is often unclear. One can then resort to Euclidean methods to perform the computations, and analytically continue afterwards. The physical interpretation as the inclusion of a state preparation makes this a rather natural procedure. The second complication is how to describe Lorentzian topology change in gravity.\footnote{The analogue of this in QFT is to understand tunneling from a purely Lorentzian point of view.}

The simplest instance of this is perhaps the formulation of QFT as a worldline \emph{gravity} theory \cite{coleman1991quantum,Anous:2020lka,Casali:2021ewu}. Interactions in the QFT are reflected by worldlines branching off. At a point where a woldline branches off, the time function $t$ is singular, because its gradient is ill defined, and the einbein $e(t)$ vanishes. One way to include topology change here is by performing a further quantisation of the worldline theory, in which one considers operators that create and end worldlines. This theory is then the usual interacting QFT, which lives in the target space of the worldline theory. 

The situation is not that much different when we replace worldlines by higher dimensional manifolds with dynamical gravity, for instance string theory has 2d gravity on the worldsheet. As in the worldline theory, when we allow for topology change of the Cauchy slice, there is no well-defined time coordinate, and the metric $g(t)$ degenerates at the times where topology change occurs. Again, this can be overcome by performing a further quantisation, usually called \emph{third quantisation} in which one allows for operators that create and destroy universes \cite{coleman1991quantum,Marolf:2020xie,coleman1988black,giddings1988loss,giddings1989baby}. The resulting theory in the case of 2d strings is called string field theory, more generally one could speak of a universe field theory.\footnote{See recently \cite{Post:2022dfi,Altland:2022xqx} for the universe field theory of JT gravity, which is a string field theory.} Fields in this theory do not live on spacetime, but on the generalization of targetspace, usually called superspace.

Because superspace is \emph{not} the spacetime (unlike in the QFT example), observers such as ourselves who live within a single universe cannot see superspace, making a description in superspace physically less desirable. Only God-like observers that can oversee many universes have access to it. 

In this paper we want to revisit the issue of topology change in Lorentzian signature. In particular we will develop a more \emph{second quantized} picture for topology change in which there are (almost) Lorentzian geometries that mediate the change of topology of a Cauchy slice. This is more akin to the description of interacting QFT with singular worldlines, and to Mandelstam's interacting string picture \cite{mandelstam1973interacting} in 2d. This means we must take seriously geometries with (mild) metric singularities \cite{Louko:1995jw,Penington:2019kki,Colin-Ellerin:2020mva,Colin-Ellerin:2021jev,Marolf:2020rpm,Marolf:2021ghr,Maxfield:2022sio,Usatyuk:2022afj}.

The question that we set out to answer is: which such singular spacetimes are Lorentzian wormholes in asymptotically AdS, replacing the role of Euclidean wormholes in real-time calculations?

Many major open questions about quantum black holes are intrinsically Lorentzian in nature, most notably the fate of an observer falling through a horizon, and relatedly the nature of black hole interiors and the supposed singularity. Topology change is expected to be important in answering these questions, see for instance \cite{Stanford:2022fdt,Almheiri:2021jwq}. Thus, a better understanding of Lorentzian wormholes seems to be a prerequisite for answering these long-standing questions about black holes. 

\subsection{Summary and structure}
We now summarize our main messages.
\begin{enumerate}
    \item The role of Euclidean wormholes is replaced in the Lorentzian path integral by \textbf{slit geometries}. These are obtained from the usual Lorentzian geometries by allowing co-dimension-$2$ singularities with opening angle $2\pi n$, which for $n=2$ creates a pair of pants type geometry locally \cite{Louko:1995jw}. Following \cite{Louko:1995jw} we will call such singularity a \textbf{crotch}. We discuss how to take into account contributions due to crotches in the real-time path integral in \textbf{section \ref{sect:constrained}}. Crotches are specified by the location of the co-dim-$2$ singular surface $\gamma$ and contribute to the action as
    \begin{equation}
        e^{-(n-1)\frac{A(\gamma)}{4G_\text{N}}}\,.
    \end{equation}
    After extremizing with respect to the embedding $\gamma$ we obtain new real-time semiclassical saddle point spacetimes where crotches are located at the dominant \textbf{extremal surface}. In \textbf{section \ref{sect:blackhole}}, we explain how to include Louko-Sorkin crotches in the context of asymptotically AdS. We claim that counting (almost) Lorentzian crotch geometries is gauge-equivalent to path integrating over all (mostly) Euclidean smooth spacetimes. The bulk of this paper (\textbf{section \ref{sect:plateau}}, \textbf{section \ref{sect:twopoint}} and \textbf{section \ref{sect:pagecurve}}) consists of three examples where we check this conjecture.\footnote{Related (almost) Lorentzian geometries were recently discussed in JT gravity in \cite{Usatyuk:2022afj}, who focused on spacetimes without asymptotic boundaries. Our focus is on spacetimes \emph{with} asymptotic boundaries, and our discussion is not limited to JT gravity. Gauge equivalence of these descriptions was a key point in \cite{Usatyuk:2022afj}. The constrained instanton method we present in section \ref{sect:constrained} then explains why these geometries can at all contribute to the JT path integral, which was not discussed in \cite{Usatyuk:2022afj}.}
    \item In the case of the \textbf{spectral form factor}, dicussed in section \ref{sect:plateau}, crotches sit at the horizon on the double cone geometry of \cite{Saad:2018bqo} (the black dots classically coincide with the red dot, here we separated them for presentation purposes) and the temporal locations of the crotches remain zero modes
    \begin{equation}
    Z_1(+\i T,-\i T)\supset \quad \begin{tikzpicture}[baseline={([yshift=-.5ex]current bounding box.center)}, scale=0.7]
 \pgftext{\includegraphics[scale=1]{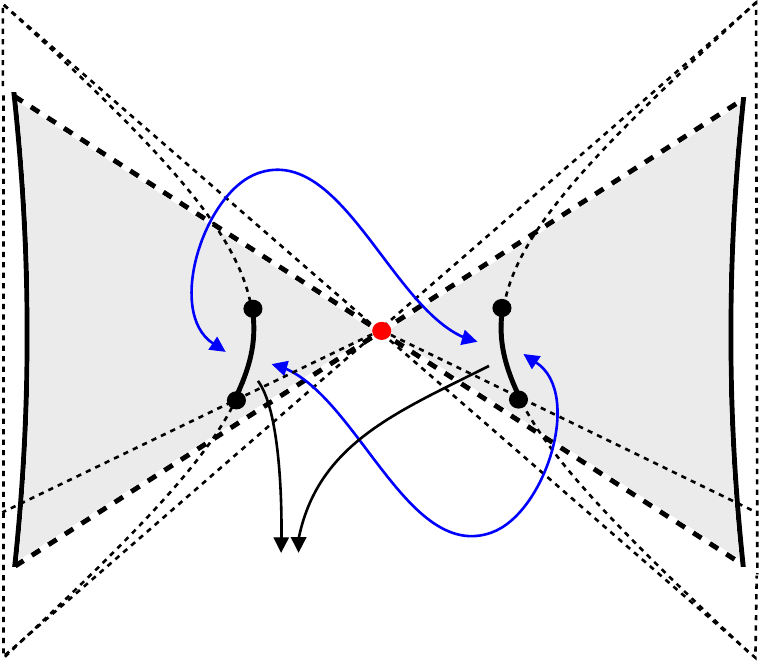}} at (0,0);
    \draw (-0.9, -2.7) node {step 1. slice};
    \draw (-0.2, 2.2) node {\color{blue}step 2. identify};
    \draw (0, -4) node {genus $1$ wormhole};
    \draw (4.4, -1.8) node {$t_1$};
  \end{tikzpicture}
\end{equation}
    They give contributions of the form \cite{Saad:2022kfe,Blommaert:2022lbh}
    \begin{equation}
        Z_g(\beta+\i T,\beta-\i T)\sim T^{2g+1} \left[\int_{\Lambda_g}^\infty \d E\, e^{-2g S(E)}\,e^{-2\beta E} + \text{negative area} \right]\,.\label{ramp-plateau}
    \end{equation}
    The first term in brackets is what we reproduce in this paper using semiclassical \textbf{saddles} at fixed energy and in particular at positive area. We have introduced a low energy cutoff $\L_g$, below which the semiclassical approximation breaks down. The constraint instanton method that we employ also allows for solutions at negative area, which are again semiclassical, but we have not studied those in detail here. Notably, in the semiclassical regime the total contribution for given energy is zero and only picks up a non-zero piece at low energies \cite{Saad:2022kfe}. A more detailed account of the exact coefficient of $T^{2g+1}$ is beyond the scope of the current paper, but see \textbf{section \ref{sect:disc}}. 
    
    Furthermore, one can easily generalize this setup to \textbf{any number of dimension}. In contrast, it is not clear how to recover this boundary prediction from a (almost) Euclidean gravity calculations in $d>2$. 
    
    \item The second example is the two-sided \textbf{two point function} at late times in dilaton gravity
    \cite{Saad:2019pqd,Blommaert:2019hjr,Blommaert:2020seb,Iliesiu:2021ari}. In \textbf{section \ref{sect:twopoint}} we show that the Euclidean gravitational path integral (or ETH from the boundary point of view) predicts a genus expansion which converges to a ramp-and plateau-type structure, analogous to \eqref{ramp-plateau} for the spectral form factor (See also \textbf{appendix \ref{sect:recap}})
    \begin{equation}
        \Tr(\mo\, e^{-(\beta/2+\i T)H}\mo\, e^{-(\beta/2-\i T)H})_g\sim T^{2g-1}\int_{\Lambda_g}^\infty \d E\,e^{-2\beta E}\, e^{-\Delta \ell(E)}\,e^{-(2g-1)S(E)}\,.\label{twopointgoalintro}
     \end{equation}
     We reproduce this structure by considering Lorentzian geometries with slits.\footnote{Again there is a negative area piece that we did not write. Situation is similar to the spectrum form factor case.} The usual Lorentzian pieces (two copies of the time-evolved TFD geometry glued at future time $T$) are replaced by the double-cone, with several additional crotches and identifications
    \begin{equation}
        \begin{tikzpicture}[baseline={([yshift=-.5ex]current bounding box.center)}, scale=0.7]
 \pgftext{\includegraphics[scale=1]{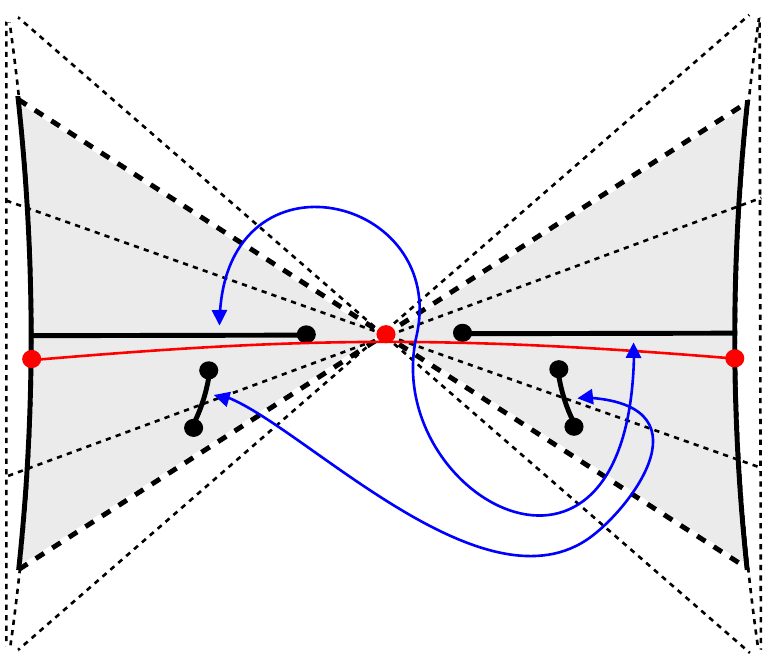}} at (0,0);
    \draw (0, -3) node {$g=2$ geometry};
    \draw (5,2) node {ket $R$};
    \draw (5,-2) node {ket $L$};
    \draw (4.4,-0.3) node {\color{red}$\mo_L$};
    \draw (-5,2) node {bra $L$};
    \draw (-5,-2) node {bra $R$};
    \draw (-4.4,-0.3) node {\color{red}$\mo_R$};
    \draw (0, 1.7) node {\color{blue}identify};
    \end{tikzpicture}\label{pic4intro}
    \end{equation}
    The distance between the left-and right boundaries is now a constant independent of time, which is precisely equal to the length of the ER bridge $\ell(E)$ in the TFD at $t=0$. So, there is a \textbf{shortcut} between the two asymptotic boundaries \cite{Saad:2019pqd}. This is discussed in \textbf{section \ref{sect:twopoint}}.
    \item In the case of gravitational matrix elements relevant for reproducing the \textbf{Page curve} in the West-Coast model \cite{Penington:2019kki}, the crotches accumulate near the black hole horizon, which results in geometries where the interiors in different copies are swapped (again the crotches in reality coincide with the red dot, representing the black hole horizon)
    \begin{equation}
    \bra{i}\ket{j}_\text{grav}\bra{j}\ket{i}_\text{grav}\supset \quad \begin{tikzpicture}[baseline={([yshift=-.5ex]current bounding box.center)}, scale=0.7]
 \pgftext{\includegraphics[scale=1]{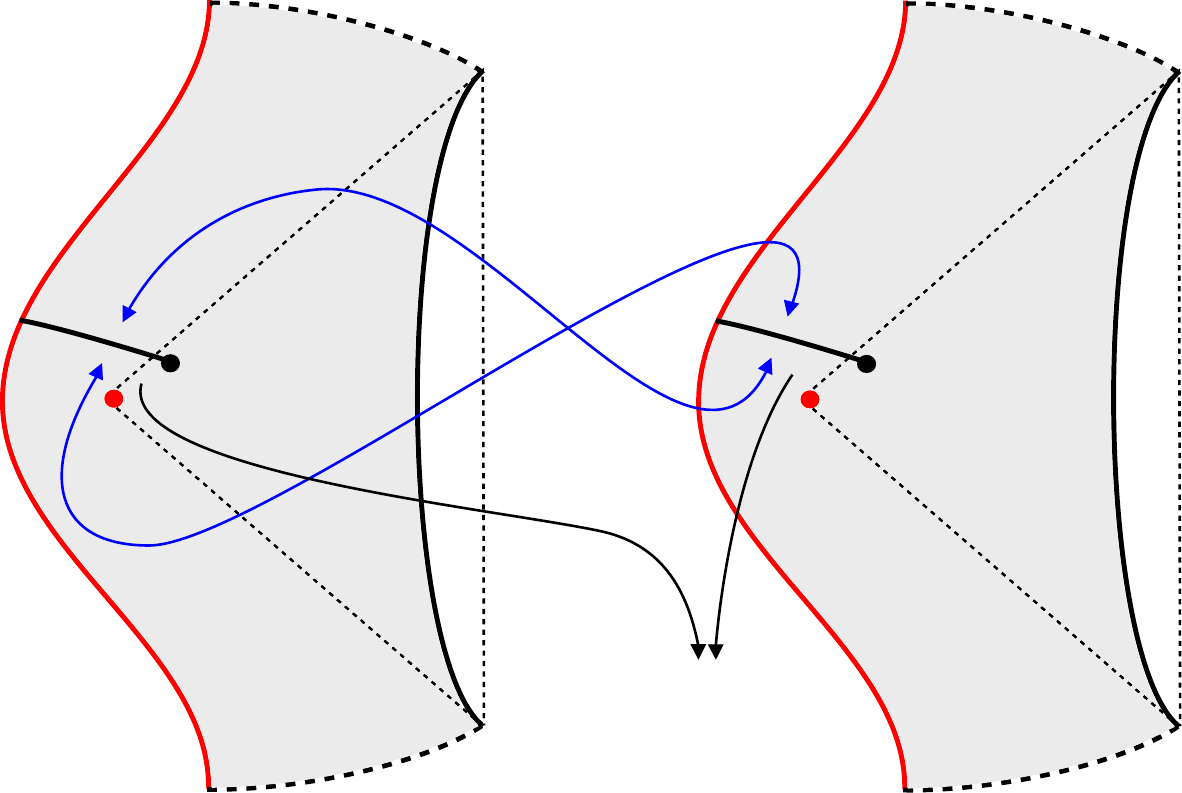}} at (0,0);
    \draw (1, -3.1) node {1. slice};
    \draw (0.5, 2.1) node {\color{blue}2. identify};
    \draw (-4.45, 3.5) node {\color{red}$i$};
    \draw (-4.45, -3.5) node {\color{red}$j$};
    \draw (2.7, -3.5) node {\color{red}$i$};
    \draw (2.7, 3.5) node {\color{red}$j$};
  \end{tikzpicture}\label{swap}
\end{equation}
    This is discussed in \textbf{section \ref{sect:pagecurve}}. That such swap geometries explain the Page curve in a Lorentzian setup is certainly not a new statement \cite{Penington:2019kki,Colin-Ellerin:2020mva,Colin-Ellerin:2021jev,Marolf:2020rpm,Marolf:2021ghr,Maxfield:2022sio}. What \emph{is} new is that there is a mechanism which explains why the crotches cling to the extremal surface. More importantly, we can calculate the semiclassical answer form the resulting Lorentzian saddles, and show that it reproduces the results of the Euclidean calculation of \cite{Penington:2019kki}.\footnote{For instance in \cite{Maxfield:2022sio} several toy models are introduced and are also found to reproduce quantitative features of the Page curve; however, as noted in the discussion of \cite{Maxfield:2022sio}, the mechanism for putting the crotches at the desired location is unclear. In this time-independent setup we do have such a mechanism, namely the classical equation of motion.}$^,$\footnote{This match between Euclidean and Lorentzian calculations was also a central point of emphasis in \cite{Colin-Ellerin:2020mva,Colin-Ellerin:2021jev}.}
\end{enumerate}

We stress that the slit geometries are not extra contributions which should be included in addition to Euclidean wormholes, rather they are \textbf{gauge equivalent} to including Euclidean wormholes. Notice that in both \eqref{twopointgoalintro} and \eqref{ramp-plateau}, the real-time gravitational path integral is surprisingly efficient at reproducing complex boundary predictions, more so than the Euclidean path integral.

The classical spacetimes which we found have \textbf{no closed baby universes} (in the sense of a closed universe propagating in time, detaching from and attaching to a parent \cite{coleman1988black,giddings1988loss}). We comment more on this in the discussion \textbf{section \ref{sect:disc}}, were we also propose some open questions.

In the case of the spectral form factor for JT gravity, the claim that counting all (almost) Lorentzian crotch geometries is gauge-equivalent to path integrating over all (mostly) Euclidean smooth spacetimes can be proven almost rigorously, even before comparing the answers. For this we use the ideas of \cite{Usatyuk:2022afj}, who pointed out that Lorentzian spacetimes with crotches appear naturally in JT gravity by going to \textbf{lightcone gauge} \cite{mandelstam1973interacting,d1987unitarity,d1988geometry,giddings1987triangulation}. We discuss this in detail in \textbf{appendix \ref{app:lightcone}}.
\section{Lorentzian topology changing geometries}\label{sect:blackhole}

Even though time in quantum gravity is only well-defined on an asymptotic boundary (where gravity is effectively turned off), the notion of a time function in the bulk spacetime is still extremely useful.\footnote{Similarly in electromagnetism we love picking a particular gauge and do calculations.} In particular, when we think about topology change in Lorentzian signature, we often have in mind that a Cauchy slice undergoes some topological transition as a function of some bulk time coordinate $t$. This means that we have picked a gauge in which $t$ is our time coordinate that runs orthogonal to our Cauchy slices and labels them. Right at the topological transition, the time coordinate is ill-defined as its gradients vanish there and the metric degenerates. In other words, the metric cannot be Lorentzian (and non-degenerate) everywhere on any spacetime manifold with topology change.

Thus we should entertain Lorentzian metrics with (mild) singularities, where the metric becomes non-invertible. There are (at least) two ways to do so. First, in the the first order (vielbein) formulation of gravity it is quite natural to not exclude configurations where the vielbein vanishes at isolated points (or surfaces)\cite{Horowitz:1991fr}.\footnote{See the discussion section \ref{sect:disc} for more about this.} Second, one can regulate the (mild) singularities by taking a limit from some metric which is Euclidean (and smooth) very close to the (would-be) singularity \cite{Louko:1995jw}. In the rest of this paper we will explore this second option. Interesting related recent work includes \cite{Usatyuk:2022afj,Colin-Ellerin:2020mva,Colin-Ellerin:2021jev,Marolf:2020rpm,Marolf:2021ghr,Maxfield:2022sio}. 

In this section, for simplicity of presentation, we will mostly consider 2d JT gravity. In later sections we often consider generic gravity models.

\subsection{We need singularities}

As argued in \cite{Louko:1995jw}, even though one cannot always find a Lorentzian metric on a 2d manifold, one can find one that is almost Lorentzian, in the sense that only at certain isolated points its signature changes. These metrics can be constructed by using a Morse function $f: \Sigma \to \mathbb{R}$ which can be physically thought of as the time (or level) function $f=t$ on the manifold. Constant $f$ defines the Cauchy slices. 

In particular, given a Euclidean metric $h_{ab}$ on $\Sigma$ and a Morse function $f$, a Lorentzian metric can be constructed as
\be 
g_{ab} = \partial_c f \partial_d f h^{cd}h_{ab} - \zeta \partial_a f \partial_b f\,,\quad \zeta>1\,.
\ee
The point is that $f$ can have critical points, and it is clear that $g_{ab}$ is singular at those points, where topology change occurs. The location of these critical points are moduli and will be important in what follows. Louko and Sorkin argued that one needs to regulate the metric by making it slightly complex at the singular points, in order to make the (almost) Lorentzian metric allowable \cite{Kontsevich:2021dmb,Witten:2021nzp}. 

Perhaps the best way to characterize the (mildly) singular points is via the Gauss-Bonnet theorem
\begin{equation}
    \chi=2-2g-n=\frac{1}{4\pi}\int_\Sigma \d x\sqrt{g}\, R+\frac{1}{2\pi}\int_\partial\d u\sqrt{h}\, K\,.
\end{equation}
This being a topological invariant, the Euler character is conserved when we analytically continue from a Euclidean metric to a complex metric, such as the (almost) Lorentzian metrics we will be interested in \cite{Witten:2021nzp}. Suppose now that we are interested in JT gravity and consider smooth Lorentzian metrics with $R+2=0$ everywhere. Because the metric is Lorentzian, $\sqrt{g} R$ is \emph{imaginary} and so there is no way to increase (real) genus $g$ using smooth Lorentzian spacetime. The regulated Lorentzian spacetimes that Louko-Sorkin described instead have $\sqrt{g}R=\i \text{ smooth }+$ delta functions, with the delta functions (with \emph{real} coefficients) located at the points of topology change. Those can decrease the Euler character, and thus allow us to have multiple asymptotic boundaries and wormholes.

In section \ref{sect:2.2}, \ref{sect:bh} and \ref{sect:crotches} we discuss a few basic examples of Lorentzian AdS geometries with such delta function sources.\footnote{For other interesting recent examples of such (almost) Lorentzian singular geometries in JT gravity, see \cite{Usatyuk:2022afj}, who focused on spacetimes without asymptotic boundaries. Our focus is on spacetimes \emph{with} asymptotic boundaries.} In particular in section \ref{sect:crotches} we describe the AdS version of \emph{crotches}, which are the key actors in our work. Then, in section \ref{sect:constrained}, we discuss the mechanism by which these singular geometries are picked up in the gravitational path integral.

\subsection{Example 1. Birth and death of baby universes}\label{sect:2.2}

JT gravity has a simple solution in Lorentzian signature that describes the birth or the death of a baby universe. Consider a compact spatial slice with the topology of a circle with size $b$, then the Lorentzian geometry describing the death (or crunch) of this spatial circle is
\be 
\d s^2 = - \d t^2 + \cos(t)^2 \, \d x^2\,,\quad x\sim x+b\,,\quad \Phi=C\sin(t)\,,
\ee
where $t$ is our time coordinate which runs from $0$ to $\pi/2$. At $t = \pi/2$ the metric is degenerate $\sqrt{g}=0$, and $\sqrt{g}R$ has a delta function source. We can use Gauss-Bonnet to determine its strength. When we include the point at $t = \pi/2$, the spacetime has the topology of a disk, with a geodesic $K=0$ boundary at $t=0$. Thus Gauss-Bonnet teaches us that $\sqrt{g}R$ must satisfy
\be 
\frac{1}{4\pi}\int_\Sigma \d^2 x \sqrt{g} R = 1=\chi_\text{disk}\label{diskGB}
\ee
Away from the singular point this metric satisfies $R+2=0$, which results (as anticipated above) in an \emph{imaginary} contribution to the Euler character\footnote{The sign follows from a more careful analysis of a regularized version of this geometry. Very close to the cone tip $t=\pi/2$ we do not see AdS curvature anymore, so the situation essentially reduces to the yarmulke singularity of \cite{Louko:1995jw}. The metric can then be regularized in the same way as they did. More low-brow one ends up using $\sqrt{g}=\i\sqrt{-g}$.}
\be 
\int_{\Sigma/x_\text{sing}} \d^2 x \sqrt{g}\,R = -2 b\,\i\,,\label{25}
\ee
with $x_\text{sing}$ the singular point at $t=\pi/2$. The full spacetime satisfies \eqref{diskGB}, thus the difference with \eqref{25} determines the strength of the source
\be 
\sqrt{g}(R+2)= 2(2\pi+\i b)\, \delta(x-x_\text{sing})\,.
\ee
This is the same type of source as the one responsible for a Euclidean trumpet spacetime \cite{Blommaert:2021fob}, we see that in Lorentzian signature this can source a baby universe death, which is cone-like.

The geometry with $t$ running from $-\pi/2$ to $0$ is the time-reversed process of the creation of a baby universe. The boundary condition should then be understood as one in the future instead of an initial slice we considered above. The source of curvature is identical.

\subsection{Example 2. Lorentzian black hole}\label{sect:bh}

Another geometry that does not satisfy $R=-2$ everywhere (which was recently also discussed in detail in \cite{Marolf:2022ybi}) is the geometry relevant for calculating $Z(\i T)$. It is the black hole geometry with Rindler time $t$ identified with period $T$. Away from the horizon $\rho=0$ (the singular point of this identification) the black hole geometry
\begin{equation}
    \d s^2=\d \rho^2-4A^2\sinh(\rho)^2\d t^2\,,\quad t\sim t+T\,,\quad \Phi=A \cosh(\rho)\,,\label{blackhole}
\end{equation}
is a solution to the JT gravity equations of motion. Here $A$ is a modulus of the solution related to the ADM energy as $A=E^{1/2}$. The strength of the conical source at $\rho=0$ is
\begin{equation}\label{RicciDisk}
    \sqrt{g}(R+2)=2(2\pi-\i 2 A T)\,\delta(x)\,,
\end{equation}
Indeed, away from the defect, the metric has everywhere $R=-2$, and the extrinsic curvature is $K = \coth \rho$. The smooth pieces of spacetime again give an imaginary contribution to Gauss-Bonnet,
\bea
\i \int_{\Sigma/x_\text{sing}} \d^2 x \sqrt{-g}\, R+2\i \int_\partial \d u \sqrt{-h}\, K=-4\i A T\int_0^{\rho_\partial} \d\rho\, \sinh(\rho) +4\i AT\cosh(\rho_\partial)=4\i AT
\eea
Combined with the contribution from the delta function in \eqref{RicciDisk} this gives the correct Euler character $\chi_\text{disk}=1$.
\subsection{Example 3. Crotch geometries}\label{sect:crotches}\label{sect:louko}

The real meat of topology change however does not come from the geometries considered thus far. For that we need a Lorentzian version of the pair of pants geometry. In flat space, such a geometry can be easily constructed using the Morse function 
\be 
f = x^2 - y^2,\quad h_{ab} = \delta_{ab}\,,
\ee
resulting in the Lorentzian metric
\be \label{crotch}
\d s^2 = (x^2 + y^2)(\d x^2 + \d y^2) - (2 \pm \i \e) (x \d x - y \d y)^2
\ee
The metric is singular at the critical point $x=y=0$ of $f$ and can be regularized to an allowable metric as \cite{Louko:1995jw}
\be 
\d s^2 = (x^2 + y^2 + i\s)(\d x^2 + \d y^2) - (2\pm \i \e) (x \d x - y \d y)^2\,,\quad \sigma\to 0\,.\label{213}
\ee
Outside of $x = 0 = y$ one can do the diffeomorphism $u + \i v = (x + \i y)^2/2$ and the metric reduces to Lorentzian flat space
\be 
\d s^2 = -(1 \pm \i \e) \d u^2 + \d v^2\,,\label{flat}
\ee
however, since the diffeomorphism is ill-defined at the origin, one finds via direct calculation a delta function there with a \emph{negative} sign
\be 
\sqrt{g}R=-4\pi\, \delta(x)\,.\label{crotchsource}
\ee
Such negative mass sources have the potential to increase the genus $g$. 

To understand what this has to do with wormholes, notice that the space \eqref{crotch} is actually a double cover of flat space \eqref{flat} in coordinates $u$ and $v$. The two covers (or two sheets) are identified along a branch cut starting at the singular point $u=v=0$ and extending out to infinity, just like the complex function $\sqrt{z}$
\begin{equation}
    \begin{tikzpicture}[baseline={([yshift=-.5ex]current bounding box.center)}, scale=0.7]
 \pgftext{\includegraphics[scale=1]{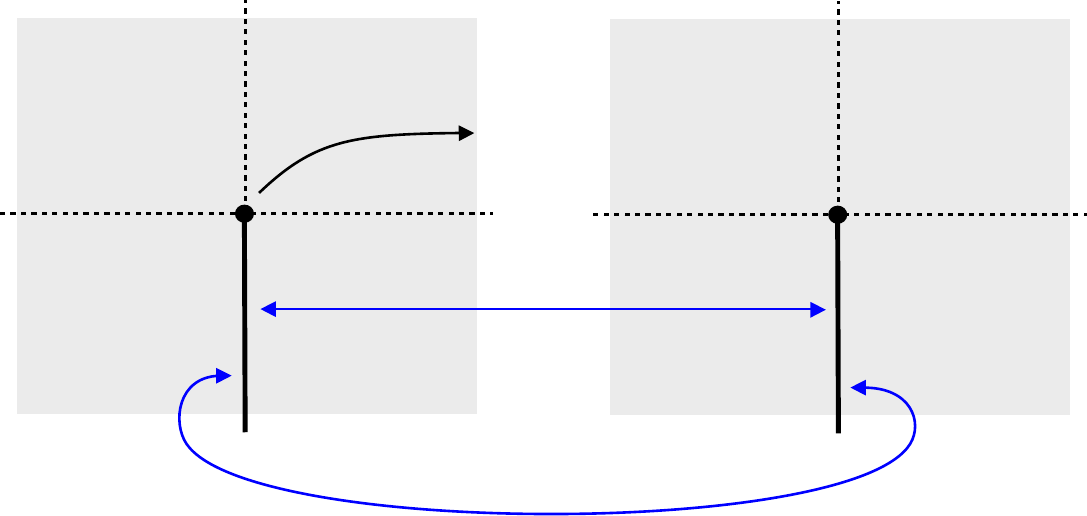}} at (0,0);
    \draw (0, -2.2) node {\color{blue}identify};
    \draw (0.4, 1.3) node {crotch};
    \draw (6.5, 0.5) node {$u=0$};
    \draw (3, 3) node {$v=0$};
    \draw (-3, 3) node {$v=0$};
  \end{tikzpicture} \label{pic}
\end{equation}
We can now make this into a pair of pants as Louko-Sorkin do by furthermore identifying the line $v=b$ on the second sheet with $v=-b$ on the first sheet, and identifying $v=a$ on the first sheet with $v=-a$ on the second sheet. Since these are geodesics ($K=0$), this cutting and gluing is a smooth operation. Louko and Sorkin called these types of singularities \eqref{crotchsource} \emph{crotches}, because in a pair of pants they are the crotch of the pants.

We can do something quite similar in JT gravity. Consider for instance the metric of the Rindler patches of the TFD in conformal gauge ($u$ is Rindler time, and the two asymptotic boundaries are at $v\sim \pm \varepsilon$)
\be
    \d s^2 = \frac{ - \d u^2 + \d v^2}{\sinh^2 v}. 
\ee
We now want to choose two semi-infinite lines on which we can cut the geometry, and then make swap identifications that implement a crotch singularity
\begin{equation}
    \sqrt{g}(R+2)=-4\pi\,\delta(x-x_\text{sing})\,.\label{218}
\end{equation}
Indeed, any point on any spacetime looks locally like flat space. This implies that the branch point of a swap identification is locally identical to the Louko-Sorkin crotch \eqref{213}, and it always has a singular source \eqref{crotchsource} (plus some smooth piece determined by the curvature of the original manifold).

However, unlike in flat space we can't identify lines at just any constant $v_1$ and $v_2$, this would give a kinked geometry because their extrinsic curvature does not add up to zero. One way to proceed is to consider two mirrored half-lines, one at $v = v_0$ and another one at $v = - v_0$. Cutting along the half-lines at these constant $v$s for $u>0$ creates four boundaries that can be glued. To see that this gives a smooth geometry, notice that when calculating the extrinsic curvature with outward point normals, the sums of the extrinsic curvatures needs to vanish. Consider $v_0 > 0$. In the case at hand we have at $v = -v_0$
\begin{align}
    K_L^{(-)} = + \cosh(v_0), \quad K_R^{(-)} = -\cosh(v_0)\, ,
\end{align}
and at $v = +v_0$ 
\begin{align}
    K_L^{(+)} = - \cosh(v_0), \quad K_R^{(+)} = +\cosh(v_0)\, .
\end{align}
So we glue the the left boundary in the negative $v$ region to the left boundary in the positive $v$ region and the same for the right boundaries, which results in a smooth geometry. In fact, this identification will be part of our construction of higher genus wormholes on the double-cone in section \ref{sect:plateau}.\footnote{There is one important difference between the identifications one can make on the TFD, and the identifications one can make on the double cone, namely the relative orientation of the slits. We clarify this in appendix \ref{app:orientation}.}

When there are multiple crotches, the branchcuts can connect and form finite sized \emph{slits}, instead of semi-infinite lines. This is desirable, because in asymptotically AdS we can not allow branchcuts going off to the asymptotic boundary (this is not allowed by the boundary conditions). These slits represent ``Lorentzian handles'', we discuss them in more detail in the next sections, for instance in \eqref{314}. Note that naively one might have thought we therefore include four crotches, which would increase the Euler character by $4$, but because of the gluing the final geometry only has two (additional) points where the metric degenerates. See also \eqref{pic}, where we have two crotches, but they are identified so we end up with a single delta function with coefficient $-4\pi$ and hence the topology of a pair-of-pants.
\subsection{Constrained instantons and extremal surfaces}\label{sect:constrained}
We have learned that we need to allow for AdS metrics with singular sources of the type \eqref{218}, in order to have Lorentzian topology change. The purpose of this section is to explain why such configurations can contribute, and with which measure. This question is sharpest in JT gravity, with action \cite{jackiw1985lower,teitelboim1983gravitation}
\begin{equation}
I_\text{JT}=-\frac{1}{2} \int_{\Sigma} \d^2 x \sqrt{g}\, \Phi\,(R+2) - \int_{\partial \Sigma} \d u \sqrt{h}\, \Phi\,(K-1) -\S\chi(\Sigma)\,.
\end{equation}
In Euclidean signature one can path integrate out the dilaton $\Phi$ and localize exactly on smooth metrics \cite{Jensen:2016pah,Maldacena:2016upp,Engelsoy:2016xyb} with
\begin{equation}
    \sqrt{g}(R+2)=0\,.\label{222}
\end{equation}
If this were to remain true in Lorentzian signature, it would pose a problem. Indeed, for instance even if we compute $Z(\i T)$ we learned in section \ref{sect:bh} that all Lorentzian black hole geometries that contribute have a conical source \eqref{RicciDisk} on the horizon \cite{Marolf:2022ybi}. Thus, if we gauge-fix the path integral to count (almost) Lorentzian spacetimes (instead of the smooth but generically complex spacetimes which solve \eqref{222}), we had better picked up geometries with conical sources.

We next present a way to pick up such contributions. Our method is similar to the constrained instanton method that picks up wormhole contributions in \cite{Cotler:2020lxj,Stanford:2020wkf,Cotler:2021cqa}. We start with resolving our confusions about the black holes contributing to $Z(\i T)$, and then generalize to crotches \eqref{218}, see also \cite{Dong:2018seb}. 

The Euclidean black hole partition function is, schematically
\begin{equation}\label{diskint}
   Z_{\text{disk}}(\beta)= \int \frac{\md g}{\text{vol}(\text{diff})}\,\md \Phi\, e^{-I_{\text{JT}}}
\end{equation}
We can insert a resolution of the identity
\begin{equation}
1= \int_{-\infty}^{\infty} \d A\frac{1}{\int \d x \sqrt{g}}\int \d x \sqrt{g}\, \delta(A-\Phi(x))\label{ideniden}
\end{equation}
and write the delta function as
\begin{equation}
\delta(A-\Phi(x))=\frac{1}{2\pi} \int_{-\infty}^{+\infty} \d\alpha\, e^{(2\pi-i\alpha)A}\, e^{-(2\pi-i\alpha)\Phi(x)}\,.\label{225}
\end{equation}
Notice that in Euclidean signature, the partition function of a disk with a
conical defect can be written as \cite{Mertens:2019tcm}\footnote{Technically speaking, this is the trumpet partition function \cite{Saad:2019lba} with a geodesic of real length $b=\alpha$. This is an analytic continuation of the defect to imaginary angles from all points of view \cite{Maxfield:2020ale,Mertens:2019tcm,Blommaert:2021fob}. In the past people have symmetrized by including $b=-\alpha$ as source to describe the trumpet, however as pointed out in \cite{Blommaert:2022lbh} that is equivalent, as the dilaton path integral is insensitive to certain dilaton insertions, see (5.36) in \cite{Blommaert:2022lbh}. The function $\gamma(x)$ is defined in \cite{Maxfield:2020ale}.}
\bea
Z(\beta,\alpha)= \int\frac{\md g}{\text{vol}(\text{diff})}\,\md \Phi\, e^{-I_{\text{JT}}} \frac{2\pi}{\gamma(1-\i \alpha/2\pi+\epsilon/2\pi)}e^{-(2\pi-\i \alpha)\Phi(x)}=\frac{e^{\Ss}}{2\sqrt{\pi\beta}}\exp(-\frac{\alpha^2}{4\beta})\,.
\eea
This means that inserting the right-hand side of \eqref{ideniden} in \eqref{diskint} we can rewrite $Z_\text{disk}(\beta)$ as\footnote{In the second step we used the fact that because of the $A$ integral, we can simplify pieces of the integrand by evaluating them on $\alpha_0=-2\pi\i$, for which $\gamma(1-\i\alpha_0/2\pi+\epsilon/2\pi)=2\pi/\epsilon$. We also used the fact that the volume of these spacetimes is given by $2\beta/\varepsilon$, see for instance equation (5.24) in \cite{Blommaert:2022lbh}.}
\begin{align}
Z_{\text{disk}}(\beta)&=\frac{1}{\int \d x \sqrt{g}}\frac{1}{2\pi}\int_{-\infty}^{+\infty}\d \alpha \,\frac{\gamma(1-\i \alpha/2\pi+\epsilon/2\pi)}{2\pi}\int_{-\infty}^{+\infty}\d A\,e^{(2\pi-\i\alpha)A}\,\frac{e^{\Ss}}{2\sqrt{\pi\beta}}\exp(-\frac{\alpha^2}{4\beta})\nonumber\\
&=\frac{1}{4\pi\beta}\int_{-\infty}^{+\infty}\d \alpha\int_{-\infty}^{+\infty}\d A\,e^{(2\pi-\i\alpha)A}\,\frac{e^{\Ss}}{2\sqrt{\pi\beta}}\exp(-\frac{\alpha^2}{4\beta})\,.
 \end{align}
Doing the integral over $\alpha$ first we obtain
\begin{equation}
Z_{\text{disk}}(\beta)=\frac{e^{\Ss}}{4\pi\beta}\int_{-\infty}^{+\infty} \d A\, e^{2\pi A-\beta A^2}=\frac{e^{\Ss}}{4\pi^{1/2}\beta^{3/2}}\,e^{\frac{\pi^2}{\beta}}\,.\label{blackholeanswer}
\end{equation}
Which is precisely the disk partition function of Euclidean path integral. On shell $A$ equals the horizon value of the dilaton $\Phi_h$, i.e. the area, as we see in \eqref{ideniden}. This area is related to the ADM energy as $A=E^{1/2}$. This procedure is thus calculating the partition function in microcanonical ensemble\cite{Dong:2018seb}, before integrate over energy with the semiclassical density of states $e^{2\pi A+\Ss}=e^{S(A)}$.

This procedure is redundant in evaluating Euclidean path integral. But as pointed out in \cite{Marolf:2022ybi}, this gives a way to think about the path integral for Lorentzian metrics. Indeed, if we consider Lorentzian boundary condition $Z(\i T)$, then after inserting the constrained instanton identity \eqref{ideniden} we have sources \eqref{225} in the path integral of precisely the type \eqref{RicciDisk} required to source Lorentzian black hole metrics \eqref{blackhole}. The path integral just picks up these contributions and we again find
\bea
Z_{\text{disk}}(\i T)=\frac{1}{2\pi \i T}\int_{-\infty}^{+\infty} \d A\, e^{2\pi A}\int_{-\infty}^{+\infty}\d\alpha\, e^{-\i\alpha A} \frac{e^{\Ss}}{2\sqrt{\pi\i T}}\exp(\i \frac{\alpha^2}{4 T})=\frac{e^{\Ss}}{4\pi\i T}\int_{-\infty}^{+\infty}\d A\,e^{2\pi A-\i T A^2}\,.
\eea
The integral over $\alpha$ is controlled by a saddle point $\alpha_0=2 A T$, which is indeed precisely the boost angle in \eqref{RicciDisk} associated with the Lorentzian black holes of area $A$ \eqref{blackhole}.

Now we generalize this method to pick up contributions from crotch geometries, in generic models of 2d dilaton gravity. For the remainder of this work we will study \emph{classical saddles}.\footnote{It would be interesting, though probably quite teadious, to evaluate these Lorentzian path integrals exactly, in particular including contributions from $\alpha\neq 0$.} We can generalize \eqref{ideniden} and \eqref{225} to
\begin{align}
    1=\frac{1}{\text{Vol}}\int_{-\infty}^{+\infty}\d A\,\int \d x_\text{sing}\sqrt{g}\,\frac{1}{2\pi}\int_{-\infty}^{+\infty}\d\alpha\,e^{-2\pi(n-1)A-\i \alpha A-(-2\pi(n-1)-\i\alpha)\Phi(x_\text{sing})}\label{1equals}
\end{align}
Classically, the $\alpha$ EOM force $A=\Phi(x_\text{sing})$.
This introduces sources of curvature for every fixed $A$ and searching for classical solutions, leads to various pairs $(\a, n)$ which are constrained by consistency with Gauss-Bonnet and the asymptotic AdS boundary conditions. We list these pairs below.
\begin{enumerate}
    \item One boundary: $\a \neq 0$ and $n = 0$. These are the geometry relevant for the calculation of $Z(\i T)$, or the birth and death of a closed universe as discussed above. 
    \item Two boundaries: The base (connected) manifold is the double cone. The new classical solutions are build using slits, which each consist of a pair of crotches. Each crotch corresponds with $n = 2$ and $\a = 0$ in the above equation.
    \item More than two boundaries: The base (connected) manifold is a Lorentzian version of the $n$-holed sphere obtained by taking $n$ identical copies of some spacetime and gluing them together cyclically, leading to an $n$-fold cover or $n$-replica geometry.\footnote{One concrete example is to take $n$ copies of the Lorentzian black hole \eqref{blackhole} but where we make cyclic identifications on the time slices $t=0$ and $t=T$ between $\rho=0$ and $\rho=\rho_\text{sing}$, whilst keeping the old identifications (making $n$ boundaries) on those slices between $\rho=\rho_\text{sing}$ and $\rho=\infty$. There is an $n$-cover source with $\alpha=0$ at $\rho=\rho_\text{sing}$, and a $n=0$ source with $\alpha=(n-1)2 A T$ at the origin $\rho=0$. Alternatively we can take $n$ copies of \eqref{blackhole}, make slits on matching lines and impose cyclic identifications on those slits. The endpoints of those slits are $n$-cover singularities with $\alpha=0$. In addition we have an $\alpha=2 A T$ and $n=0$ singularity at the horizon in each copy. Both constructions indeed reproduce the Euler character of a $n$-holed sphere.} Higher genus versions of these wormholes can be constructed by decorating with $n=2$ crotches. 
\end{enumerate}

From this list we see that we find metrics with swap-or replica-type identifications when $\alpha=0$.\footnote{The explicit examples presented in section \ref{sect:plateau} and section \ref{sect:pagecurve} should clarify these statements, should they sound confusing. This comment also holds for the next few sentences.} The one remaining modulus is the location of the singularity $x_\text{sing}$. We get a contribution to the on-shell action from the piece $-2\pi(n-1)A$ with on-shell $A=\Phi(x_\text{sing})$ and the contribution $-(n-1)\Ss$ from evaluating the Einstein-Hilbert action on the metric with singular source (generalizing \eqref{218})
\begin{equation}
    \sqrt{g}R\supset-4\pi(n-1)\,\delta(x-x_\text{sing})\,.\label{coversing}
\end{equation}
Aside from that, the on-shell action is that of the original geometry (before inserting crotches). Hence, semiclassically each $n$-cover crotch contributes a factor
\begin{equation}
    \int \d x_\text{sing}\sqrt{g}\,e^{-(n-1)\frac{A(x_\text{sing})}{4G_\text{N}}}\,,\quad \frac{A(x_\text{sing})}{4G_\text{N}}=\Ss+2\pi\Phi(x_\text{sing})\,.\label{231}
\end{equation}
Finally we impose the equation of motion associated with the location of the crotch $x_\text{sing}$. This implies that the classical solution is that the crotch singularities sit at an extremal surface\footnote{We will be a bit more careful below and include the $\sqrt{g}$ factor.}
\begin{equation}
    \frac{\d}{\d x_\text{sing}}A(x_\text{sing})=0\quad \Leftrightarrow\quad x_\text{sing}=x_\text{extr}\,,\label{233}
\end{equation}
so that contribution to the on-shell action is the classical entropy associated with that surface
\begin{equation}
    e^{-S_\text{inst}}=e^{-(n-1)\frac{A(x_\text{extr})}{4G_\text{N}}}\,.
\end{equation}
The most important contribution will come from crotches at the \emph{dominant} extremal surface.

This discussion has been on the level of pure 2d dilaton gravity, but it generalizes to any gravitational theory. The on-shell gravity action for these singular geometries is always \cite{Marolf:2022ybi} $-(n-1)A(\gamma_\text{sing})/4G_\text{N}$, with $\gamma_\text{sing}$ a co-dim 2 surface where $\sqrt{g}=0$, generalizing the crotches of Louko and Sorkin.\footnote{In cases where the slit between two crotches is spacelike, matter fields contribute the $n$-th Renyi entropy $-(n-1)S_n(\gamma_\text{sing})$ \cite{Maxfield:2022sio} and one must extremizing the generalized $n$-th Renyi entropy \cite{Lewkowycz:2013nqa,Faulkner:2013yia,Dong:2016hjy,Dong:2017xht} to find the classical location of the crotch.}

As compared to earlier similar discussions \cite{Colin-Ellerin:2020mva,Colin-Ellerin:2021jev,Marolf:2020rpm,Marolf:2021ghr,Maxfield:2022sio}, we stress that at least within 2d dilaton gravity \eqref{1equals} is an exact identity. We are not changing the value of the gravitational path integral by inserting several copies of the identity \eqref{1equals}. Rather, we should view this as part of a contour rotation, analogous to the construction of \cite{Saad:2021rcu}. This contour rotation involved, as a first step, making a gauge choice. There are (at least) two choices, and depending on the choice made, one should use the left-hand side or the right-hand side of \eqref{1equals} to pick up the metrics on the integration contour that we have gauge-fixed to.

For clarity, let us explain these two gauge choices within JT gravity.
\begin{enumerate}
    \item Working with the left-hand side of \eqref{1equals} the JT path integral localizes exactly on
    \begin{equation}
        \sqrt{g}(R+2)=0\,.
    \end{equation}
    This is appropriate when we gauge-fix to smooth but in general highly complex spacetimes, and obviously gives the same answer as the standard procedure (where one starts with a completely Euclidean path integral and analytically continues the answer).
    \item Working with the right-hand side, the path integral localizes on spacetimes with singular sources such as \eqref{coversing}, in this case the \emph{classical} solutions are spacetimes which are Lorentzian everywhere, and where the crotch singularities are located at (either of the) extremal surfaces
    \begin{equation}
        \sqrt{g}(R+2)=-4\pi\sum_{i}(n_i-1)\,\delta(x-x_{\text{sing}\,i})\,,\quad x_{\text{sing}\,i}\in x_\text{extr}\,.\label{coversingbis}
    \end{equation}
    This is appropriate when we gauge-fix to Lorentzian spacetimes, as we want to do in this paper. We claim that this reproduces the same answer as the Euclidean calculations, because we inserted \eqref{1equals}.\footnote{This is good, because Euclidean calculations with wormholes and replica wormholes have been very successful in recent years in reproducing complex predictions from the boundary dual, see for instance \cite{Saad:2018bqo,Saad:2019pqd,Blommaert:2019hjr,Iliesiu:2021ari, Kruthoff:2022voq,Penington:2019kki,Stanford:2022fdt,Blommaert:2022lbh,Saad:2022kfe,Almheiri:2019qdq}.}
\end{enumerate}
The remainder of this paper consists of three examples where we provide evidence for this claim.

For 2d dilaton gravity, the fact that these two options are different gauge choices can be made quite precise on an abstract level by (mildly) modifying the ideas of \cite{Usatyuk:2022afj}, who pointed out that these singular real-time spacetimes appear naturally when going to lightcone gauge. We explain how our analysis for JT gravity in section \ref{sect:plateau} is consistent with that picture in appendix \ref{app:lightcone}. The modifications as compared to \cite{Usatyuk:2022afj} include the application to asymptotically AdS spacetimes, and the above-explained constrained instanton method that explains why there can uberhaupt be contribution from singular spacetimes in the JT path integral, especially (constrained) saddles. 

\section{Spectral form factor}\label{sect:plateau}
Here we show how the ramp-plateau structure \eqref{ramp-plateau} is explained by semiclassical Lorentzian wormhole geometries where $2$-replica crotches accumulate at the (would-be) horizon of the double-cone of \cite{Saad:2018bqo}. In section \ref{sect:2dplateau} we consider JT gravity \cite{jackiw1985lower,teitelboim1983gravitation,Maldacena:2016upp,Jensen:2016pah,Engelsoy:2016xyb}, and 2d dilaton gravities with a more general action \cite{Witten:2020wvy,Maxfield:2020ale,Almheiri:2014cka}. This is the most controlled setup, where our counting of zero modes for the crotch locations is supported by the lightcone gauge description of \cite{Usatyuk:2022afj}, as we discuss in appendix \ref{app:lightcone}. We then extrapolate this, and count the same configurations in higher dimensional models, reproducing again \eqref{ramp-plateau}, in section \ref{sect:higherplateau}.\footnote{The assumption behind the extrapolation seems physically mild, namely that each crotch occurs at an identical instance of time in both wedges. In 2d, this is made rigorous via lightcone gauge, where essentially the time coordinate in the bulk is gauge-fixed, and the time coordinate of the crotch is the time at which the strings interact \cite{mandelstam1973interacting,d1987unitarity,d1988geometry,giddings1987triangulation,Usatyuk:2022afj}. One should think of the single time coordinate for each crotch as the time at which one baby universe is created or annihilated \cite{coleman1988black, giddings1988loss}. Notice in particular also the factor $T\,e^{-S_\text{inst}}$ associated with each such occurrence in equation (3.6) of \cite{giddings1988loss}, which we also have in \eqref{ramp-plateau}.} We recap how \eqref{ramp-plateau} is derived in appendix \ref{app:sff}.

\subsection{Double cone}
Before discussion the semiclassical Lorentzian wormhole geometries that account for the plateau, which (as we will demonstrate) are obtained by introducing slits (or crotches) in the double-cone, let us recap the semiclassical double-cone geometry itself \cite{Saad:2018bqo}. This is the ``canvas'' on which we will built higher genus Lorentzian wormholes.

In JT gravity, the double-cone solution \cite{Saad:2018bqo} with ADM energy $E$ \cite{Cotler:2020lxj,Stanford:2020wkf,Cotler:2021cqa} is
\begin{equation}
    \d s^2=\d\rho^2-4 E \sinh(\rho-i\varepsilon)^2\d t^2\,,\quad t\sim t+ T\,,\quad \Phi=E^{1/2}\cosh(\rho)\,.\label{doublecone}
\end{equation}
This is an allowable \cite{Louko:1995jw,Kontsevich:2021dmb,Witten:2021nzp} (almost) Lorentzian spacetime since the metric determinant has a positive real part everywhere
\begin{equation}
    \sqrt{g}=2 E^{1/2}\varepsilon\cosh(\rho)+2E^{1/2}\i\sinh(\rho)\,.\label{allowable}
\end{equation}
With this regulator one finds that the metric is perfectly smooth everywhere, including at the would-be horizon $\rho=0$ (see also appendix \ref{app:orientation} for a visualization of this regulator, and more on the double cone topology)
\begin{equation}
    \sqrt{g}(R+2)=0\,.
\end{equation}
With this understanding, we can drop the regulator $\varepsilon$ in all equations henceforth. Note that aside from the different regularization procedure, this is precisely two copies of a Lorentzian black hole \eqref{blackhole} with the same ADM energy $E$, a fact which will be crucial for what follows. The holographic boundaries \cite{Maldacena:2016upp} are located at $\rho=\pm \rho_c \pm\ell/2$ with $E=e^{-\ell}$ where indeed $\Phi=e^{\rho_c}/2$ and $\d s=\pm \i e^{\rho_c}\, \d t$. This geometry thus looks like (we exaggerated the separation between the holographic and coordinate boundaries)
\begin{equation}
     \begin{tikzpicture}[baseline={([yshift=-.5ex]current bounding box.center)}, scale=0.7]
 \pgftext{\includegraphics[scale=1]{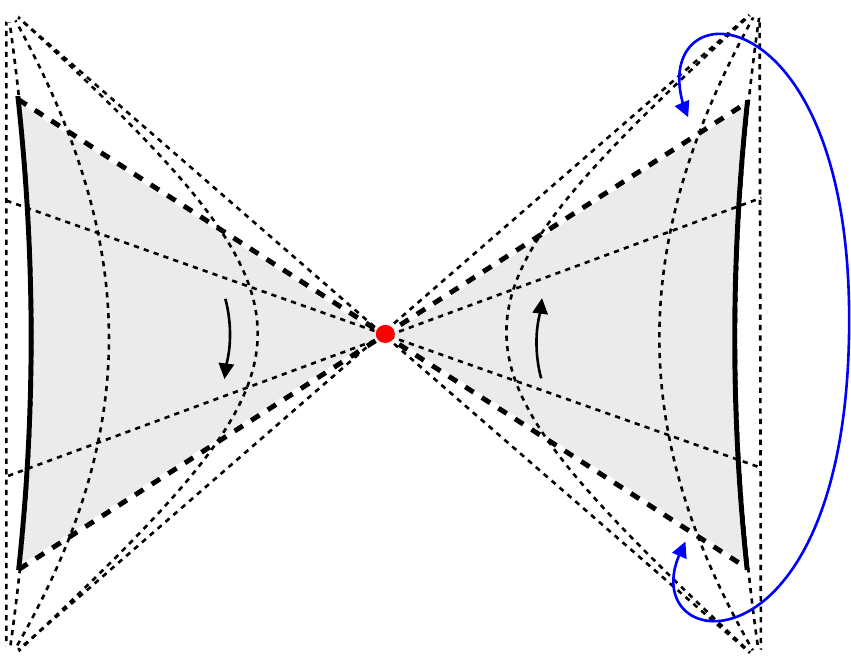}} at (0,0);
    \draw (-2.8, 0) node {$t$ flow};
    \draw (5.6, 0) node {\color{blue}identify};
  \end{tikzpicture}\label{34}
\end{equation}

Naively these are not solutions to the JT equations of motion \cite{Stanford:2020wkf}, however they \emph{do} become solutions when one introduces a Lagrange multiplier that fixes $\ell$ \cite{Stanford:2020wkf,Cotler:2020lxj} or equivalently the ADM energy $E$ in a manner that is very similar to the trick we used to get singular sources in the Lorentzian path integral, see \eqref{ideniden} and \eqref{1equals}. For every $E$, \eqref{doublecone} is then a full-fledged solution to the equations of motion \cite{Stanford:2020wkf}, and one ends up integrating over all these solutions, analogous to how we had an $A$ integral at the end in \eqref{blackholeanswer}.\footnote{In a more detailed calculation, which we will leave out, the integration modulus $\ell$ replaces the integration modulus $b$ in the Euclidean calculation \cite{Saad:2019lba,Saad:2019pqd} in a very precise manner with $b=2Te^{-\ell/2}$.} So, for any choice of boundary time $T$, there is a one-parameter family of (almost) saddles labeled by $E$.

As Saad-Shenker-Stanford \cite{Saad:2018bqo} explained the moduli space of saddles is in fact two-dimensional, the other classical modulus is a twist mode associated with rotating the two copies of the Lorentzian black hole relative to one another at the horizons. The volume of these twists is $T$ (because twisting with $T$ returns the original setup).\footnote{This two-dimensional classical phase space (twists and length) is the same as the one discussed by Harlow-Jafferis \cite{Harlow:2018tqv}.}

The on-shell action vanishes for these geometries \cite{Stanford:2022fdt} (because the ADM energy $E$ is equal on both sides, and the total boundary length vanishes identically). Introducing finite temperature results in a non-zero on-shell action $e^{-2\beta E}$. For the purposes of this paper we will think about finite $\beta$ as associated with a Euclidean preparation region, as one usually does for the thermal state (this is analogous to for instance how Saad \cite{Saad:2019pqd} thought about finite $\beta$). In this case, the preparation looks like a gutter, or the bottom half of the double trumpet \cite{Saad:2019lba} (the boundary locations $\rho$ are the same as in \eqref{doublecone})
\begin{equation}
    \d s^2=\d\rho^2+4E\cosh(\rho)^2\d \tau^2\,,\quad \tau\sim \tau+2\beta\,,\quad \Phi=E^{1/2}\cosh(\rho)\,.\label{gutter}
\end{equation}
The part between $\tau=0$ and $\tau=\beta$ is the gutter, and one can glue this smoothly in between the $t=T/2$ and $t=-T/2$ slices of the double-cone \eqref{doublecone}, as these slices are geodesics $K=0$ with the same length $\ell+2\rho_c$, and with identical dilaton profiles $\Phi$. This gutter will not play any role in the remainder of this work, except for providing the on-shell action $e^{-2\beta E}$, since we will make the gauge choice to have all topology change occur in the purely real-time piece of the geometry (not in the preparation region).

Since the action does not depend on the twist, their volume simply comes out, and we end up with the semiclassical contribution from the double-cone resulting in the integral \cite{Saad:2018bqo}\footnote{The twist mode is slightly less obvious for finite $\beta$ and after a more careful analysis \cite{Saad:2018bqo} one eventually finds the volume $(\beta^2+T^2)^{1/2}$. As explained in the beginning of this section, we are interested only in the double scaling limit $T\to\infty$ and $e^{\Ss}\to\infty$ with fixed ratio; with $\beta$ remaining finite the volume then indeed reduced to $T$ again. Essentially as compared to the Lorentzian piece of geometry, the Euclidean region is so tiny that it can be ignored for the twisting argument.}
\begin{equation}
    Z_0(\beta+\i T,\beta-\i T)_{\text{conn}}\sim T\int_{\Lambda_0}^\infty \d E\,e^{-2\beta E}\,.\label{ramp}
\end{equation}

One of the key insights in \cite{Saad:2018bqo} is that this construction is universal, meaning that we can take $\L_0$ all the way to zero. Namely, for any gravity model in any number of dimensions one can always create a double-cone by compactifying time and choosing a contour for the radial coordinate that avoids the conical singularity where the two copies of the (rolled-up) Lorentzian black holes meet.

The zero mode factor $T$ will clearly always be there, and as clarified in more detail in \cite{Cotler:2021cqa} the same goes for the on-shell action $e^{-2\beta E}$ and the integration modulus $E$ (now associated with a volume that was constrained first). Thus, semiclassically the double-cone \emph{universally} gives \eqref{ramp}.\footnote{In higher dimensions it is not obvious to prove that the $E$ measure is flat. We will not address that problem here, since we are interested in presenting a semiclassical approximation only. See \cite{Cotler:2021cqa}.}

The goal of this section is to find a similarly-universal generalization of the double-cone geometry which gives the whole semiclassical ramp-plateau structure \eqref{ramp-plateau}. We will first consider the 2d gravity case.

\subsection{Dilaton gravity models}\label{sect:2dplateau}
Instead of specializing to JT gravity, we consider the generalized models of 2d dilaton gravity \cite{Almheiri:2014cka,Witten:2020ert,Witten:2020wvy,Maxfield:2020ale} characterized by a dilaton potential $W(\Phi)$ and with action
\begin{equation}
    \exp\bigg(\frac{1}{2}\int \d t\, \d r \sqrt{g} (\Phi R+W(\Phi))+\int_\partial \d t \sqrt{h} \Phi(K-1)\bigg)\,.\label{acgen}
\end{equation}
Here the coordinate $t$ is the \emph{physical} coordinate used by boundary observers \cite{Maldacena:2016upp,Engelsoy:2016xyb}. Indeed, in gravity bulk diffeo's are redundancies but boundary coordinates are physical. We consider these models with general $W(\Phi)$ for two reasons. First, all of them have precise matrix integral duals \cite{Maxfield:2020ale,Witten:2020wvy}, so that there are precise boundary \cite{Saad:2022kfe} and Euclidean bulk \cite{Blommaert:2022lbh} calculations which reproduce the ramp-and plateau structure \eqref{ramp-plateau}. So in these cases it is a sharp constraint that the purely Lorentzian gravity description should reproduce it too, because it should be gauge-equivalent to the Euclidean one.\footnote{Of course, as already emphasized, with black holes being quantum chaotic, the boundary calculation using e.g. periodic orbits should apply in any number of dimensions, hence the boundary gives a sharp prediction in any case.}

Second, these models are more akin to higher dimensional cases, in that the importance of classical saddles is more clear. In JT gravity with $W(\Phi)=2\Phi$, one can simply integrate out the dilaton exactly, which obscures the importance of \emph{on-shell} dilaton configurations. In realistic theories we have no such powers, and we should resort back to saddles and expanding around them. The 2d models with generic $W(\Phi)$ can sometimes still be quantized exactly, by expanding the potential perturbatively around $2\Phi$, but such an expansion is \emph{not} close to the classical behavior of this system, and this is \emph{not} what we have in mind here. We want to introduce a \emph{universal} phenomenon that depends on saddles, and not on the specific (enormous) amount of control that one has in the JT gravity setup.

In these models the Lorentzian black hole solutions replacing \eqref{blackhole} are (the $\rho$ coordinate we introduced for future purposes) \cite{Witten:2020ert}
\begin{equation}
    \d s^2=-4F(r)\d t^2+\frac{1}{F(r)}\d r^2\,,\quad \Phi=r=\Phi_h\cosh(\rho)\,,\quad F(r)=\int_{r_h=\Phi_h}^r\d x\,W(x)\,.\label{gensol}
\end{equation}
These satisfy the boundary conditions $\Phi= r_c$ and $\d s=\i r_c\,\d t$, and as in the JT case they are sourced by a conical singularity at the horizon
\begin{equation}
    \sqrt{g}R\subset 4\pi\bigg(1-\frac{1}{\beta(\Phi_h)}\,\i \int \d t\bigg)\delta(x-x_\text{extr})\,,\quad \beta(\Phi_h)=\frac{2\pi}{W(\Phi_h)}\,,\quad \int\d t=T\,,
\end{equation}
which can be introduced in the path integral using the same constrained instanton trick \eqref{ideniden}.\footnote{In Euclidean signature these are fixed area states \cite{Dong:2022ilf,Dong:2018seb}.} Using the relation between the near-boundary metric and the ADM energy 
\begin{equation}
    F(r_c)-\frac{1}{4}r_c^2=-E(\Phi_h)\,,\quad  E(\Phi_h)=\Phi_h^2-\int_{\Phi_h}^\infty \d \Phi (W(\Phi)-2\Phi)\,,\label{310}
\end{equation}
and plugging this into the action \eqref{acgen}, one recovers the semiclassical approximation to the black hole answer \eqref{218} (We have renamed $A=\Phi_h$.)\cite{Witten:2020ert}
\begin{equation}
    Z(\i T)\sim \int\d A\,e^{2\pi A-\i T E(A)}\sim \int_{\dots}^\infty\d E\,e^{S(E)}\,e^{-\i T E}\,.
\end{equation}
In particular to recover the $e^{2\pi A}$ it is important to take into account the contribution from the singular source. One also checks that $\d E(\Phi_h)=1/\beta(\Phi_h)\,\d S(\Phi_h)$ and by inverting this relation to $S(E)$ one can match $\rho(E)=e^{S(E)}$ with the density of states for general dilaton gravities given in (1.4) of \cite{Witten:2020wvy}.

The double cone is (as always) two copies of this spacetime \cite{Saad:2018bqo,Cotler:2021cqa} where one replaces $\rho$ by $\rho+\i\epsilon$ and considers $-\infty<\rho<+\infty$. By inspecting the solution \eqref{gensol} for $-1\ll \rho\ll 1$
\begin{equation}
    \d s^2= -2\Phi_h W(\Phi_h)\,(\rho-\i\varepsilon)^2\d t^2+\frac{2\Phi_h}{W(\Phi_h)}\d\rho^2\,,\label{312}
\end{equation}
we indeed see that this makes the metric allowable, with an imaginary part of $\sqrt{g}$ that changes sign at $\rho=0$ as in \eqref{allowable}, and one checks that with this $\varepsilon$ regulator there is no singular source in $\sqrt{g}R$ anymore.

We now look for solutions to the same equations of motion and the same boundary conditions, but where we allow for two Louko-Sorkin-type crotch singularities
\begin{equation}
    \sqrt{g}R\supset-4\pi\sum_{i=1}^2\delta(x-x_{\text{sing}\,i})\,.\label{313}
\end{equation}
We explained in section \ref{sect:constrained} that such configurations are picked up via a modification of the fixed area mechanism, with the extra rule that one should extremize over the locations $x_{\text{sing}\,i}$ in the end. One can construct such configurations by taking the usual double cone, slicing it open along two mirroring \emph{slits}, and swap-identifying the sides of the slits as follows:
\begin{equation}
    \begin{tikzpicture}[baseline={([yshift=-.5ex]current bounding box.center)}, scale=0.7]
 \pgftext{\includegraphics[scale=1]{slit1.pdf}} at (0,0);
    \draw (-0.9, -2.7) node {step 1. slice};
    \draw (-0.2, 2.2) node {\color{blue}step 2. identify};
    \draw (0, -4) node {genus $1$ wormhole};
    \draw (4.4, -1.8) node {$t_1$};
    \draw (-4.8, -1.8) node {$T-t_1$};
  \end{tikzpicture}\label{314}
\end{equation}
The endpoints of the right slit are $(t_1,\rho)$ and $(t_2,\rho)$, whereas for the left slit they are $(T-t_1,-\rho)$ and $(T-t_2,-\rho)$. The reason for this particular choice will become clear. We strongly emphasize that away from these endpoints (which as we will explain are the locations of the crotches), this cutting an gluing does not change the smooth solution of the metric and dilaton. To be very clear, the metric and dilaton on this spacetime is everywhere (except at the singular points, see below) still given by \eqref{gensol}.
\begin{equation}
    \d s^2=-4F(r)\d t^2+\frac{1}{F(r)}\d r^2\,,\quad \Phi=r=\Phi_h\cosh(\rho)\,,\quad F(r)=\int_{\Phi_h}^r\d x\,W(x)\,.\label{gensolbis}
\end{equation}
Indeed, smoothly cutting and gluing geometries in gravity requires two things, namely that the surfaces (here lines) on which we glue have the same induced metric (here length along the line) and the opposite extrinsic curvature. Because the two slits are each other's mirror image, and because the two sides of the double cone are two copies of the same black hole (with the same energy $E$), this is satisfied. Very concretely, the curvature and induced metric on these fixed $r$ slices is ($+$ for right copy, $-$ for left copy).
\begin{equation}
    K=\pm \frac{W(r)}{2 F(r)^{1/2}}\,,\quad \d s=\pm \i\, 2 F(r)^{1/2}\, \d t\,,
\end{equation}
The signs of $K$ tell us to identify the left outside with the right inside, and vice versa. This should be quite intuitive, if one comes in from the left, space is getting smaller, and it should keep getting smaller after crossing this slit, in order to have a smooth identification.\footnote{The relative orientation by which we glue on the slits is determined by the fact that we should end up with an orientable metric, this is discussed in more detail in appendix \ref{app:orientation}.}

Close to each of the endpoints of the slits, which are pairwise identified, the geometry is exactly the Lorentzian double-cover of Louko-Sorkin \cite{Louko:1995jw} which we described in section \ref{sect:louko}\footnote{The left picture was flipped left-to-right as compared to the example in section \ref{sect:louko}, but locally that does not affect the topology.}
\begin{equation}
    \begin{tikzpicture}[baseline={([yshift=-.5ex]current bounding box.center)}, scale=0.7]
 \pgftext{\includegraphics[scale=1]{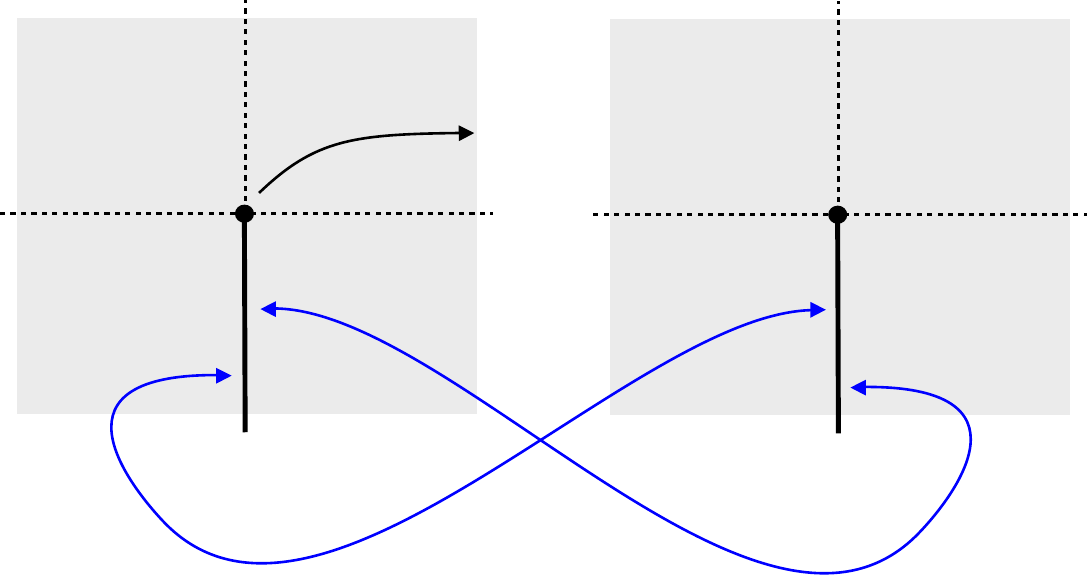}} at (0,0);
    \draw (0, -2.8) node {\color{blue}identify};
    \draw (0.4, 1.5) node {crotch};
    \draw (6.5, 0.7) node {$t=t_1$};
    \draw (-7, 0.7) node {$t=T-t_1$};
    \draw (3, 3.2) node {$r=r_1$};
    \draw (-3, 3.2) node {$r=r_1$};
  \end{tikzpicture} \label{picbis}
\end{equation}
As explained there, the metric on these geometries is indeed unaffected except for mild singularities at the endpoints on the slits with strength $-4\pi$ \eqref{313}.

Because of the opposite sign of $\sqrt{g}$ on both sides of the double cone \eqref{312}, we see that the smooth part of $\sqrt{g} R$ is exactly opposite on both sides of the cone. The same argument holds for the boundary contribution to Gauss-Bonnet, which therefore picks up only the contributions from the singular sources
\begin{equation}
    \chi=\frac{1}{4\pi}\int \d^2 x \sqrt{g}\, R+\frac{1}{2\pi}\int_\partial \d t \sqrt{h}\, K=-2\,.
\end{equation}
This confirms mathematically that we have created a genus $g=1$ wormhole.\footnote{To avoid all confusion, there is no such thing as ``Lorentzian Gauss-Bonnet''. There is simply Gauss-Bonnet, computing a topological invariant, which makes sense for all allowable metrics \cite{Witten:2021nzp}. The well-defined thing to write for all allowable metrics is $\sqrt{g}$.} Physically, one should see it clearly in \eqref{314}: cutting the double-cone open on the slits creates two ``holes'' with the topology of a circle, and identifying those ``circles'' makes a handle. (If this is not yet clear enough, the pictures in appendix \ref{app:lightcone} should help.)

We note that, much like for branchcuts of complex functions, the actual trajectory of the slits is not physical, the only thing that determines the geometry is actually the location of the crotch singularities. Two crotches connecting to form one slit (the ``top'' crotch and the ``bottom'' crotch) also do not need to sit at identical $\rho$ coordinates (but for each individual crotch, we do need the $\rho_i$ coordinates left-and right to be opposite, to satisfy the gluing rules).

Thus we have found solutions to the metric and dilaton equations with singular sources with weight $-4\pi$ \eqref{313} for all locations $x_{\text{sing}\,i}$ of the crotches. As explained around \eqref{1equals}, each crotch is weighed by an action
\begin{equation}
    e^{-\Ss}\,e^{-2\pi \Phi(x_\text{sing})}\,,\quad \Phi(x_\text{sing})=\Phi_h \cosh(\rho_\text{sing})\,.
\end{equation}
Away from the crotches the metric and the dilaton are still those of the double-cone \eqref{gensolbis} and we simply have the usual double-cone contribution $e^{-2\beta E}$ which we explained around \eqref{ramp}. The only remaining step to find classical saddles, as explained again around \eqref{1equals}, is to vary the area (or the dilaton) with respect to its location
\begin{equation}
    \frac{\d}{\d\rho_\text{sing}} \left(\log \sqrt{g(\rho_\text{sing})} - 2\pi \Phi_h \cosh(\rho_\text{sing})\right)=0\quad \Leftrightarrow\quad  \rho_\text{sing} \approx \frac{1}{(2\pi \Phi_h)^{1/2}}\,,\label{322}
\end{equation}
such that the crotches sit classically (when $\Phi_h$ is large) at the would-be horizon of the double cone. Thus our saddles look more like
\begin{equation}
    \begin{tikzpicture}[baseline={([yshift=-.5ex]current bounding box.center)}, scale=0.7]
 \pgftext{\includegraphics[scale=1]{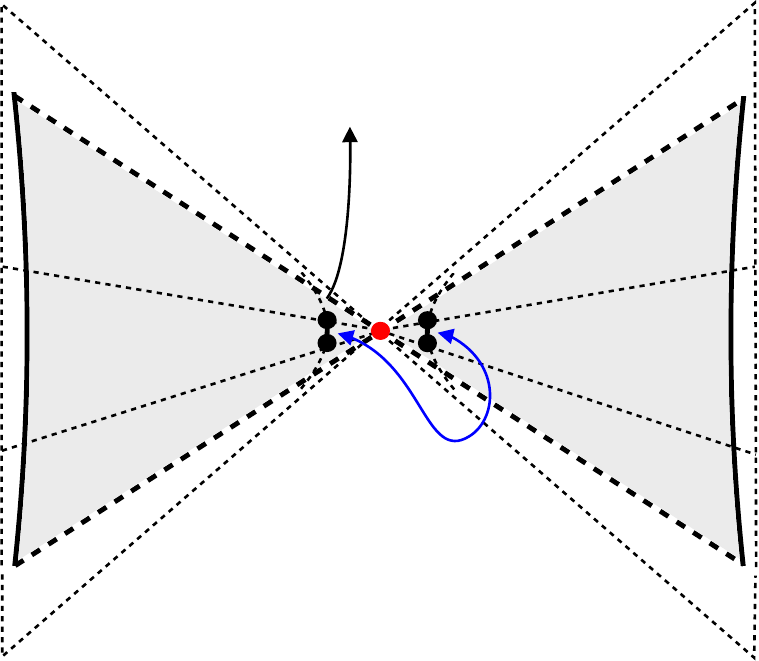}} at (0,0);
    \draw (0, -3) node {genus $1$ saddle};
    \draw (4.4, -1.3) node {$t_1$};
    \draw (4.4, 0.8) node {$t_2$};
    \draw (0, 2.5) node {$\rho_\text{sing} \approx 0$};
   \end{tikzpicture}\nonumber
\end{equation}
Another important point is that since $\rho_{\text{sing}} > 0$ due to a one-loop effect, it does not secretly disappear into the regulated Euclidean region of the double cone and causes an order of limits issue.\footnote{We thank Don Marolf for asking about this.}
Moreover, notice that the temporal coordinate is an exact \emph{zero mode}
\begin{equation}
    \frac{\d}{\d t_\text{sing}} \Phi_h \cosh(\rho_\text{sing})=0\,,
\end{equation}
so we have a saddle-point manifold parameterized by the crotches' temporal locations, quite analogous to the twist saddle-point manifold that one has for the empty double cone (which is of course also still there). Nothing else in the problem depends on these time coordinates, thus we just integrate over this saddle-point manifold. This (crucially!) produces a volume factor $T$ for each crotch. 

Since all classical saddles have $\Phi(x_\text{sing})=\Phi_h$ and since $\Ss+2\pi\Phi_h=S(E)$ the on-shell contribution of each crotch becomes
\begin{equation}
    T\,e^{-S(E)}\,.\label{crotchweight}
\end{equation}
Thus we find that, in this more refined version of the Lorentzian baby-universe picture \cite{coleman1988black, giddings1988loss}, we should identify the instanton action with the area of the extremal surface at which the topology change takes place
\begin{equation}
    \frac{A(\gamma_\text{extr})}{4G_\text{N}}=S_\text{inst}\,.
\end{equation}
This provides a natural mechanism for topological suppression of other topologies in higher dimensions too, as long as we consider classical black holes (with large areas), we will discuss this below in section \ref{sect:higherplateau}. Putting together the pieces we get the following semiclassical contribution at $g=1$
\begin{equation}
    Z_1(\beta+\i T,\beta-\i T)_{\text{conn}}\sim T^{3}\int_{\L_1}^\infty \d E\, e^{-2 S(E)}\,e^{-2\beta E}\,,\label{ramp-plateaug1}
\end{equation}
matching (part of) the boundary prediction \eqref{ramp-plateau}\footnote{As mentioned in the introduction it the full semiclassical result should give a vanishing coefficient of the power of time. The contribution that cancels this one is a semiclassical geometry with negative area, i.e. one ends the geometry at $-\Phi_h$ instead of $\Phi_h$. We will discuss this more in section \ref{sect:disc}. We thank Adam Levine for discussions on this.  Similar formulas like this will appear below also and should be understood in the same way. We decided to write the expressions this way because it emphasizes that our method also reproduces the structure of the integrand.}. To be clear, we still have the double-cone moduli and on-shell action \eqref{ramp}, we have simply \emph{included} into this the crotch moduli and on-shell action \eqref{crotchweight}.

The generalization to higher genus is obvious. There is no constraint on how many crotches one can include in the Lorentzian path integral, by simply inserting the identity \eqref{1equals} an arbitrary number of times. For every even number of crotches there are classical solutions of the type discussed above. For instance, for genus $g=2$ we have the following geometries (classically the crotches all sit at $\rho_{\text{sing}\,i}=0$, but that does not make for the clearest figures, so we have pictured generic off-shell $\rho_{\text{sing}\,i}$ here)
\begin{equation}
    \begin{tikzpicture}[baseline={([yshift=-.5ex]current bounding box.center)}, scale=0.7]
 \pgftext{\includegraphics[scale=1]{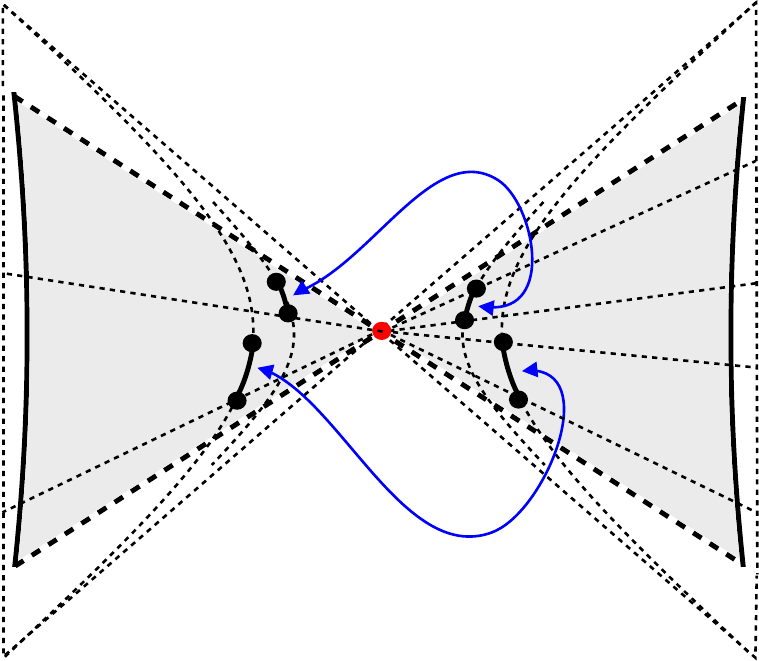}} at (0,0);
    \draw (4.4, 1.8) node {$t_4$};
    \draw (4.4, 0.5) node {$t_3$};
    \draw (4.4, -0.4) node {$t_2$};
    \draw (4.4, -2) node {$t_1$};
    \draw (0, -3) node {genus $2$ wormhole};
    \end{tikzpicture}\label{327}
\end{equation}
with modulo the identifications still the same metric and dilaton as on the original double cone. Nothing changes, except that now we have 4 temporal locations of the crotches marking zero modes, each crotch still contributes \eqref{crotchweight} and we obtain almost immediately
\begin{equation}
    Z_2(\beta+\i T,\beta-\i T)_{\text{conn}}\sim T^{5}\int_{\L_2}^\infty \d E\,e^{-4 S(E)}\,e^{-2\beta E}\,.\label{ramp-plateaug2}
\end{equation}
Higher genus is a trivial generalization at this point and one reproduces the boundary prediction \eqref{ramp-plateau}
\begin{equation}
    Z_g(\beta+\i T,\beta-\i T)_{\text{conn}}\sim T^{2g+1}\int_{\L_g}^\infty \d E\,e^{-2g S(E)}\,e^{-2\beta E}\,.\label{ramp-plateaugg}
\end{equation}

So, what do we learn from this? First, the fact that we recover the boundary prediction is evidence that we have indeed identified the correct Lorentzian version of wormhole geometries. Second, we note that as compared to the Euclidean calculations \cite{Saad:2022kfe,Blommaert:2022lbh} it was simpler to get the answer at semiclassical regime, except for one-loop factor that doesn't growth with time. As always with gauge choices, depending on the question, one choice can lead to more physically intuitive pictures than another (although the final answers should match). What we are advocating for is that for (at least some) real-time questions, it is maybe more efficient to think about Lorentzian metrics with crotches, than mostly Euclidean metrics.

One could wonder why we do not allow one crotch to have two temporal locations (meaning backing off from using the same parameter $t_i$ for the locations of the singular point on the left-and right copy). Physically this restriction seems sensible, one thinks of the birth and death of a baby universe as taking place at one instance of time. Mathematically, it is part of the gauge choices that we make with choosing a Morse function. This can be made precise in these 2d gravity models \cite{Usatyuk:2022afj,d1987unitarity,d1988geometry,giddings1987triangulation}. 

To appreciate that, one could think of the 2d Lorentzian spacetimes as 2d string worldsheets. Then one can reformulate the spacetimes that we are considering as the lightcone diagrams in Mandelstam's interacting string picture \cite{mandelstam1973interacting}. Now the (unique) interaction time $t_i$ is the lightcone time at which two open strings interact in a two-to-two open string scattering process. Such lightcone diagrams are gauge equivalent to integrating over the moduli space of smooth constant curvature Riemann surfaces. In appendix \ref{app:lightcone} we detail the lightcone gauge description of JT gravity \cite{Usatyuk:2022afj} and apply it to our current setup with Lorentzian asymptotically AdS boundaries. We show how the parameters in our spacetimes map to the moduli of Mandelstam's lightcone diagrams, and we show in more detail that the temporal locations of the crotches are zero modes for late times (beyond the current on-shell approximation). In particular this is important to show that we are not over-or undercounting the number of saddle-point geometries (though the match \eqref{ramp-plateau} is already strong evidence).

\subsection{Higher dimensional gravity}\label{sect:higherplateau}
The previous construction generalizes in a rather simple way to arbitrary gravity models in higher dimensions. As mentioned around \eqref{ramp}, for any gravity model there is a double-cone saddle made up out of two copies of the Lorentzian black hole (with an $\i\varepsilon$ prescription which removes the would-be conical singularity at the horizon) \cite{Saad:2018bqo}, leading to the universal structure \eqref{ramp} from a saddle-point analysis \cite{Saad:2018bqo,Cotler:2021cqa}.

We can represent one side of this double-cone (so the Lorentzian black hole) as a higher dimensional ``hollow doughnut''
\begin{equation}
    \begin{tikzpicture}[baseline={([yshift=-.5ex]current bounding box.center)}, scale=0.7]
 \pgftext{\includegraphics[scale=1]{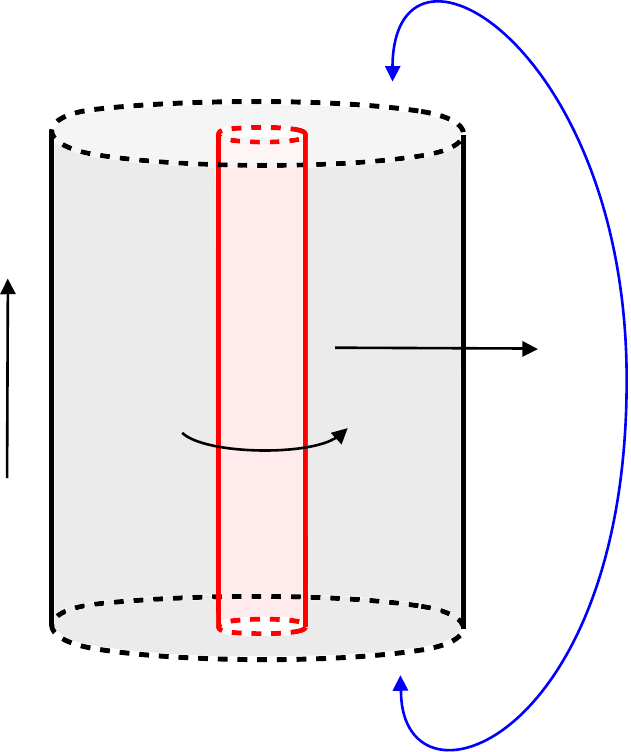}} at (0,0);
    \draw (1, -0.1) node {$\rho$};
    \draw (-3.5, 0) node {$t$};
    \draw (0.5, -1) node {$x^\parallel$};
    \draw (4.2, -2) node {\color{blue}identify};
    \draw (-1, -3.5) node {black hole};
    \end{tikzpicture}
\end{equation}
The red would-be horizon surface is null along the $t$ direction (the picture should be thought of as using Rindler-type coordinates), and has an area along the $x^\parallel$ directions equal to the black hole entropy for fixed charges $J$ (and as always we consider fixed ADM energy $E$)
\begin{equation}
    \frac{A(\gamma^\parallel_\text{extr})}{4 G_\text{N}}=S(E,J)\,.
\end{equation}
The real-time identification introduces some singularity at the horizon \cite{Marolf:2022ybi}, though this will not be our concern here.

Now we can consider the effects of allowing double-cover singularities in the Lorentzian spacetime. Now, double-covers are associated with some co-dimension-$2$ singular surface $\gamma_\text{sing}$.\footnote{Similarly, emitting a baby-universe at a real-time slice $t_i$ requires the metric to vanish (at least one component) on such a co-dimension-$2$ surface. For instance, in 3d we require the metric to vanish on a circle, such that a sphere is emitted.} The cut associated with such a singular co-dim-$2$ surface can end at another singular co-dim-$2$ surface, forming a co-dim-$1$ slit. That slit should then be swap-identified with an identical slit in a second copy of the geometry, in order to qualify as a double-cover geometry. 

Concretely, the single slit geometries (analogous to $g=1$ in 2d) in which we're interested look like\footnote{One could of course consider other topologies, for instance $\gamma_\text{sing}$ does not need to be homologous to the horizon, it could be some contractible circle outside of the horizon. However, those will not give rise to saddles, because such contractible loops have no extremal area. Similarly, the surface does not need to be this symmetric obviously, but the extremal ones (the saddles \eqref{233}) are.}
\begin{equation}
    \begin{tikzpicture}[baseline={([yshift=-.5ex]current bounding box.center)}, scale=0.7]
 \pgftext{\includegraphics[scale=1]{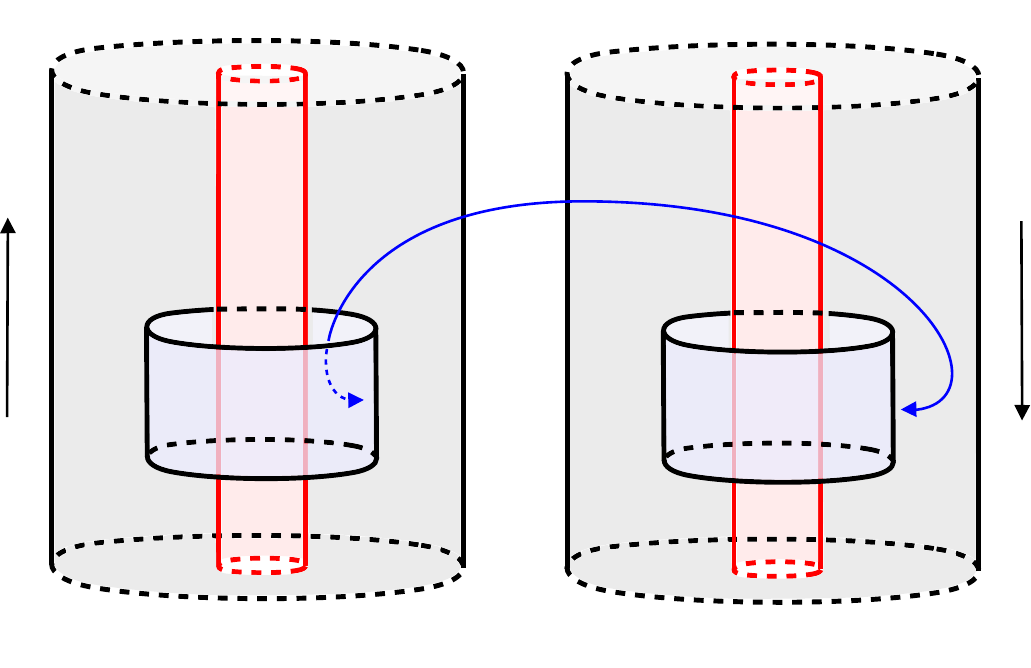}} at (0,0);
    \draw (-3.9, 0.6) node {$\gamma_{\text{sing}\,2}$};
    \draw (1.35, 0.6) node {$\gamma_{\text{sing}\,2}$};
    \draw (-5.5, 0) node {$t$};
    \draw (5.5, 0) node {$t$};
    \draw (-3.9, -1.8) node {$\gamma_{\text{sing}\,1}$};
    \draw (1.35, -1.8) node {$\gamma_{\text{sing}\,1}$};
    \draw (0, -3.5) node {$1$ slit wormhole};
    \draw (4.9, 1.5) node {\color{blue}swap insides};
    \end{tikzpicture}
\end{equation}
If we consider again mirror-configurations, where the slits have precisely the same positions in the two copies, then the metric (and all other fields for that matter) will again be smooth across the surface of identification, and the Einsteins equations (and other equations of motion) will be satisfied everywhere, except potentially at the singular surfaces $\gamma_{\text{sing}\,i}$. Notice that we are gluing always outside-to-inside in this double cone setup. The physical picture seems to be that you might think you are falling into your black hole, but you might end up in someone else's interior: you might have been swapped close to the horizon. In an evaporating setup, this might mean you are teleported by a swap to the horizon of the black hole created by the quantum computed that acts on the radiation \cite{Penington:2019kki}. 

But back to our technical problem now. The key point, explained by Marolf in \cite{Marolf:2022ybi} (following earlier work \cite{Colin-Ellerin:2020mva,Colin-Ellerin:2021jev,Marolf:2020rpm,Marolf:2021ghr,Maxfield:2022sio}), is that this procedure again does not change the on-shell action of the spacetime, except for an additional weight for each crotch-type singularity $\gamma_{\text{sing}\,i}$
\begin{equation}
    e^{-\frac{A(\gamma_\text{sing})}{4G_\text{N}}}
\end{equation}
This follows because roughly speaking \cite{Marolf:2022ybi}
\begin{equation}
    \frac{1}{16\pi G_\text{N}}\int \d x \sqrt{g} R+\dots = -\frac{A(\gamma_\text{sing})}{4G_\text{N}}+\text{double-cone answer}\,,
\end{equation}
which is explained by the fact that the metric away from the crotches is identical to that on the double cone, and because roughly speaking the type of curvature singularity is essentially the same as in 2d \cite{Marolf:2022ybi}
\begin{equation}
    \sqrt{g}R=-4\pi\sum_{i=1}^{2g}\delta(x^\perp-x^\perp_{\text{sing}\,i})+ \text{double-cone answer}
\end{equation}
As before the last step to get saddlepoints is to extremize with respect to the embedding $\gamma_\text{sing}$ (one can again include an analogous one-loop piece coming from the integral over all codimension two surfaces.)
\begin{equation}
    \frac{\d}{\d\gamma_\text{sing}}A(\gamma_\text{sing})=0\quad \Leftrightarrow\quad \gamma_\text{sing}=\gamma_\text{extr}\,.
\end{equation}
In our case the only classically extremal such surface is the spacelike co-dimension-$2$ surface that limits to the horizon of each of the copies of the Lorentzian black hole. In other words, the slits attach to the horizon, like heat-shrink tubing
\begin{equation}
    \begin{tikzpicture}[baseline={([yshift=-.5ex]current bounding box.center)}, scale=0.7]
 \pgftext{\includegraphics[scale=1]{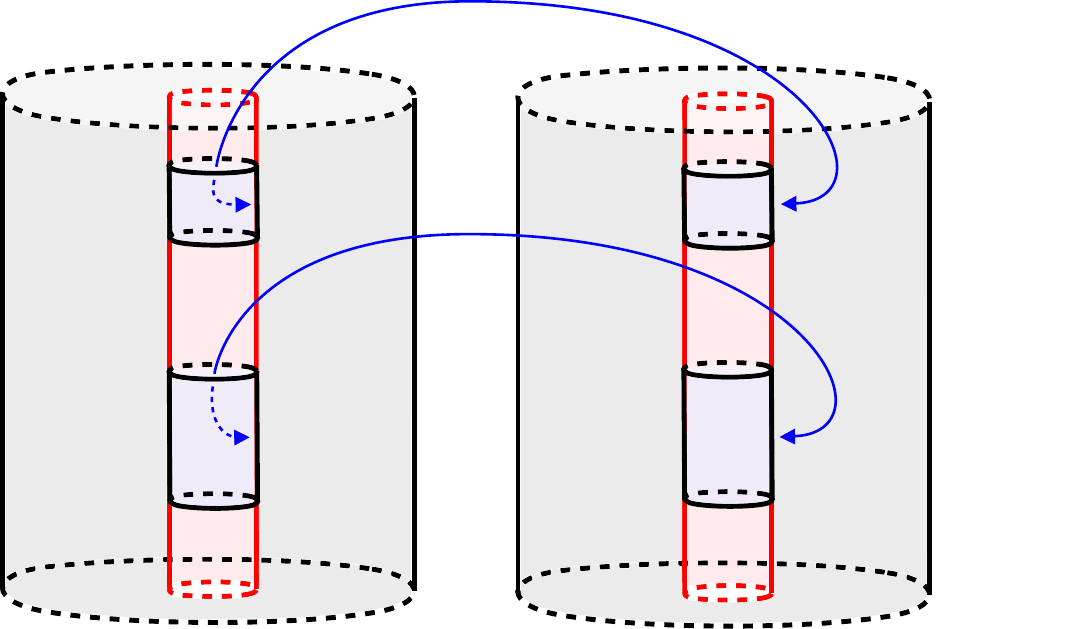}} at (0,0);
    \draw (-2.5, 1.5) node {$t_4$};
    \draw (-2.5, 0.8) node {$t_3$};
    \draw (-2.5, -0.6) node {$t_2$};
    \draw (-2.5, -1.9) node {$t_1$};
    \draw (0, -4) node {$2$ slit saddle};
    \draw (4.1, 0.5) node {\color{blue}swap insides};
    \end{tikzpicture}\label{exam}
\end{equation}
The area of each of the curves is then equal to the area of the black hole horizon at fixed ADM energy and fixed charges, such that even in higher dimensions the instanton action equals the black hole entropy
\begin{equation}
    S(E,J)=\frac{A(\gamma_\text{extr})}{4G_\text{N}}=S_\text{inst}\,.\label{supr}
\end{equation}
Moreover the temporal location of the crotches at the horizon remains a zero mode, so the weight of each crotch grows with time
\begin{equation}
    T\,e^{-S(E)}\,.
\end{equation}
That means we recover almost immediately the boundary prediction \eqref{ramp-plateau}
\begin{equation}
    Z_g(\beta+\i T,\beta-\i T)_{\text{conn}}\sim T^{2g+1}\int_{\Lambda_g}^\infty \d E\, e^{-2g S(E)}\,e^{-2\beta E}\,.
\end{equation}
Physically this seems to mean that for late times these swap-slits clinging to the horizon proliferate.\footnote{It would be interesting to study the statistics of these configurations in some more detail. Intuitively, one might expect the swap probability to saturate at $1/2$. Indeed, in the lightcone gauge description of appendix \ref{app:lightcone}, two slits never overlap, meaning we ought to restrict to $t_4>t_3>t_2>t_1$ in the example \eqref{exam}. However, because of the top-bottom identification there is a second option that should be included, where we connect the crotch at $t_2$ and $t_3$ with a slit, and the crotch at $t_4$ and $t_1$ by another slit (which goes through the top-bottom identification). In \eqref{exam} this means one should also include the configuration where the blue and red colors are interchanged. This means that for $g>0$ the swap probability is exactly $1/2$ at any fixed instance of time (the $g=0$ double cone is the exception because it involves no swaps).\label{fn:34}}

So, what new things did we learn from this? Generally speaking, the lesson seems to be that higher dimensional gravity may not be so out of control as we might have thought. We learned that there is a topological suppression, because topology changing processes (at least in this setup) are exponentially suppressed in the black hole entropy \eqref{supr}.

Second, and perhaps relatedly, we recover the boundary prediction from summing a countable set of bulk geometries. This is surprising even for 3d gravity. In that case, the double-scaled spectral form factor gives a precise prediction for the power series. The spectrum that one should insert in \eqref{a4} is
\begin{equation}
    \rho(E,J)=e^{\Ss(J)}\frac{2}{J^{1/2}E^{1/2}}\sinh((\pi E)^{1/2} b)\sinh((\pi E)^{1/2} /b)\,,\quad \Ss(J)=(\pi J)^{1/2}\bigg(b+\frac{1}{b}\bigg)\,,
\end{equation}
which has indeed an expansion in half-integer powers $E^{n+1/2}$.\footnote{To see quantum chaos one should look at fixed charges \cite{Haake:1315494,Mehta_1994}, this also includes fixed descendant labels here \cite{Cotler:2021cqa}. The applicability of chaotic universality demands that we consider large charges $e^{\Ss(J)}\to\infty$, this simplifies the BTZ spectrum to this form with $E$ the energy above extremality. In this double-scaling limit the other handle-bodies (or modular images) do not contribute since they give contributions that scale with a lower power of $e^{\Ss (J)}$ \cite{Maxfield:2020ale}. We use the usual
\begin{equation}
    c=1+6\bigg(b+\frac{1}{b}\bigg)^2\,.
\end{equation}
} It would be interesting to understand which Euclidean geometries are analytic continuations of our saddles, the guess would be that they are just $\Sigma_{g,2}\times S_1$ with all modular images. Do these give finite path integrals reproducing these answers?\footnote{It was suggested in \cite{Eberhardt:2022wlc} that they diverge, but perhaps there is some clever way to avoid this as for the wormhole in \cite{Cotler:2020ugk}.} Then what about the other $3$-manifolds? If we do not need them, then perhaps one should simply not include them. Notice that there are toy-examples of 3d gravity where indeed only a very restricted set of $3$-manifolds appears in the bulk description \cite{Maloney:2020nni,Afkhami-Jeddi:2020ezh}.
\section{Two-point correlation function}\label{sect:twopoint}
The purpose of this section is to demonstrate that similar Lorentzian crotch (or slit) geometries explain the semiclassical behavior of the two-point function
\begin{equation}
    \Tr(\mo\, e^{-(\beta/2+\i T)H}\mo\, e^{-(\beta/2-\i T)H})\,,
\end{equation}
in the same late-time double scaling limit where $T\to\infty$ and $e^{\Ss}\to\infty$ with $Te^{-\Ss}$ fixed. This had better been true, since the raison d'etre of the spectral form factor $\Tr(e^{-(\beta+\i T)H})\Tr(e^{-(\beta-\i T)H})$ is essentially to serve as a toy-model for this two-point function \cite{Cotler:2016fpe,Papadodimas:2015xma}. We consider operators $\mo$ that are dual in the bulk to a particle of mass $\Delta$ and are interested in the probe approximation, where the particle does not backreact on the geometry. In this regime, we are just computing the expectation value of
\begin{equation}
    e^{-\Delta \ell}\,,
\end{equation}
with $\ell$ the (regularized) distance between the two boundary points where operators $\mo$ are inserted.In this section we will consider 2d dilaton gravity exclusively. 

Following similar logic as for the spectral form factor \cite{Saad:2022kfe,Blommaert:2022lbh}, we demonstrate in appendix \ref{app:twopoint} that the two-point function behaves in the $\tau$-scaling limit as
\begin{align}\label{TrOOMMMain}
    \Tr(\mo\, e^{-(\beta/2+\i T)H}\mo\, e^{-(\beta/2-\i T)H})= \frac{T}{2\pi} Q_{\D}(\b) -\sum_{g>1}^\infty\frac{(T/2\pi)^{2g-1}}{(2g-2)(2g-1)}\oint_0\frac{\d E}{2\pi \i}\, e^{-\beta E}\, \frac{M_{\D}(E)}{\rho_0(E)^{2g-2}e^{\Ss}}\,.
\end{align}
with 
\be 
Q_\D(\b) = \int_0^{\infty} \d E\, e^{-\b E} e^{-\Ss} M_{\D}(E)
\ee
and $M_{\D}(E)$ given in \eqref{DSMElem} \footnote{Here we worked with $E_0 = 0$, but one can easily generalize to non-zero $E_0$.}. This has a similar ramp and plateau structure as the spectral form factor, as predicted by Saad \cite{Saad:2019pqd}. 

We will identify the semiclassical Lorentzian geometries that contribute to this observable, making the gauge choice that topology change takes place in the Lorentzian region. The slit geometries which we identify share the property that $\ell$ has an order one (not growing with $T$) value, and the probability of the geometries themselves grows as $T^{2g-1}e^{-(2g-1)\Ss}$. This should be contrasted with the disk where $\ell$ grows linearly with $T$ \cite{Harlow:2018tqv,Yang:2018gdb,Iliesiu:2021ari,Hartman:2013qma,Susskind:2014rva}, causing exponential decay of the correlator. 

The physical picture is that all $g>0$ Lorentzian geometries which we are led to include, have the property that the $t=T$ slice is reminiscent of the $t=0$ slice of the original TFD geometry, or a fixed Rindler-time slice of the double-cone (which is identical). The order one $\ell$ is the size of the $t=0$ ER bridge. This is reminiscent of statements made in \cite{Stanford:2022fdt,Susskind:2015toa,Saad:2019pqd} that wormholes could rejuvenate the TFD, but with different quantitative results. Our Lorentzian wormholes all rejuvenate to $t=0$. This happens because we compute $\average{e^{-\Delta\ell}}$, which essentially projects out older TFDs.\footnote{More specifically in the $\tau$-scaling limit any $\ell\sim T$ projects out. In this sense it is not technically true that our observable does not backreact on the geometries. For the same reason we do not have a sum over paths \cite{Saad:2019pqd}, only the shortest geodesic contributes. So, we do not have contributions from the other geodesics which are responsible for the quasinormal modes.} This does not happen when one would compute $\average{\ell}$, which we do not pursue here, but see comments in the discussion section \ref{sect:disc}.

For our Lorentzian setup, we are interested in reproducing the semiclassical features of this formula. So we will consider large black holes $E\gg 1$ and probe matter that does not backreact on the geometry $\Delta\ll E^{1/2}$. Using Stirling to approximate $M_\Delta(E)$ \eqref{DSMElem} this results in the semiclassical approximation (again without the negative area contribution)
\begin{equation}
     \Tr(\mo\, e^{-(\beta/2+\i T)H}\mo\, e^{-(\beta/2-\i T)H})_g\sim T^{2g-1}\int_{\L_g}^\infty \d E\, e^{-\beta E}\, e^{-\Delta \ell(E)}\,e^{-(2g-1)S(E)}\,,\label{twopointgoal}
\end{equation}
where we introduced
\begin{equation}
    \ell(E)=-\log(4E)\,,\quad S(E)=2\pi E^{1/2}\,.
\end{equation}
Here $\ell(E)$ is (see below) the renormalized length of the ER bridge in the (fixed energy) TFD at $T=0$. This form of equation \eqref{twopointgoal} remains valid for generic dilaton gravity models, where one should replace $\ell(E)$ with the boundary-to-boundary length in the metric \eqref{gensol} and the entropy becomes $2\pi \Phi_h(E)$ with the relation \eqref{310}.\footnote{Whilst we do not yet know the matrix elements $\mo_{E_1E_2}\mo_{E_2E_1}$ for generalized dilaton gravity, we do know via ETH that they are smooth functions of $\omega$ so in this sense \eqref{46} and what follows remains true. The fact that $\ell(E)$ (being the saddle of $\mo_{E E}\mo_{E E}$) takes on the value of the boundary-to-boundary geodesic in the TFD at $T=0$ can be understood from the generalization of the discussion of the semiclassical wavefunction around (2.13), (2.14) and (2.18) in \cite{Stanford:2022fdt}.} 

The two-point function in which we are interested is the two-sided two-point function in the thermofield double state of the boundary theory
\begin{equation}
     \Tr(\mo\, e^{-(\beta/2+\i T)H}\mo\, e^{-(\beta/2-\i T)H})=\bra{\text{TFD}_{\beta}}e^{+\i T/2 (H_L+H_R)}\mo_L\,\mo_R\,e^{+\i T/2 (H_L+H_R)}\ket{\text{TFD}_{\beta}}\label{bc}
\end{equation}
The corresponding $g=0$ geometry is obtained by first preparing the TFD two-sided geometry at $t=0$ and then Lorentzian time-evolving it to $t=T/2$, at which point two operators $\mo_L$ and $\mo_R$ are inserted on both boundaries to probe the geometry. The same preparation and Lorentzian time evolution applies to the bra, and the bra-and ket geometries are glued together on the trajectory of the probe particle.\footnote{One could evolve further on both sides, but such additional pieces of geometry would cancel in the bra-ket path integral.} The glued Lorentzian pieces of geometry make a tent-shape:
\begin{equation}
    \begin{tikzpicture}[baseline={([yshift=-.5ex]current bounding box.center)}, scale=0.7]
 \pgftext{\includegraphics[scale=1]{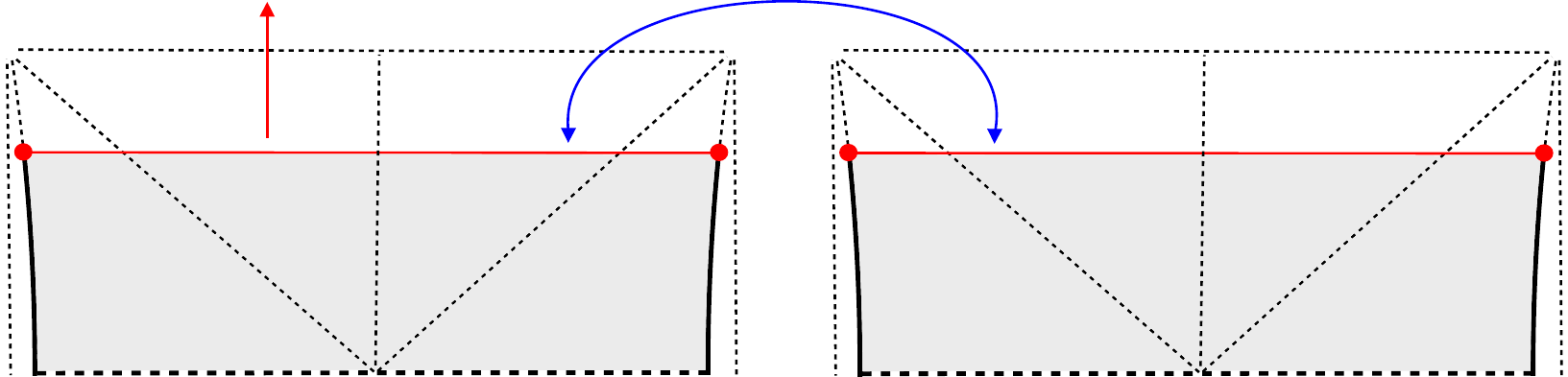}} at (0,0);
    \draw (0, 0.5) node {\color{red}$\mo_R$};
    \draw (8.8, 0.5) node {\color{red}$\mo_L$};
    \draw (-8.8, 0.5) node {\color{red}$\mo_L$};
    \draw (-4.3, -2.5) node {ket};
    \draw (9.3, 1.3) node {$X=\frac{\pi}{2}$};
    \draw (9.3, -1.95) node {$X=0$};
    \draw (4.3, -2.5) node {bra};
    \draw (4.3, 2) node {$\sigma=\frac{\pi}{2}$};
    \draw (0, -3) node {$g=0$ geometry};
    \draw (0, 2.5) node {\color{blue}identify};
    \draw (-5.5, 2.5) node {\color{red}particle};
    \end{tikzpicture}\label{pic1}
\end{equation}
The metric and dilaton on these Lorentzian slices are
\begin{equation}
    \d s^2=\frac{\d\sigma^2-\d X^2}{\sin(\sigma)^2}\,,\quad \Phi=E^{1/2}\frac{\cos(X)}{\sin(\sigma)}\,,\label{geom}
\end{equation}
with the diverging boundary trajectory parameterized by boundary time $t$
\begin{equation}
    \tan(X/2)=\tanh(E^{1/2} t)\,,\quad \sigma=\pi-\varepsilon\, \frac{\d X}{\d t}\,,
\end{equation}
The operators $\mo_L$ and $\mo_R$ are inserted on this boundary trajectory at $t=T/2$ and the shortest geodesic between them is spacelike at constant global time $X$. For $T E^{1/2}\gg 1$ this length of the ER bridge grows linearly in time
\begin{equation}
    \ell(E,T)=-2\log(2\varepsilon)-\log(4 E)+2E^{1/2}T\,.
\end{equation}
On the classical saddle this gives the usual exponential decay $\exp(-2\pi \Delta T /\beta)$ of two-sided correlation. Note for future reference that the (regularized) ER bridge length at $T=0$ is precisely $\ell(E)=-\log(4 E)$, with classically $E=\pi^2/\beta^2$. The Lorentzian geometry is cut off at $X=0$ and glued to the Euclidean preparation region of the TFD, which is half of the disk
\begin{equation}
    \d s^2=\d\rho^2+4E\sinh(\rho)^2\d \tau^2\,,\quad \tau\sim \tau+\beta\,,\quad \Phi=E^{1/2}\cosh(\rho)\,.
\end{equation}
One indeed checks the metric and dilaton glue smoothly, because at $X=0$ we have $\cosh(\rho)=1/\sin(\sigma)$.

Next we want to construct the Lorentzian $g=1$ geometry, which ought to be the real-time equivalent of the handled disk. To do so we will use two guidelines. First, we know that we can view the Euclidean handled disk as a wormhole \emph{with an extra identification} of a (geodesic) segment of the two boundaries of the wormhole. Second, comparing equations we see that the same physics underlies the ramp-and plateau structure of the spectral form factor and the two-point function. 

With Lorentzian boundaries, the wormhole is replaced by the double-cone \eqref{34} glued onto a gutter-shaped Euclidean preparation region \eqref{gutter}. The above logic then suggests that the handled disk should be replaced by the same double-cone but \emph{with an extra identification} between the two boundaries. We therefore are led to consider the following $g=1$ geometry
\begin{equation}
    \begin{tikzpicture}[baseline={([yshift=-.5ex]current bounding box.center)}, scale=0.7]
 \pgftext{\includegraphics[scale=1]{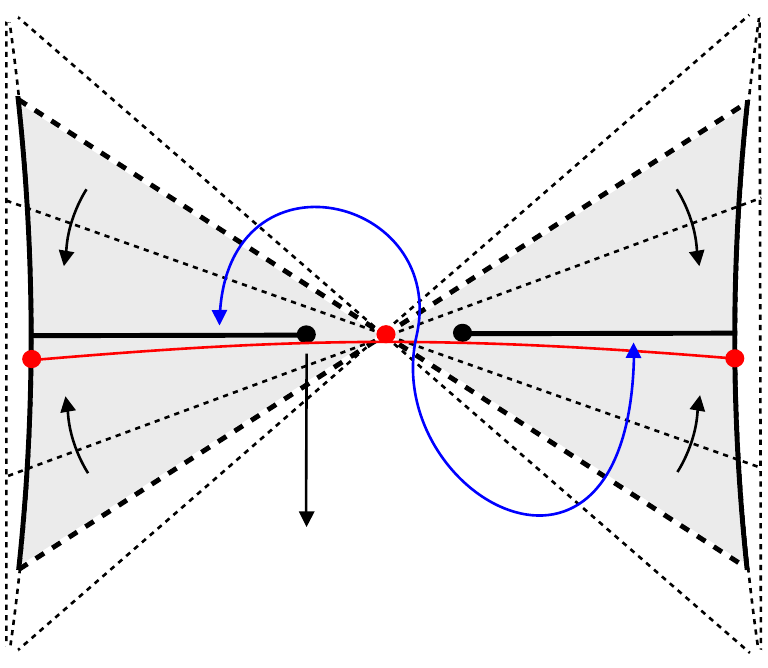}} at (0,0);
    \draw (0, -4) node {$g=1$ geometry};
    \draw (5,2) node {ket $R$};
    \draw (5,0.8) node {time flow};
    \draw (-0.8,-2.5) node {crotch};
    \draw (5,-2) node {ket $L$};
    \draw (4.4,-0.3) node {\color{red}$\mo_L$};
    \draw (-5,2) node {bra $L$};
    \draw (-5,-2) node {bra $R$};
    \draw (-4.4,-0.3) node {\color{red}$\mo_R$};
    \draw (0, 1.7) node {\color{blue}identify};
    \end{tikzpicture}\label{pic2}
\end{equation}
Here the diagonals are glued to the Euclidean gutter-shaped preparation region \eqref{gutter}, just like for the double cone. The only difference with the double cone is that we've inserted one additional crotch, the branchcut of which now extends out to the asymptotic boundary. Classically, as always, the crotch will be located close to the would-be horizon (the red dot). The swap identifications at the slit implement correctly the gluing of the left wedge in the ket to the left wedge in the bra, and vice versa for the right wedges. When one then inserts the operators $\mo_R$ and $\mo_L$ near the location where the slits intersect the asymptotic boundaries, we see that this geometry indeed satisfies the boundary conditions implemented by the two-sided two-point function \eqref{bc}.

In this geometry, one should think about a fixed bulk time slice $t=t_1$ in the context of the two-point function as the two slices $t=+t_1$ and $t=-t_1$ in the double cone \eqref{doublecone}, these combine into a ``cross'' in \eqref{pic2}. Bulk time flow corresponds to that cross becoming sharper, as indicated in \eqref{pic2} by the little black arrows. One salient feature of this geometry is the fact that all these time slices are identical, in particular the distance between the left-and right boundaries is now a constant independent of time, which is precisely equal to the length of the ER bridge $\ell(E)=-\log(4E)$ in the TFD at $t=0$. So, this $g=1$ geometry has created a shortcut between the two asymptotic boundaries \cite{Saad:2019pqd}. The semiclassical evaluation of the probe matter correlator on this geometry thus simply gives
\begin{equation}
    \average{e^{-\Delta\ell}}\approx e^{-\Delta \ell(E)}\,,\quad \ell(E)=-\log(4E)\,.
\end{equation}
The on-shell action of this geometry is identical to that of the double-cone, except for an extra factor $e^{-S(E)}$. This factor is the standard contribution \eqref{crotchweight} of the crotch, which sits close to the (would-be) horizon. Since we have the same metric and dilaton as for the double-cone away from the crotch, the remainder of the on-shell action can indeed be copied. 

To make it more obvious that this construction is the Lorentzian analogue of the two-point function on a handled disk, we can choose to draw $\mo_L$ on the other side of the identification. This makes it look as if we make the identification \emph{on} the particle trajectory (red)\footnote{One can check that the red trajectory behaves semiclassically as a geodesic, this is the dominant contribution to a single particle path integral within some topological class and the action equals its length $\ell(E)$ times the mass $\Delta$ of the particle. See the end of appendix \ref{app:determinants}.}
\begin{equation}
    \begin{tikzpicture}[baseline={([yshift=-.5ex]current bounding box.center)}, scale=0.7]
 \pgftext{\includegraphics[scale=1]{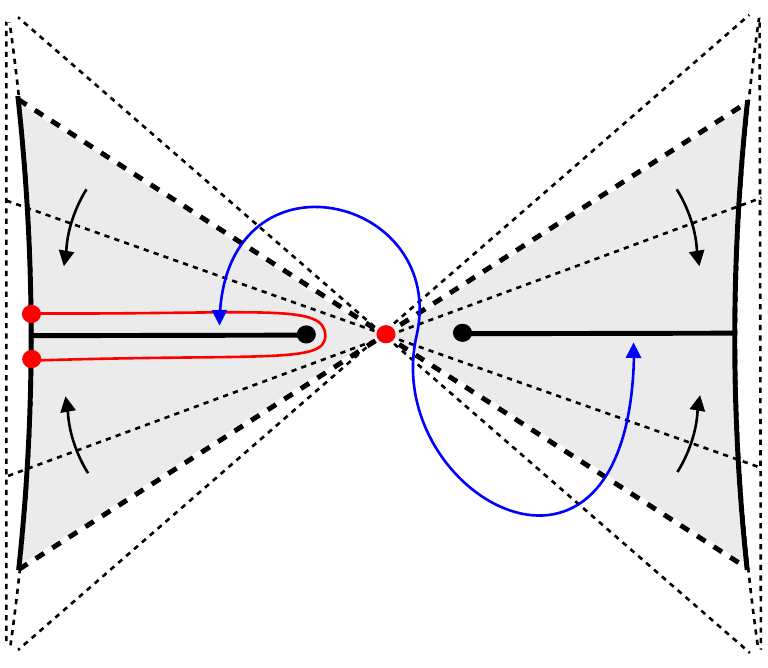}} at (0,0);
    \draw (5,2) node {ket $R$};
    \draw (5,0.8) node {time flow};
    \draw (5,-2) node {ket $L$};
    \draw (-4.4,0.3) node {\color{red}$\mo_L$};
    \draw (-5,2) node {bra $L$};
    \draw (-5,-2) node {bra $R$};
    \draw (-4.4,-0.3) node {\color{red}$\mo_R$};
    \draw (0, 1.7) node {\color{blue}identify};
  \end{tikzpicture}\label{pic3}
\end{equation}
This is the generalization of the Euclidean handled disk to a geometry that is ``as Lorentzian as possible''. Indeed if we cut the handled disk on the particle trajectory, we get a geometry that looks like a wormhole with two boundaries, where on each boundary there is an asymptotic segment and a geodesic segment, with a $\pi/2$ angle (in the no-backreaction probe limit) between the two sections (as we have here too).

Another way to visualize this spacetime, which makes it more clear what has changed as compared to the $g=0$ tent geometry \eqref{pic1} is as follows
\begin{equation}
    \begin{tikzpicture}[baseline={([yshift=-.5ex]current bounding box.center)}, scale=0.7]
 \pgftext{\includegraphics[scale=1]{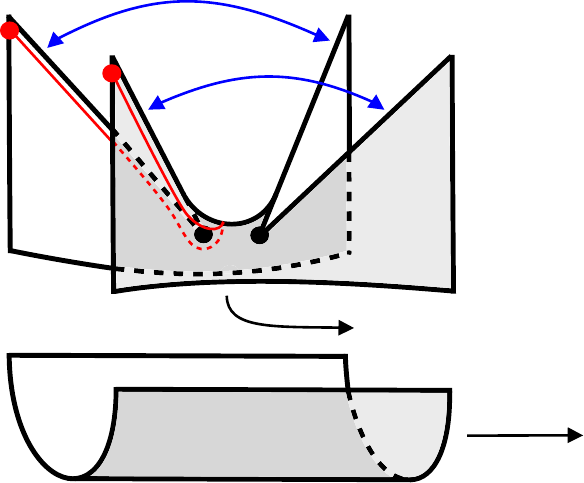}} at (0,0);
    \draw (2.4,0.5) node {ket};
    \draw (2.7,-0.9) node {tiny wormhole};
    \draw (4,-2) node {gutter};
    \draw (-3.5,2.2) node {\color{red}$\mo_L$};
    \draw (-3.7,0.5) node {bra};
    \draw (-1.2,2) node {\color{red}$\mo_R$};
    \draw (-1, 2.9) node {\color{blue}identify};
    \draw (0,-3.5) node {$g=1$ geometry};
  \end{tikzpicture}\label{picnew}
\end{equation}
Here we've over-exaggerated to size of the tiny Euclidean wormhole in the double cone at $\rho=0$, and we have included the Euclidean gutter-shaped preparation region (to which one should glue the Lorentzian spacetime). The length of the (red) particle trajectory does not grow with time $T$. When we make the blue identification we obtain topologically indeed a ($g=1$) handled disk (with one boundary).

So a portion of double-cone time evolution can be used to steal length of the ER bridge.

Just like the double cone, this geometry has two zero modes. Those are related with the fact that both in the bra-and the ket on the right hand side of \eqref{bc}, we can choose to redistribute the amount of time evolution with $H_L$ and with $H_R$, as long as the total remains $\pm T$. In the $g=0$ geometry \eqref{pic1} both these modes are redundancies, as they are Rindler boosts in either the bra-or ket two-sided black hole. On the double cone, only the mode associated with $t$ translation in \eqref{34} is redundant \cite{Saad:2018bqo}, which in \eqref{pic3} would move the left crotch down and the right crotch up. The mode associated with moving both crotches up simultaneously is \emph{physical}, and labels new solutions \cite{Saad:2018bqo}.\footnote{For instance in \eqref{327} this also corresponds with moving all crotches up simultaneously, or rotating the right side relative to the left side.}

In summary, we have a two-dimensional phase space of classical solutions. The twists give a volume factor proportional to $T$ and the energy integral is weighed by $e^{-S(E)}e^{-\beta E}$. Thus we recover the $g=1$ version of the boundary prediction \eqref{twopointgoal}
\begin{equation}
     \Tr(\mo\, e^{-(\beta/2+\i T)H}\mo\, e^{-(\beta/2-\i T)H})_1\sim T\int_{\L_1}^\infty \d E\,e^{-\beta E}\, e^{-\Delta \ell(E)}\,e^{-S(E)}\,.
\end{equation}

At this point the higher genus generalization should be quite obvious. We can introduce mirror slits which either connect the right bra to the left ket, or the left bra to the right ket. For instance, at $g=2$ one contributing geometry is
\begin{equation}
    \begin{tikzpicture}[baseline={([yshift=-.5ex]current bounding box.center)}, scale=0.7]
 \pgftext{\includegraphics[scale=1]{twopoint4.pdf}} at (0,0);
    \draw (0, -3) node {$g=2$ geometry};
    \draw (5,2) node {ket $R$};
    \draw (5,-2) node {ket $L$};
    \draw (4.4,-0.3) node {\color{red}$\mo_L$};
    \draw (-5,2) node {bra $L$};
    \draw (-5,-2) node {bra $R$};
    \draw (-4.4,-0.3) node {\color{red}$\mo_R$};
    \draw (0, 1.7) node {\color{blue}identify};
  \end{tikzpicture}\label{pic4}
\end{equation}
The crotches still cling to the horizon \eqref{322}, giving them an action $e^{-S(E)}$, and their temporal locations are still zero modes. Thus we almost trivially recover the prediction \eqref{twopointgoal}
\begin{equation}
     \Tr(\mo\, e^{-(\beta/2+\i T)H}\mo\, e^{-(\beta/2-\i T)H})_g\sim T^{2g-1}\int_{\L_g}^\infty \d E\, e^{-\beta E}\, e^{-\Delta \ell(E)}\,e^{-(2g-1)S(E)}\,.
\end{equation}

We remark that we can also think about fixed time slices of the double-cone as fixed time slices of an eternal traversable wormhole \cite{Maldacena:2018lmt}. We furthermore note that for $g>1$ there are configurations in \eqref{pic2} where one of the slits associated with the extra crotches intersects with the slit that ends on the asymptotic boundary, as shown here:
\begin{equation}
    \begin{tikzpicture}[baseline={([yshift=-.5ex]current bounding box.center)}, scale=0.7]
 \pgftext{\includegraphics[scale=1]{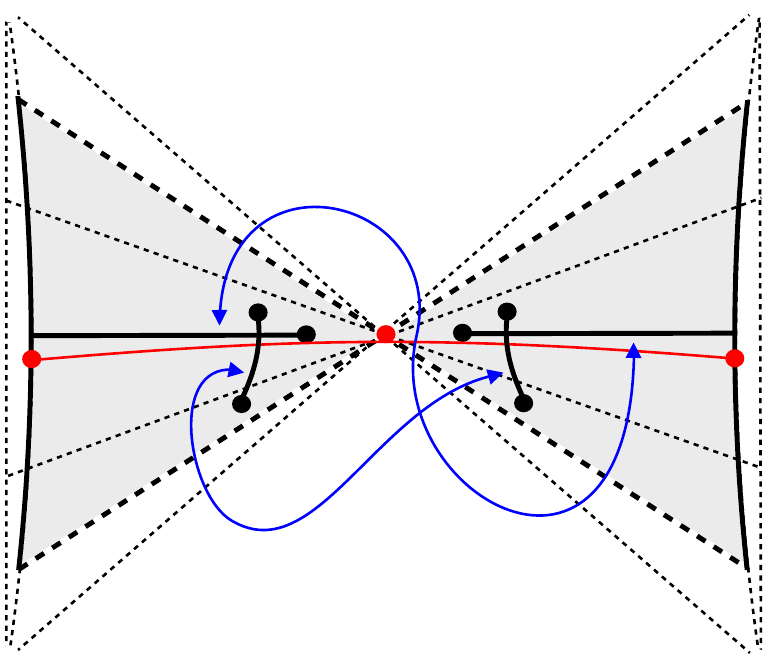}} at (0,0
    \draw (0, -3) node {$g=2$ geometry};
    \draw (5,2) node {ket $R$};
    \draw (5,-2) node {ket $L$};
    \draw (4.4,-0.3) node {\color{red}$\mo_L$};
    \draw (-5,2) node {bra $L$};
    \draw (-5,-2) node {bra $R$};
    \draw (-4.4,-0.3) node {\color{red}$\mo_R$};
    \draw (0, 1.7) node {\color{blue}identify};
  \end{tikzpicture}\label{pic5}
\end{equation}
One checks that this indeed also makes a smooth higher genus geometry. So there are geometries where the probe particle goes through slits, and geometries where it does not. For $g>1$, precisely $1/2$ of the geometries have the particle going through some slit. One way to appreciate this is that when we draw $2g$ crotches on the double cone with a certain time ordering, we are supposed to connect subsequent crotches by slits. But because of the (approximate for finite $\beta$) rotation symmetry, this means that for any choice of $2g$ crotch times $t_i$ we have two options; namely we first choose one crotch and connect it to either the next, or to the previous crotch. This fixes how to connect the others. See footnote \ref{fn:34}.

It would be interesting to understand what the physical implications are of the particle going through slits or not, naively it appears to be quite harmless, but dynamical matter might not behave as nicely around the crotches.
\section{Gravitational matrix elements}\label{sect:pagecurve}
As a third application of these slit geometries for Lorentzian physics, let us consider West-Coast replica wormholes \cite{Penington:2019kki}. In that paper they consider the following entangled state as function of $k$
\begin{equation}
    \ket{\Psi}=\sum_{i=1}^k \ket{i}_\text{grav}\otimes \ket{i}_\text{ref}\,,
\end{equation}
where $\ket{i}_\text{grav}$ is a state associated with a black hole with EOW brane of mass $\Delta$ and flavor $i$ behind the horizon, with a Euclidean preparation time $\beta/2$ which fixes the temperature of the black hole. More in particular, they consider JT gravity with action
\begin{equation}
    \exp\bigg(\Ss\chi+\frac{1}{2}\int\d^2 x\sqrt{g}\,\Phi(R+2)+\int_\text{asym}\d u\sqrt{h}\,\Phi (K-1)-\Delta\int_\text{brane}\d u\sqrt{h}\bigg)\,.\label{actioneow}
\end{equation}
The real-time EOW brane geometry associated with a pure state $\ket{i}_\text{grav}\bra{i}$ was described in detail in \cite{Gao:2021tzr}
\begin{equation}
    \rho_\text{grav}=\ket{i}_\text{grav}\bra{i}\quad \Leftrightarrow\quad \begin{tikzpicture}[baseline={([yshift=-.5ex]current bounding box.center)}, scale=0.7]
 \pgftext{\includegraphics[scale=1]{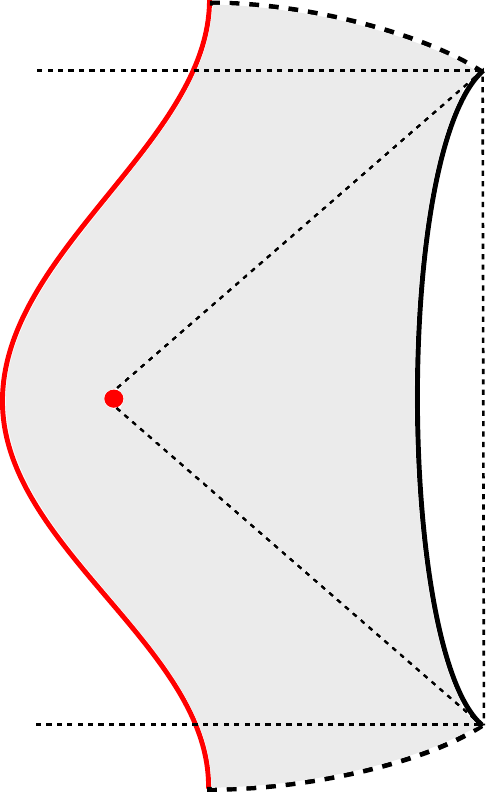}} at (0,0);
    \draw (1, 4.5) node {$t=+\infty$};
    \draw (1, -4.5) node {$t=-\infty$};
    \draw (3.5, 0) node {$\sigma=\pi$};
    \draw (-1.3, -2.8) node {\color{red}$i$};
    \draw (-1.3, 2.8) node {\color{red}$i$};
    \draw (-2.5, -3.8) node {$T=-\pi/2$};
    \draw (-2.5, 3.8) node {$T=+\pi/2$};
  \end{tikzpicture}\label{realgeom}
\end{equation}
Here the metric and dilaton are the same as in \eqref{geom}. The EOW brane trajectory is known explicitly
\begin{equation}
    \frac{\cos(\sigma)}{\cos(T)}=\frac{\Delta}{(\Delta^2+E)^{1/2}}\,,
\end{equation}
but in the regime where this EOW matter particle is just a probe $\Delta\ll E^{1/2}$, this simplifies to $\sigma=\pi/2$.

To understand a bit better how this Lorentzian geometry is related with the Euclidean preparation we can think about computing some real-time observable in the pure state \eqref{realgeom}
\begin{equation}
    \Tr(\rho_\text{grav}\,\mo)=\bra{i}\mo\ket{i}_\text{grav}
\end{equation}
We can think of this path integral as consisting of three parts. The first piece is a Euclidean preparation followed by back-in-time evolution to $t=-\infty$, which we associate with the ket. Then there is the real-time geometry \eqref{realgeom} which goes from $t=-\infty$ to $t=+\infty$ and in which we can carry out all experiments and measurements, in this case measuring $\mo$. Finally in order to implement the bra (or final state) there is back-in-time evolution from $t=+\infty$ to zero, where one glues onto the second Euclidean preparation region. For the purposes of this discussion we want to make the \emph{gauge choice} that no topology change can occur in the preparation regions, or conversely that all of the topology change is restricted to occur in the real-time geometry \eqref{realgeom}, which interpolates between some initial and some final configurations at $t=-\infty$ respectively $t=+\infty$.

The point of \cite{Penington:2019kki} is that, despite appearances, states $\ket{i}_\text{grav}$ and $\ket{j}_\text{grav}$ with $i\neq j$ are not orthogonal (or not independent). In realistic theories of quantum gravity (which are not dual to ensembles but to individual systems \cite{Blommaert:2021fob,Blommaert:2021gha,Saad:2021rcu}) this means that there are dynamical processes happening in the real-time geometry \eqref{realgeom} which change the flavor of the brane behind the horizon \cite{Blommaert:2021etf}, such that $\braket{i}{j}_\text{grav}\neq 0$.

The model \eqref{actioneow} is too crude to directly see such non-zero overlaps.\footnote{In the language of this paper, the slit saddle-point geometries that we are introducing always require (at least) a second identical copy of the geometry to swap-identify with. One can evolve from an initial state $\ket{i}\otimes \ket{j}$ to the final state $\ket{j}\otimes \ket{i}$ by swapping interiors, but not from $\ket{i}$ to $\ket{j}$, since the total charge is not conserved. Dynamical branes as in \cite{Gao:2021tzr} are not sufficient to resolve this, because brane nucleation creates a pair with total charge zero.} Instead, this model of gravity is an ensemble of theories, and it realizes the next best thing, namely the expectation value of off-diagonal matrix elements vanishes $\braket{i}{j}_\text{grav}=0$ but their variance does not $\braket{i}{j}_\text{grav}^2\neq 0$. In the Euclidean setup, this variance (and higher moments) is explained because of replica wormholes. Here we want to show how slit geometries accomplish the same feat from a direct Lorentzian point of view.\footnote{The contents of this section should be read as a \emph{minor} improvement on (and application of) earlier discussions in \cite{Penington:2019kki,Colin-Ellerin:2020mva,Colin-Ellerin:2021jev,Marolf:2020rpm,Marolf:2021ghr,Maxfield:2022sio}.}

When we compute the variance $\braket{i}{j}_\text{grav}^2$ and we introduce (as we are used to by now) mildly singular points in the JT gravity path integral by inserting the identity \eqref{1equals}, we can find an exact solution to
\begin{equation}
        \sqrt{g}(R+2)=-4\pi\,\delta(x-x_{\text{sing}})\,,
\end{equation}
by inserting just one crotch on two copies of the geometry \eqref{realgeom}, as follows:
\begin{equation}
    \bra{i}\ket{j}_\text{grav}\bra{j}\ket{i}_\text{grav}\supset \quad \begin{tikzpicture}[baseline={([yshift=-.5ex]current bounding box.center)}, scale=0.7]
 \pgftext{\includegraphics[scale=1]{replica1.pdf}} at (0,0);
    \draw (1, -3.1) node {1. slice};
    \draw (0.5, 2.1) node {\color{blue}2. identify};
    \draw (-4.45, 3.5) node {\color{red}$i$};
    \draw (-4.45, -3.5) node {\color{red}$j$};
    \draw (2.7, -3.5) node {\color{red}$i$};
    \draw (2.7, 3.5) node {\color{red}$j$};
  \end{tikzpicture}\label{swapbis}
\end{equation}
Here we can make the double-cover slice on a geodesic that starts on the crotch and ends (for instance) orthogonally on the EOW brane, effectively splitting the EOW brane in half. If we make the same slice on two identical copies of this geometry \eqref{geom} (which notably enforces the ADM energies to be identical $E_1=E_2$ for both copies!) then the configuration where we make swap-identifications as shown in \eqref{swapbis} remains a smooth solution with metric and dilaton exactly \eqref{geom}, except for the singular source that we see very close to the crotch.

To compute the classical actions of these configurations, we first note that (because aside from the crotches nothing changes) the action will be twice the on-shell action (at fixed energy $E$) for one matrix element, augmented with the by-now standard contribution from the crotch(es)
\begin{equation}
    e^{-\Ss}\int\d x_\text{sing}\sqrt{g}\,e^{-2\pi \Phi(x_\text{sing})}\,,\quad \Phi(x_\text{sing})=E^{1/2}\,\frac{\cos(T_\text{sing})}{\sin(\sigma_\text{sing})}\,.
\end{equation}
The saddle-point equations that result from this have in this case a unique solution, the crotch(es) are located at the black hole horizon\footnote{One could argue that there is a zero mode in the location with volume proportional to $\beta$, but we are not trying to track such subleading factors here. Also notice that the $\sqrt{g} = 1/\sin^2\s$ does not change the saddle point nor its value.}
\begin{equation}
    \frac{\d}{\d x_\text{sing}}\Phi(x_\text{sing})=0\quad\Leftrightarrow\quad T_\text{sing}=0\,,\quad\sigma_\text{sing}=\frac{\pi}{2}\,.
\end{equation}
Therefore the spacelike region being swapped is the entire black hole interior. It is of course no accident that this region is identical to the \emph{island} \cite{Penington:2019npb,Almheiri:2019hni,Almheiri:2020cfm} in this scenario \cite{Penington:2019kki,Marolf:2020rpm,Marolf:2021ghr,Maxfield:2022sio}. Because the crotch is located near the black hole horizon, the suppression of this single-slit geometry is
\begin{equation}
    e^{-S(E)}.
\end{equation}
As we will see this suppression matches on the nose with the semiclassical limit of the Euclidean replica wormhole amplitudes in \cite{Penington:2019kki}.

Before proceeding we note that we should of course also allow configurations with multiple crotches, creating a series of swaps. In the case of the variance $\braket{i}{j}_\text{grav}^2$, an even number of crotches results in a new ``diagonal'' contribution, and an odd number of crotches results in an ``off-diagonal'' contribution (with one net swap). These corrections correspond with higher genus configurations (involving handles) and are subleading in the regime discussed in \cite{Penington:2019kki}, but nonetheless they might still be important.

When we compute higher moments of matrix elements, other saddles appear which are identical in construction as \eqref{swapbis}, except that they involve crotches which are $n$-fold covers \eqref{coversing}. Nothing much changes except that these geometries have all energies equal $E_1=E_2=\dots=E_n$ and that the $n$-cover crotches end up having an on-shell action
\begin{equation}
    e^{-(n-1)S(E)}\,.
\end{equation}
These $n$-cover crotches can also be viewed as a combination of $(n-1)$ standard double-cover crotches, the distinction is likely semantics.

The computation of the on-shell action is largely a Euclidean exercise, since (asides from the crotch contributions) the Lorentzian pieces cancel. We present it as an application of fixed area states \cite{Dong:2022ilf,Dong:2018seb}. The combination of the two Euclidean preparation regions is half of a disk with metric
\begin{equation}
    \d s^2=\d\rho^2-4E\sinh(\rho)^2\d \tau^2\,,\quad \Phi=E^{1/2}\cosh(\rho)\,.
\end{equation}
Here $\tau$ runs from $0$ to $\beta$ on half of the disk. The half disk is cut off at the equator by a particle following a geodesic trajectory of length $\ell(E)=-\log(4E)-2\log(2\varepsilon)$, which intersects the boundary with a $\pi/2$ angle. Because of those straight angles, Gauss-Bonnet for this geometry reads
\begin{equation*}
    \frac{1}{2}\int\d^2 x\sqrt{g}\, R+\int_\text{asym}\d u\sqrt{h}\, K=\pi\,.
\end{equation*}
Calculating the contributions due to the smooth part of the metric (as for the black hole in section \ref{sect:bh}) we find that this geometry has a conical singularity which is roughly half as strong as that for the full disk in \eqref{RicciDisk}
\begin{equation*}
    \sqrt{g}R\supset (2\pi-4 E^{1/2}\beta )\,\delta(x).
\end{equation*}
Combining the elements this leads to the following on-shell action for each bra-ket combination
\begin{equation*}
    e^{\pi E^{1/2}}\,e^{-\Delta \ell(E)}\,e^{-\beta E}\,,
\end{equation*}
with \emph{half} of the entropy appearing, because we had half of the defect. In total this gives for the $n$-replica geometry
\begin{equation}
    Z_n\sim e^{\Ss}\int_{\L_g}^\infty \d E\, e^{(1-n)\pi E^{1/2}}\,e^{-n\,\Delta \ell(E)}\,e^{-n\beta E}\,,\label{516}
\end{equation}
which reproduces the exact Euclidean results of \cite{Penington:2019npb} after using Stirling on the Gamma functions. 
\section{Concluding remarks}\label{sect:disc}
Combining the results of section \ref{sect:plateau} and \ref{sect:twopoint} with \eqref{516} gives us confidence that we have identified correctly the moduli space of semiclassical Lorentzian wormhole geometries, which was the main goal of our work. By itself it is not per se that interesting \emph{that} one can reproduce the Euclidean answers by counting only Lorentzian geometries, rather we believe the interesting question is \emph{how}. These comparisons, also with the boundary predictions, are meant to teach us what the rules are for real-time gravity. Our evidence suggest that this ``moduli space of slit geometries'' is part of those (universal) rules.

It is noteworthy, and perhaps surprising, that this space does not include any bona-fide closed baby universes. One reason to have expected this is that closed AdS universes crunch in real time, as we saw in section \ref{sect:2.2}. This being said, we have not ruled out the possibility to have contributions from closed universes to the path integral when considering $\alpha\neq 0$ in \eqref{1equals} \cite{Usatyuk:2022afj}. However, if they exist, it is likely as off-shell contributions, over which we have currently little control in general gravity models anyway. Topologically of course, our slit geometries \emph{do} have closed cycles, but in this gauge choice they are \emph{not} (classical) closed universes propagating in time, detaching from (and attaching to) parents \cite{coleman1988black,giddings1988loss}.

To end this work we list some (potentially) doable, interesting open problems.

\begin{enumerate}
    \item The slit geometries seem to be quite efficient in reproducing ``complicated'' boundary predictions. It would be nice to now use them to make \emph{new} predictions. For instance, one natural application would be to try and predict the firewall probability in the setup of \cite{Stanford:2022fdt} at higher genus, and sum that series. There would now be contributions from geometries which look like \eqref{pic3}, but where the double-cone is evolved for a time $T_1$, at which point two $t=0$ TFD geometries are glued into the crotches (after opening them up), and evolved for a time $T_2$ (with $T_1+T_2=T$), before gluing the TFD in the bra to the TFD in the ket. The length now takes the value $\ell(E,T_2)$,\footnote{These geometries do not contribute to the two-point function in the double scaling limit, because the on shell action of the matter probe $e^{-\Delta\ell(E,T_2)}\to 0$ when $T_2\to\infty$. It seems natural to gauge fix the interior length to be the \emph{unique} purely spatial bulk geodesic in this case. What this choice corresponds to in the Euclidean setup is unclear, but by construction it is a physical observable, since we are working in a gauge-fixed setting and within that setting this observable is well-defined.} hence the TFD can be rejuvenated by any amount of time by wormholes (as in \cite{Stanford:2022fdt}).
    \item In similar spirit it might be valuable to revisit bulk reconstruction with these geometries in mind. One approach would be to use the light-cone construction of \cite{Blommaert:2019hjr}, which seems natural for these spacetimes. Or even more low-brow, one could attempt to compute a semiclassical approximation to the infalling two-point function at all genus, and (attempt to) sum that series. One could also ask from the more algebraic point of view how QFTs would behave around such peculiar causal structures.
    \item Throughout the text we mentioned that we reproduced the boundary prediction. Strictly speaking we focused on the semiclassical positive area solutions. However, we know (see for instance \eqref{a4}) that for the spectral form factor and the two-point function semiclassically (i.e. high energy) the coefficient of $T^{2g+1}$ for $g > 0$ should vanish.\footnote{We thank Steve Shenker and Douglas Stanford for asking about this.}
    
    It is plausible that one can account for this cancellation by including the semiclassical solution at $A < 0$. To see this note that when going from the $A$ integral to an energy integral, one goes from a contour on the real axis to one that wraps around the positive real (energy) axis. Semi-classics on the contour above the axis gives the results quoted in the main text and following the same analysis, the contour below the real axis should give (for JT)
    \be 
    \sim T\int^{-\sqrt{\L_g}}_{-\infty} A \d A\, e^{-2\b A^2} \left( T e^{+2\pi A-\Ss} \right)^{2g} = - T^{2g+1} \int_{\L_g}^{+\infty} \d E e^{-2\b E} e^{-2g S(E)}
    \ee
    and indeed cancels with the positive area contribution.
    
    However, a more detailed analysis which includes both branches of the spectral density, and one-loop corrections, should result in replacing $e^{-2g S(E)}$ with the full disk spectral density $\rho_{0}(E)^{-2g}$, as in \eqref{a4}. Whilst the contributions still cancel at high energies, we see that the contour in the complex $E$ plane needs to go around the $E = 0$ region and we get a non-zero contribution from the key-holed region in the contour integral similar to \cite{Saad:2022kfe}. It would be worthwhile to try to work all this out in detail. Notice that for the double cone $g = 0$, there are no constraint instantons and so this issue does not arise and we just get the universal answer.
    
    \item{It would be interesting to see the relation between the solution in this paper with the semiclassical encounter contribution in \cite{Saad:2022kfe}. In particular, it would be interesting to compare the boundary to boundary correlation pattern from both sides.}

    \item Finally it would be interesting to understand the first order formulation better, where one allows non-invertible vielbeins. By enlarging the gauge group of allowed diffeomorphisms one can think of such configurations as gauge-equivalent to non-singular configurations \cite{Horowitz:1991fr}.\footnote{This should be equivalent to the claim we want to make, that including all Lorentzian geometries with crotch singularities is gauge-equivalent to path integrating over all (mostly) Euclidean smooth spacetimes, see also \cite{Usatyuk:2022afj}.} In the tractable settings of the 3d Chern-Simons formulation of AdS$_3$ gravity (or of the 2d BF formulation of JT gravity) allowing crotches naively would seem to correspond to having contributions from other Euler classes (besides the maximal one) for the gauge connection. Could one gauge-fix to another set of connections that is \emph{not} the Teichmuller component \cite{Eberhardt:2022wlc,Blommaert:2018iqz,Mertens:2022ujr}? This approach would also be natural in defining a canonically quantized theory of gravity that allows for topology change.
\end{enumerate}
\section*{Acknowledgments}
We thank Luca Iliesiu, Don Marolf, Thomas Mertens, Mehrdad Mirbabayi, Pratik Rath, Ronak Soni, Douglas Stanford, Stephen Shenker, Mykhaylo Usatyuk and Ying Zhao  for useful discussions. AB was supported by the ERC-COG Grant NP-QFT No. 864583 and by INFN Iniziativa Specifica GAST, and thanks Stanford's SITP and Berkeley's BCTP where part of this work was completed. JK is supported by NSF grant PHY-2207584 and PHY-1748958 and would like to thank KITP where part of this work was completed.
\appendix


\section{Observables in the tau-scaling limit}\label{sect:recap}

Here we review and present a derivation of the $\tau$-scaling limit of the spectral form factor as was done in \cite{Okuyama:2018gfr} and expanded on in \cite{Saad:2022kfe,Okuyama:2020ncd,Blommaert:2022lbh,Weber:2022sov}. The basis idea is to take a Lorentzian observable, for instance by analytically continuing a Euclidean one, depending on a time $T$ and sending $T$ to infinity together with $e^{\Ss}$ while keeping the ratio $\tau = T/e^{\Ss}$ fixed.

\subsection{Spectral form factor}\label{app:sff}
In \cite{Saad:2022kfe,Okuyama:2018gfr,Okuyama:2020ncd,Blommaert:2022lbh} it was shown that in this limit the spectral form factor exactly reduces to the following folding integral
\bea
Z(\beta+iT,\beta-iT)= \int_0^{\infty} \d E\, e^{-2\beta E} \text{min} (T/2\pi,\rho_0(E))\,,
\eea
in which the min function is nothing but Fourier transform of the sine-kernel
\bea
\rho(E_1,E_2)=\delta(E_1-E_2)\rho_0(E)-\frac{\sin^2(\pi\rho_0(E)(E_1-E_2))}{\pi^2(E_1-E_2)^2}\,,
\eea
which is universal for all quantum chaotic system, when the energy $E_1$ and $E_2$ come close \cite{Haake:1315494}. Doing this folding integral predicts a genus expansion of the spectrum form factor with a non-zero radius of convergence, which reproduces the ramp-and plateau structure \cite{Saad:2022kfe,Blommaert:2022lbh}
\bea
Z(\beta+iT,\beta-i T)=\frac{T}{4\pi \beta}+\sum_{g=1}^{\infty} P_{g-1}(\beta)\, T^{2g+1}\,,
\eea
where the degree $g-1$ polynomial $P_{g-1}(\beta)$ depends on the spectral curve
\begin{equation}
    P_{g-1}(\beta)=-\frac{1}{(2\pi)^{2g+1}(2g)(2g+1)}\oint_R\d E \,\rho_0(E)^{-2g}\,e^{-2\beta E}\,.\label{a4}
\end{equation}
This contour around the real axis $R$ can be reduced to a circle around the origin. In the microcanonical ensemble, with energy high enough so the path integral is still semiclassically well controlled, all higher genus contribution cancel with each other, which is consistent with the fact that microcanonical spectral form factor has a perfect linear ramp \cite{Saad:2022kfe}. However, there are some contribution at low enough energy, where the path integral is not dominant by semiclassical contribution, that give a non-cancelling answer. Those contributions come into play when we integrate over energy.

We aim in section \ref{sect:2dplateau} to reproduce semiclassical features of \eqref{a4} using Lorentzian spacetimes. We will only partially succeed in this, in particular we will find the semiclassical (large $E$) approximation
\begin{equation}
    P_{g-1}(\beta)\sim \int_{\Lambda_g}^\infty \d E\, e^{-2g S(E)}\,e^{-2\beta E}\,.
\end{equation}
For order one or small energies $E$, the semiclassical approximation in gravity is no longer reliable, and one has to do the full path integral again.

\subsection{Two-point function}\label{app:twopoint}

In JT gravity the two point function takes the exact form \cite{Saad:2019pqd,Blommaert:2019hjr,Blommaert:2020seb,Iliesiu:2021ari}
\begin{align}
    &\Tr(\mo\, e^{-(\beta/2+\i T)H}\mo\, e^{-(\beta/2-\i T)H})\label{43}\\\nonumber &\qquad\qquad=\int_{-\infty}^{+\infty}\d E_1\, e^{-(\beta/2+\i T)E_1}\int_{-\infty}^{+\infty}\d E_2\, e^{-(\beta/2-\i T)E_2}\,\rho(E_1,E_2)\,e^{-\Ss}\,\frac{\Gamma(\Delta\pm \i E_1^{1/2}\pm\i E_2^{1/2})}{2^{2\D + 1}\Gamma(2\Delta)}\,,
\end{align}
where $\rho(E_1,E_2)$ is the spectral two-point function of some random matrix theory with a potential fine-tuned for JT gravity \cite{Saad:2019lba,mehta2004random}. In the Euclidean gravity calculation, this result is obtained by summing over wormhole geometries, and matching those results order per order with the random matrix theory expansion \cite{Saad:2019pqd,Blommaert:2019hjr,Blommaert:2020seb,Iliesiu:2021ari}. 

From the point of view of the dual quantum mechanics the shape of this equation follows from the fact that the theory is chaotic, this implies in some sense both random matrix statistics for the energies, as well as the ETH ansatz \cite{Srednicki:1994mfb,deutsch1991quantum} for operator matrix elements \cite{Saad:2019pqd,Foini:2018sdb,Blommaert:2020seb,Belin:2020hea,Belin:2020jxr}. The details require some input from gravity, namely the spectrum and the kernel for the two-point function \cite{Yang:2018gdb, Saad:2019pqd,Mertens:2017mtv,Mertens:2018fds,Kitaev:2018wpr,Blommaert:2018iqz,Iliesiu:2019xuh,Blommaert:2018oro}\footnote{We follow the convention of \cite{Mertens:2017mtv} for the signs in the Gamma functions, so this is a product of four Gamma functions.}
\begin{equation}
    \mo_{E_1 E_2}\mo_{E_2E_1}=e^{-\Ss}\,\frac{\Gamma(\Delta\pm \i E_1^{1/2}\pm\i E_2^{1/2})}{2^{2\D + 1}\Gamma(2\Delta)}\,.
\end{equation}
For generalized dilaton gravities this kernel is not known,\footnote{It was claimed in \cite{Iliesiu:2021ari} that this kernel is universal, but one can even see semiclasically that this gives the wrong answer. The technical reason is that they did not include contributions from the particle winding around the defects in the gas \cite{Maxfield:2020ale}.} but fortunately here we will only need its semiclassical approximation, and that we do know (see below). Once this kernel gets computed on disk level, the genus expansion will work similarly and one will recover \eqref{43}, but with a different $\rho_0(E)$ and operator kernel. This fact also follows from the boundary ETH prediction.

We want to study the $\tau$-scaling limit of \eqref{43}. The late time Fourier transform in \eqref{43} localizes on the least analytic features of $\rho(E_1,E_2)\,\mo_{E_1 E_2}\mo_{E_2E_1}$ as function of $\omega=E_1-E_2$ (we furthermore introduce $2E=E_1+E_2$). Those features are universal \cite{mehta2004random}, one ends up approximating $\rho(E_1,E_2)$ by the sine kernel
\begin{equation}
    \rho(E_1,E_2)_\text{eff}=\delta(E_1-E_2)\rho_0(E)-\frac{\sin^2(\pi\rho_0(E)(E_1-E_2))}{\pi^2(E_1-E_2)^2}\,.\label{sinekernel}\,,
\end{equation}
and simply evaluates the operator matrix elements on the stationary phase saddle $\omega=0$ to $\mo_{E E}\mo_{E E}$. Fourier transforming the sine kernel \cite{Cotler:2016fpe} one then arrives at the exact double scaled answer
\begin{equation}
    \Tr(\mo\, e^{-(\beta/2+\i T)H}\mo\, e^{-(\beta/2-\i T)H})=\int_{-\infty}^{+\infty}\d E\,e^{-\beta E}\,\text{min}(\rho_0(E),T/2\pi)\,e^{-\Ss}\,M_{\D}(E)\,.\label{46}
\end{equation}
with 
\begin{align}\label{DSMElem}
    M_{\D}(E) = \frac{\Gamma(\Delta)^2}{2^{2\D + 1}\Gamma(2\Delta)}\,\Gamma(\Delta\pm 2 \i E^{1/2})
\end{align}
the $\tau$-scaled matrix elements. One can also do the $\omega$ integral more rigorously without first approximating $\mo_{E_1E_2}\mo_{E_2E_1}$ by its saddle, using contour deformation. The pole at $\omega=\i\varepsilon$ of the sine kernel gives the above contribution, and one checks that the poles from the $\Gamma$ functions give contributions that decay in time, which thus indeed do not contribute in this double scaling limit where $T\to\infty$.\footnote{Furthermore, in Euclidean gravity, contributions from the disk topology or from cases where we have handles on either side of the worldline of the particle (but where the handle does not bridge over the particle) can also be checked to decay in time, as power laws, see for instance \cite{Bagrets:2017pwq,Mertens:2017mtv,Blommaert:2019wfy}. Only connected topologies, where wormholes connect both sides of the particle's worldline (analogous to the Euclidean geometries contributing to the spectral form factor) survive. This is why we end up with the same kernel $\text{min}(\rho_0(E),T/2\pi)$ in this double scaling limit. To be clear, the contributions that survive are exclusively those of the third type in figure 17 of \cite{Blommaert:2019hjr}, for a relation with the SFF geometries see figure 19 in \cite{Blommaert:2019hjr}.} 

Thus, at fixed energy, we have a sharp ramp-to-plateau transition for the two-point function as well. But, just like for the spectral form factor \cite{Saad:2022kfe,Blommaert:2022lbh,Weber:2022sov} this sharp transition is smoothed out in the canonical ensemble, and we obtain a convergent genus expansion in $Te^{-\S}$. To find that expansion, one can start by modifying the steps between (2.5) and (2.6) in \cite{Blommaert:2022lbh}
\begin{equation}\label{A11}
     \Tr(\mo\, e^{-(\beta/2+\i T)H}\mo\, e^{-(\beta/2-\i T)H})= \int_0^{T/2\pi}\d \rho_0\int_{E(\rho_0)}^\infty \d E\, \,e^{-\beta E}\,\,e^{-\Ss}\,M_{\D}(E)\,.
\end{equation}
Taylor series of functions can be computed by contour integrals around the origin, therefore we obtain
\begin{align}
   \nonumber&\int_{E(\rho_0)}^\infty \d E\, \,e^{-\beta E}\,\,e^{-\Ss}\,M_{\D}(E)\\\nonumber&\qquad=\sum_{n=0}\rho_0^{2n}\,\frac{1}{2\pi\i }\oint_0\frac{\d \rho_0}{\rho_0^{2n+1}}\,\int_{E(\rho_0)}^\infty \d E\, \,e^{-\beta E}\,\,e^{-\Ss}\,M_{\D}(E)\\&\qquad= Q_{\D}(\b) -\sum_{n=0}\rho_0^{2n}\,\frac{1}{2n}\frac{1}{2\pi\i }\oint_0\frac{\d E}{\rho_0(E)^{2n}}\,e^{-\beta E}\,\,e^{-\Ss}\,M_{\D}(E)\,,
\end{align}
where in the first step we used the fact that this function has only even Taylor coefficients in $\rho_0$ and in the second step we used integration by parts and defined the $n=0$ term as
\be 
Q_\D(\b) = \int_{E_0}^{\infty} \d E\, e^{-\b E} e^{-\Ss} M_{\D}(E)
\ee
The $\rho_0$ integrals in \eqref{A11} result in the expansion 
\begin{align}\label{TrOOMM}
    \Tr(\mo\, e^{-(\beta/2+\i T)H}\mo\, e^{-(\beta/2-\i T)H})= \frac{T}{2\pi}Q_{\D}(\b) -\sum_{g>1}^\infty\frac{(T/2\pi)^{2g-1}}{(2g-2)(2g-1)}\oint_0\frac{\d E}{2\pi \i}\, e^{-\beta E}\, \frac{M_{\D}(E)}{\rho_0(E)^{2g-2}e^{\Ss}}\,.
\end{align}

Let us now see that this formula agrees with the gravitational calculation in the $\tau$-scaling limit and we can indeed interpret $g$ as the genus. As mentioned above we can focus on geometries where the wormhole bridges over the particle's worldline. Using the formulas from \cite{Saad:2019pqd}, we can write the genus $g$ answer as
\begin{align}\label{TrOOgrav}
    \Tr(\mo\, e^{-(\beta/2+\i T)H}\mo\, e^{-(\beta/2-\i T)H})_g &= e^{-(2g-1)\Ss} \int_0^{\infty} b_1 \d b_1 b_2 \d b_2 \Vol_{g-1,2}(b_1,b_2) \nonumber\\
    &\quad \times \int_{-\infty}^{\infty} \d \ell e^{\ell} \psi_{{\rm Tr}, \b/2 + \i T}(b_1, \ell)\psi_{{\rm Tr}, \b/2 - \i T}(b_2, \ell) e^{-\D \ell} 
\end{align}
It is convenient to write the wavefunctions $\psi_{\rm Tr}$ as inverse laplace transforms of the trumpet partition function,
\begin{align}
    \psi_{{\rm Tr}, \b/2 - \i T}(b_1, \ell) = \int_0^{\infty} \d E e^{-(\b/2 - \i T)E} \psi_E(\ell) \int_{-\i\infty}^{\i \infty} \d \b e^{\b E} Z_{\rm Tr}(\b, b_1)
\end{align}
Plugging this back into \eqref{TrOOgrav} we see that we have two integrals over $b_1$ and $b_2$, two integrals over auxiliary temperatures (lets call them $\b_1$ and $\b_2$), two energy integrals $E_1$ and $E_2$ and an $\ell$ integral. The $\ell$ integral gives the usual factor of gamma functions and since we are interested in large $T$ these should be evaluated at coincident energies $E_1 = E_2$. For the remaining integrals we see that the $b_i$ integrals together with the volumes and the trumpet partition functions give the two boundary partition function $Z_{g-1}(\b_1, \b_2)$ at genus $g-1$,
\begin{align}
     \Tr(\mo\, e^{-(\beta/2+\i T)H}\mo\, e^{-(\beta/2-\i T)H})_g &= e^{-(2g - 1)\Ss}\int_0^{\infty} \d E e^{-\b E} \frac{\Gamma(\D)^2 \G(\D \pm 2\i E^{1/2})}{2^{2\D + 1}\G(2\D)} \nonumber\\
     &\quad \times \frac{1}{2}\int_{-\infty}^{\infty} \d \w \, \int_{-\i \infty}^{\i \infty} \frac{\d \tilde{\b}}{2\pi \i} \int_{-\infty}^{\infty} \frac{\d \tilde{T}}{2\pi} e^{\tilde{\b}E + \i (\tilde{T} - T)\w } Z_{g}(\tilde{\b} + \tilde{T}, \tilde{\b} - \tilde{T})\, ,
\end{align}
where we went to coordinates $2\b_i = \tilde{\b} \pm \tilde{T}$. The integral over $\w$ gives a delta function setting $\tilde{T}$ equal to $T$ and so we arrive at,
\begin{align}
\Tr(\mo\, e^{-(\beta/2+\i T)H}\mo\, e^{-(\beta/2-\i T)H})_g &= \frac{1}{2}\int_0^{\infty} \d E e^{-\b E} \frac{\Gamma(\D)^2 \G(\D \pm 2\i E^{1/2})}{2^{2\D + 1}\G(2\D)}\int_{-\i \infty}^{\i \infty} \frac{\d \tilde{\b}}{2\pi \i} e^{\tilde{\b}E} Z_g(\tilde{\b} + T, \tilde{\b} - T)
\end{align}
For the final integral over $\tilde{\b}$ we can use the expression in the $\tau$-scaling limit as given in D.2 in \cite{Saad:2022kfe}, which boils the genus $g$ contribution to the two point function in the $\tau$-scaling limit down to,
\begin{align}
    \Tr(\mo\, e^{-(\beta/2+\i T)H}\mo\, e^{-(\beta/2-\i T)H})_g = -\frac{(T/2\pi)^{2g -1}}{(2g-2) (2g - 1)} \oint_0 \frac{\d E}{2\pi \i} \frac{e^{-\b E}}{\rho_0(E)^{2g-2}e^{\Ss}} \frac{\Gamma(\D)^2 \G(\D \pm 2\i E^{1/2})}{2^{2\D + 1}\G(2\D)}
\end{align}
This indeed matches exactly with \eqref{TrOOMM} and justifies the replacement of $n$ with the geometric quantity $g-1$.

\section{Lightcone gauge calculations in JT gravity}\label{app:lightcone}
In this appendix, we consider the light-cone gauge formulation of JT gravity \cite{Usatyuk:2022afj} for the geometries of section \ref{sect:plateau}, explaining detailed procedures and discussing some possible subtleties.\footnote{As compared to \cite{Usatyuk:2022afj} we add several new ingredients, notably the application to spacetimes with holographic boundaries (as compared to closed spacetimes in \cite{Usatyuk:2022afj}) and the constrained instanton method, which explains why the JT path integral can pick up contributions from spacetimes with (mildly) singular curvature sources.} This appendix will contain in particular
\begin{enumerate}
\item Formulating the JT path integral in lightcone gauge in section \ref{app:general}. For this we'll use the second order formalism and fix to flat gauge (except at some isolated points), and present different factors coming out of such gauge choice; namely a ratio of determinants and a Liouville action.

\item Explaining the moduli space of lightcone diagrams in \ref{app:moduli}. In particular we conjecture the domain of integration by relating it to the closed string string lightcone diagrams of \cite{giddings1987triangulation}.

\item Showing that (for the configurations in which we were interested in the main text) the Liouville action does not contribute significantly in section \ref{app:Liouville}.

\item Showing that the ratio of determinants becomes independent of $T$ in the tau-scaling limit $T\to\infty$.

\end{enumerate}

\subsection{General set up}\label{app:general}

Consider the JT path integral over metrics at fixed genus $g$ (and ignore the boundary and topological part of the JT action for now)
\begin{equation}
    \mathcal{F}_g = \int \frac{\md g}{\text{vol}(\text{diff})}\,\md \Phi\,\exp\bigg(\frac{1}{2}\int d^2 x \sqrt{g}\,\Phi(R+2)\bigg)
\end{equation}
We can manipulate this path integral by first putting in constraint instantons, namely we insert several factors of \eqref{1equals} into the path integral. After that, we can gauge-fix the diffeos by choosing a particular slice of metrics and including the Jacobian. More concretely, we can fix to a conformal gauge $g=e^{2\omega}\hat g$. After such gauge fixing there is the usual Liouville term and the action becomes
\begin{align}
\mathcal{F}_g =& \prod_i \frac{1}{\Vol}\int \d x_{\text{sing}\,i}\sqrt{g}\, \frac{1}{2\pi}\int_{-\infty}^{+\infty} \d A_i \int_{-\infty}^{+\infty}\d\alpha_i e^{(2\pi-\i \alpha_i)A}  \int\d (\text{moduli})\times \text{Jacobian}
\nonumber\\&\int \md \omega\, \md \Phi\exp\bigg(-26I_\text{L}(\omega)+\frac{1}{2}\int d^2 x\sqrt{\hat{g}}\,\Phi \bigg(\hat R-2\hat\Delta_0 \omega +2e^{2\omega}\bigg)+(2\pi-\i\alpha_i)\Phi(x_{\text{sing}\,i})\bigg) \label{gauge_fixed_action} 
\end{align}

One common gauge choice is to fix to Euclidean $\hat R=-2$ metrics, then we integrate over the Weil-Petersson moduli and the Jacobian is known, see for instance \cite{Saad:2019lba}.\footnote{
 When gauge-fixing to Euclidean $\hat R=-2$ metrics $\hat g=(dx^2+d y^2)/y^2$, $\omega_0$ will localize to $0$. Integrate over $\omega$ and dilaton $\Phi$ gives
\begin{equation}
    \int_{\mathcal{F}} \d (\text{Weil-Petersson})\frac{\det(-\hat \Delta_1+2)}{\det(-\hat \Delta_0+2)}\,,
\end{equation}
where $\hat\Delta_0$ is the scalar Laplacian $y^2(\partial_x^2+\partial_y^2)$ on Euclidean AdS$_2$. In the Weil-Petersson setup one would not have introduced the sources in \eqref{gauge_fixed_action} of course, those were introduced with lightcone gauge in mind.} Light-cone gauge \cite{mandelstam1973interacting} is a different choice where one fixes to \emph{flat} metrics $\hat{g}$ with $\hat R=0$. The advantage of this gauge is is that there is no complicated fundamental domain, one just integrates over all light-cone diagrams (as will be discussed more below). It was shown in \cite{d1987unitarity,d1988geometry,giddings1987triangulation} that those diagrams cover moduli space precisely once. 

The disadvantage, when it comes to JT gravity, is that the Jacobian is more complicated, because the JT path integral always localizes on $R=-2$, such that the solution for $\w$ is nonzero. In lightcone gauge we can consider $\hat{g}$ using coordinates $\d x^2+ \d y^2$ and the Jacobian from the gauge-fixing is a scalar determinant,\footnote{Alternatively, the square root of the vector Laplacian is just the scalar determinant $\det(-\hat{\Delta}_1)^{1/2} = \det(-\hat{\Delta}_0)$.} this can be thought of as due to the ghosts that in bosonic string theory cancel $2$ of the naively $26$ families of oscillator modes. 

Importantly, in the usual light-cone formulation \cite{giddings1987triangulation} all the curvature is coming from the conformal factor $\w$ and $\hat{R} = 0$ \emph{everywhere}, but in our formulation where we work with constraint instantons, this is slightly different. Because the $\Phi(x_i)$ introduce delta functions in the curvature, it is more convenient to gauge fix to metrics which are flat everywhere \emph{except} at these $x_i$. This ensures that the term $\sqrt{\hat{g}}\, \Phi \hat{R}$ in \eqref{gauge_fixed_action} cancels with the $\sum_i (2\pi - \i\a_i)\Phi(x_i)$ term. This simplifies the conformal factor $\w_0$ massively, on the saddle $\alpha_i=0$ it becomes identical for \emph{all} diagrams that we consider in section \ref{sect:plateau} (we discuss this in more detail in section \ref{sect:detsonslits})
\begin{equation}
    e^{2\omega_0}=\frac{1}{\sinh(x)^2}\,.
\end{equation}
After doing the $\Phi$ path integral which localizes on metrics $g$ satisfying \eqref{coversing} we are left with a factor\footnote{The ratio of determinants compares a massless scalar particle on the light-cone diagram with a massless particle on the same light-cone diagram but with some nontrivial potential $e^{2\w_0}$.}
\begin{equation}
     \int_{\text{crotches at } x_i} \d (\text{lightcone})\frac{\det (-\hat\Delta_0)}{\det (-\hat\Delta_0+e^{2\omega_0})}\,e^{-26 I_\text{L}(\omega_0)}\label{38}\, ,
\end{equation}
where we left implicit the integrals over $x_i$, $A_i$ for clarity. The integral over $\a_i$ has been done using a saddle point approximation which sets $\a_i = 0$, because in the main text we were interested in classical solutions. Off-shell, $\alpha_i\neq 0$ could be relevant \cite{Usatyuk:2022afj}, but we will not consider it here (see also the discussion section \ref{sect:disc}).
\subsection{Moduli space}\label{app:moduli}
We start this section by explaining what's our coordinate system of the moduli space. We will show that the positions of the crotches gives a natural triangulation of moduli space, and discuss their measure. This should be viewed as a conjecture by an analog between the light-cone moduli for closed strings \cite{giddings1987triangulation} and our diagrams in section \ref{sect:plateau}.

We explain this by treating the $g=2$ example \eqref{327}. The $g=2$ crotch spacetimes are characterized by $4$ interaction times $t_{1,i}<t_{1,f}<t_{2,i}<t_{2,f}$ at which crotches are inserted\footnote{One can imagine configurations with $t_{2,i}>t_{2,f}$, but those are actually indistinguishable from some of the configurations that we count here. Indeed, upon exchanging $r_{2,i}$ and $r_{2,f}$ as well as $t_{2,i}$ and $t_{2,f}$ one obtains an identical Riemann surface. Therefore we can limit ourselves to the ordering discussed here $t_{1,i}<t_{1,f}<t_{2,i}<t_{2,f}$.}
\begin{equation}
    \begin{tikzpicture}[baseline={([yshift=-.5ex]current bounding box.center)}, scale=0.7]
 \pgftext{\includegraphics[scale=1]{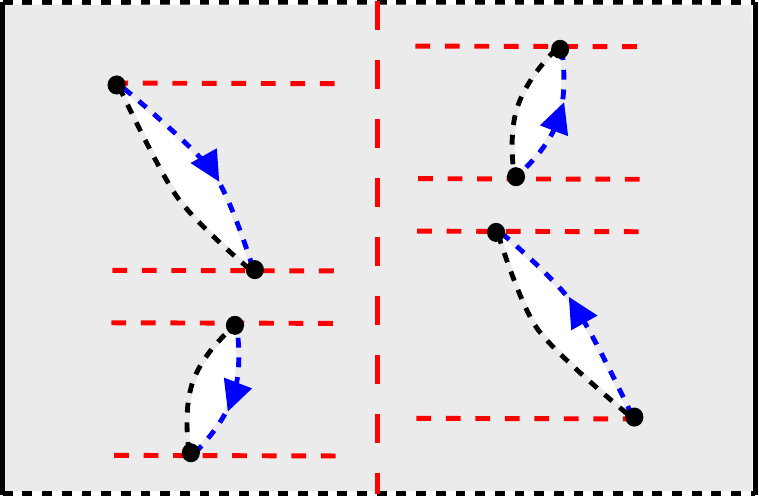}} at (0,0);
    \draw (-0.04, -3) node {genus $2$};
      \draw (-3.3, 1.85) node {$t_{1,i}$};
      \draw (-3.3, 0) node {$t_{1,f}$};
      \draw (-3.3, -0.7) node {$t_{2,i}$};
      \draw (-3.3, -2) node {$t_{2,f}$};
      \draw (3.3, -1.85) node {$t_{1,i}$};
      \draw (3.3, 0) node {$t_{1,f}$};
      \draw (3.3, 0.7) node {$t_{2,i}$};
      \draw (3.3, 2) node {$t_{2,f}$};
  \end{tikzpicture}\label{genus2diag}
\end{equation}
The relation of this geometry with \eqref{327} is explained in appendix \ref{app:orientation}. The locations of the interaction vertices on this diagrams are fixed by several constraints. The constraint that the lengths and curvatures on the slits in \eqref{b17} matches, fixes $t_{j,f}-t_{j,i}$ and also constrains the radial coordinates to satisfy 
\begin{equation}
   r_{n,i}+\bar{r}_{n,i}=0\,,\quad r_{n,i}<0\quad \text{and}\quad  r_{n,f}+\bar{r}_{n,f}=0\,,\quad r_{n,f}<0\,.\label{constraint}
\end{equation}
Here the un-barred coordinates are in the right-wedge of the spacetime. We glue the black dotted lines to each others as well as the blue ones, the orientation (arrows) is fixed by orientability of the resulting geometry, see also appendix \ref{app:orientation}. This orientation forces us to order interaction times \emph{oppositely} in the two wedges. We can choose thus for instance $T-t_\text{int}$ for the left interaction times and $t_\text{int}$ for the right interaction times. That one variable $t_\text{int}$ labels the left-and right times of the crotch (instead of having two independent times) is part of the conjectured analogy with the lightcone diagrams of \cite{giddings1987triangulation} which we will now detail.\footnote{In the language of Louko-Sorkin \cite{Louko:1995jw} light-cone gauge is indeed like choosing a global time coordinate (or Morse function). We are leaving a twist freedom \cite{Saad:2018bqo} implicit here that shifts the origins of left-and right times relative with respect to each other.}

Now we try to conjecture a integration measure and fundamental domain of our diagram, by viewing them as open string diagrams\cite{giddings1987triangulation}. Given the constraints on the orientation in \eqref{genus2diag} we are led to believe that the analogy with lightcone diagrams works by thinking of the left-and right wedges of the double cone as \emph{two} separate open strings, which one obtains by cutting \eqref{genus2diag} on the horizon.\footnote{This seems in phase with the fact that also in the computation of the determinants on the double cone in appendix \ref{app:determinants} one can treat both wedges as essentially decoupled.}

We then have two open string diagrams of $\infty$ (spatial) length (since the conformally flat $r$ coordinate runs from $-\infty$ to $0$ in the right wedge) which are each others mirrored image (to make this analogy we have here flipped the left wedge, such that both sides share common interaction times $t_\text{int}$)

\begin{equation}
    \begin{tikzpicture}[baseline={([yshift=-.5ex]current bounding box.center)}, scale=0.7]
 \pgftext{\includegraphics[scale=1]{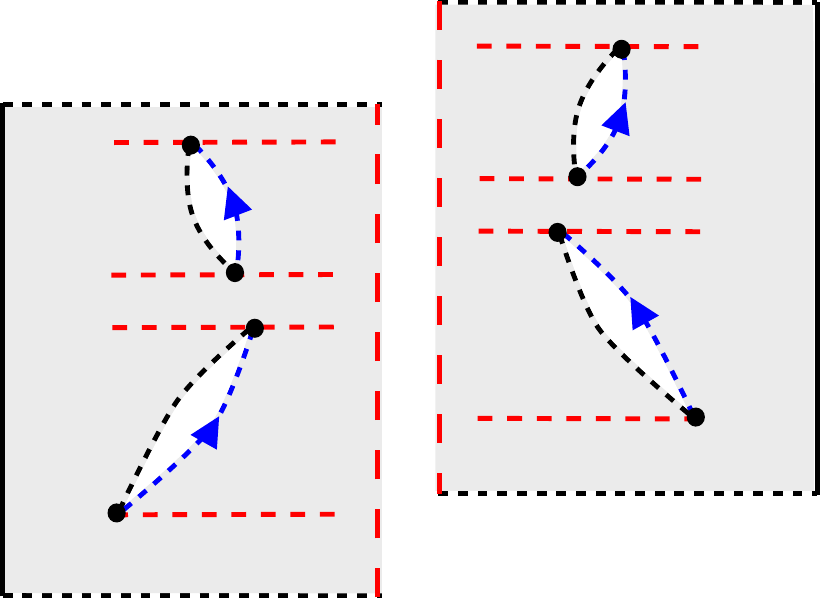}} at (0,0);
      \draw (3.6, -1.35) node {$t_{1,i}$};
      \draw (3.6, 0.5) node {$t_{1,f}$};
      \draw (3.6, 1.2) node {$t_{2,i}$};
      \draw (3.6, 2.5) node {$t_{2,f}$};
  \end{tikzpicture}\label{529}
\end{equation}
We should think of this as just a special case of two open strings with identical lengths $a/2$ with $a\to\infty$. As mentioned in \cite{giddings1987triangulation}, the moduli space and measure of such open string diagrams can be obtained by ``cutting open'' closed string lightcone diagrams. Assuming that this analogy we've made is correct, we will follow this procedure to find the measure and integration range on our diagrams. Let us therefore briefly recap the closed string lightcone moduli.

Associated with every extra wormhole (the top hole in the picture below) there are $6$ extra light-cone moduli. The two obvious ones are the interaction times $t_i$ and $t_f$. Besides this there is the circumference $a_1$ of one of the closed strings during the period between the interactions (the second modulus is fixed $a_2=a-a_1$ because the total length is a conserved quantity). The remaining $3$ moduli are twists $\tau$, $\tau_1$ and $\tau_2$ along the three closed strings that are involved.
\begin{equation}
    \begin{tikzpicture}[baseline={([yshift=-.5ex]current bounding box.center)}, scale=0.7]
 \pgftext{\includegraphics[scale=1]{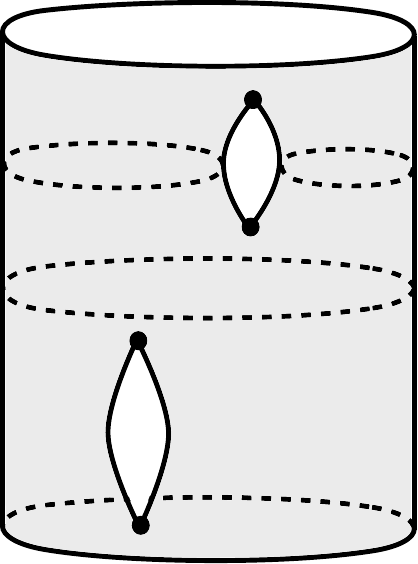}} at (0,0);
    \draw (-1, 1.8) node {$\tau_1$};
    \draw (1.3, 1.8) node {$\tau_2$};
    \draw (-1.5, -0.7) node {$\tau$};
  \end{tikzpicture}
\end{equation}
As was shown in \cite{d1987unitarity,d1988geometry,giddings1987triangulation}, to cover the moduli space of metrics modulo diffeos, one should count all such closed string diagrams once with the flat measure
\begin{equation}
    \d t_i\,\d t_f\, \d a_1\, \d \tau\, \d \tau_1\,\d\tau_2\,.
\end{equation}

To map this to open string moduli one cuts these diagrams along the seem (which means we follow a timelike geodesic). Let us first consider $\tau_1=\tau_2=0$. Then open string diagrams with $\tau=0$ respectively finite $\tau$ are obtained by cutting the two diagrams below on the red lines (the seems)
\begin{equation}
    \begin{tikzpicture}[baseline={([yshift=-.5ex]current bounding box.center)}, scale=0.7]
 \pgftext{\includegraphics[scale=0.8]{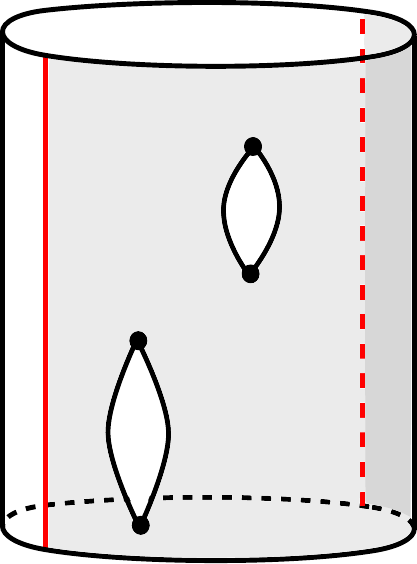}} at (0,0);
  \end{tikzpicture}\quad \begin{tikzpicture}[baseline={([yshift=-.5ex]current bounding box.center)}, scale=0.7]
 \pgftext{\includegraphics[scale=0.8]{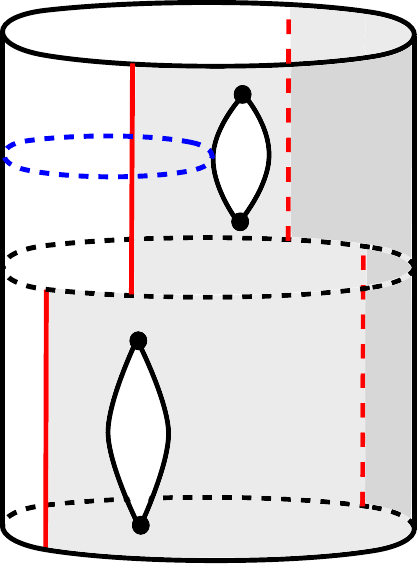}} at (0,0);
    \draw (-1.2, 1.5) node {\color{blue}$a_1$};
    \draw (-1.05, -0.6) node {\color{red}$\tau$};
  \end{tikzpicture}\quad \overset{\text{cut}}{\to} \quad
      \begin{tikzpicture}[baseline={([yshift=-.5ex]current bounding box.center)}, scale=0.7]
 \pgftext{\includegraphics[scale=0.8]{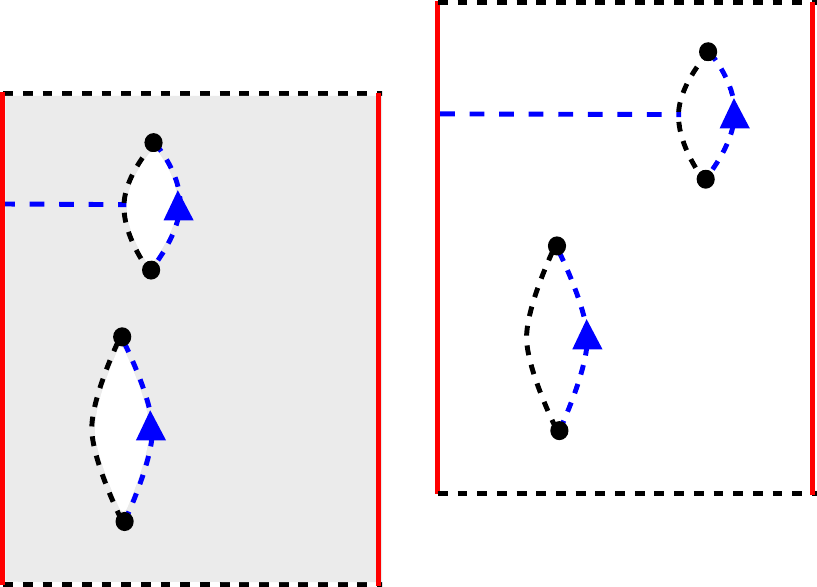}} at (0,0);
  \end{tikzpicture}
\end{equation}
We see that the role of the moduli $a_1$ and $\tau$ is to determine the spatial locations $r_{i}$ and $\bar{r}_i$ of the crotches on the open string (with the current choice $\tau_1=\tau_2=0$ the final locations of the crotches are identical to the initial ones). One recovers a flat measure on $r_i$ and $\bar{r}_i$
\begin{equation}
    r_i=\frac{a_1}{2}+\tau\,,\quad \bar{r}_i=\frac{a_1}{2}-\tau\quad \Rightarrow \quad \d a_1\, \d \tau = \d r_{i}\,\d \bar{r}_i\,.
\end{equation}
In our solution space $\alpha_i=0$ there are only AdS$_2$ solutions when $r_i+\bar{r}_{i}=0$ as mentioned in \eqref{constraint}, thus one can think of the JT path integral over $\Phi$ as introducing a factor $\delta(r_i+\bar{r}_{i})$.

What is left is to determine the roles of $\tau_1$ and $\tau_2$ in terms of open strings. Consider first the axial twists $\tau_1+\tau_2=0$. One outgoing open string gets shortened by $\tau_1-\tau_2$ and the other has increased its length by $\tau_1-\tau_2=L$. AdS$_2$ solutions (for $\alpha_i=0$) exist only for $L=0$ so one can think of the JT path integral over $\Phi$ as introducing a factor $\delta(L)$. The final mode $\tau_1=\tau_2$ has the effect of tilting the slit, as we see below
\begin{equation}
    \begin{tikzpicture}[baseline={([yshift=-.5ex]current bounding box.center)}, scale=0.7]
 \pgftext{\includegraphics[scale=1]{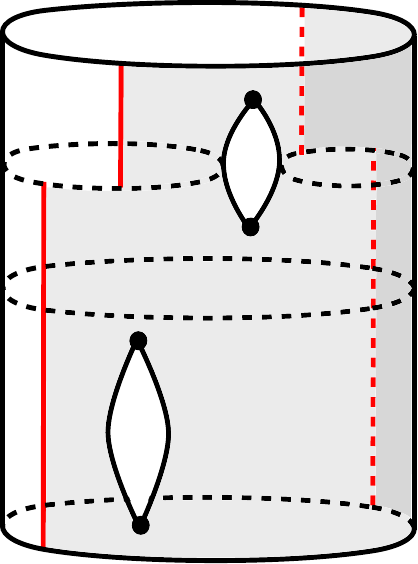}} at (0,0);
  \end{tikzpicture}\qquad\overset{\text{cut}}{\to}\qquad \begin{tikzpicture}[baseline={([yshift=-.5ex]current bounding box.center)}, scale=0.7]
 \pgftext{\includegraphics[scale=1]{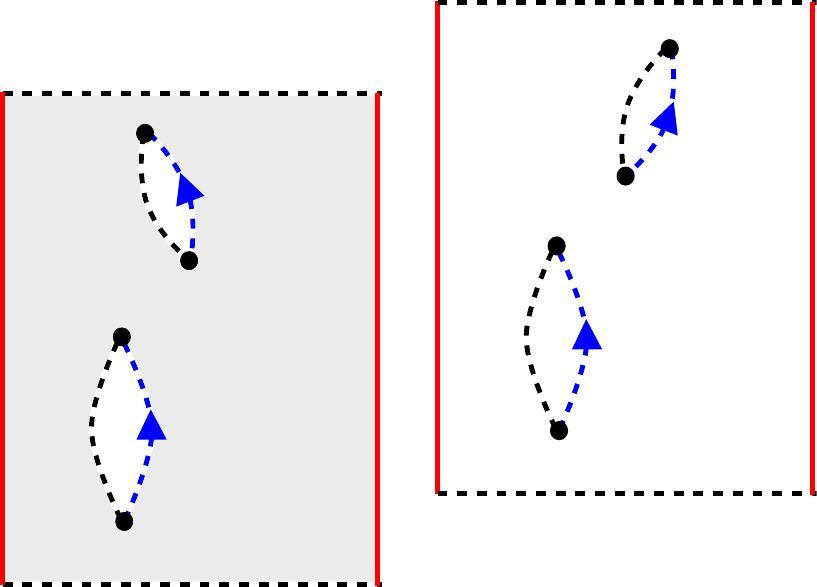}} at (0,0);
  \end{tikzpicture}
\end{equation}
One finds $\d \tau_1\, \d \tau_2=\d r_f\,\d L$ and so combining all the elements we see that the measure for our crotches-on-the-double-cone diagrams of section \ref{sect:plateau} is \emph{flat}
\begin{equation}
    \d(\text{moduli})=\d t_i\,\d t_f\,\d r_i\,\d r_f\,.
\end{equation}
We were told \cite{d1987unitarity,d1988geometry,giddings1987triangulation} to count all of these diagrams to cover the moduli space of metrics modulo diffeos once. This flat measure was expected intuitively for the locations where baby universes are born and die in for instance \cite{coleman1988black,giddings1988loss}. The $\Phi$ path integral only fires when the coordinates of these singularities in the lightcone diagrams match with the coordinates $x_{\text{sing}\,i}$ of the dilaton sources in \eqref{gauge_fixed_action}. The final result is therefore simply an integral over possible locations $x_{\text{sing}\,i}$ of crotches as we did in the main text.

Of course, in principle one still has to deal with the ratio of determinants and the Liouville action in \eqref{38} as an extra factor in the integrand. In the remainder of this appendix we will show that neither contributes significantly for the purpose of our discussion in the main text.

\subsection{Liouville action}\label{app:Liouville}
First we discuss the Liouville factor in \ref{38}. For the AdS double cone geometry, we have the metric\footnote{Another coordinate system we use in the main text is \eqref{doublecone}
\begin{equation}
    \d s^2=\d\rho^2-\sinh(\rho+i\varepsilon)^2\d t^2\,,\quad t\sim t+ T\,.
\end{equation}
They are equivalent upon the coordinate transformation $\sinh(\rho)=-1/\sinh(r)$. We have absorbed $2 E^{1/2} T\to T$. Since we are also interested in semiclassical black holes $E\gg 1$ this redefinition is harmless for our purposes here.
}
\begin{equation}
    \d s^2=\frac{1}{\sinh(r)^2}\left[\d r^2-(1+i\epsilon\cosh(r)\,\text{sgn}(r))^2\d t^2\right]=\frac{1}{\sinh(r)^2}\,\d \hat{s}^2 \,.\label{b17}
\end{equation}
Note that we put the $i \varepsilon$ prescription explicitly in this section in order to choose the right branch when calculating $\sqrt{-\hat{g}}$. The allowability condition \cite{Louko:1995jw, Witten:2021nzp} tells us that the branch choice one should make is
\bea
\begin{aligned}
\sqrt{-\hat{g}}&= 1+i \epsilon \cosh(r)\,,\quad r>0\,(\text{right})\\
&=-1+i \epsilon \cosh(r)\,,\quad r<0\,(\text{left})
\end{aligned}
\eea
We should compute the Liouville action in both wedges and sum them. For convenience, we redefine the radial coordinate in the left wedge $\tilde{r}=-r>0$. We can thus write out the Liouville action as\footnote{There are no boundary terms because we have Dirichlet boundary conditions on $\w_0$ and since $\hat{R}$ is flat (except at isolated points) the constant $r$ slices close to the boundary have $\hat{K} = 0$.}
\begin{align}
    I_\text{L}(\omega_0)&=\,\frac{1}{2}\int_0^T\d t\int_{0}^\infty \d r\,\sqrt{-\hat{g}}\,(\hat{g}^{\mu\nu}\partial_\mu\omega_0\partial_\nu\omega_0+\hat{R}\,\omega_0)\quad\text{(right)}\nn\\&\qquad+\frac{1}{2}\int_0^T\d t\int_{0}^\infty\d \tilde{r}\,\sqrt{-\hat{g}}\,(\hat{g}^{\mu\nu}\partial_\mu\omega_0\partial_\nu\omega_0+\hat{R}\,\omega_0)\quad\text{(left)}
\end{align}
The conformal factor is identical in both wedges, indeed $\omega_0=-\ln \sinh(r)$ (right) and $\omega_0=-\ln \sinh(\tilde{r})$ (left). However $\sqrt{-\hat{g}}$ is opposite in both wedges. We have furthermore
\begin{equation}
    \sqrt{-\hat{g}}\hat{R}=-2i\varepsilon \cosh(r) \text{sgn}(r)\,.
\end{equation}
One can then safely take $\varepsilon\to 0$ everywhere. Because of the opposite signs of $\sqrt{-\hat{g}}$ in both wedges, one finds that the total Liouville action vanishes on the double cone
\begin{equation}
    I_\text{L}(\omega_0)=0\,.
\end{equation}
On the crotch geometries one finds some mild dependence on $\omega_0(r_{\text{sing}\,i})$ from the source terms in $\sqrt{\hat{g}}\hat{R}$, but it does not significantly affect the classical locations of the crotches nor their on-shell actions, both of which were important in the main text. We expect to have similarly mild dependence on $r_{\text{sing}\,i}$ from the ratio of determinants in \eqref{38}, to which we turn next.

The key point is that there is no \emph{time} $T$ dependence coming out of either one, in the tau-scaling limit where $T\to\infty$.
\subsection{Determinants} \label{app:determinants}
Here we compute the ratio of determinants that appears as integration kernel in \eqref{38} for the spacetimes relevant in section \ref{sect:plateau}. In this section we always work with the flat metric $\hat{g}$, we'll drop all the hats for notational comfort below.
\subsubsection{Warming up with the double-cone}
We start by discussing the calculation of determinants on the double cone geometry, when there are no crotches. The metric is \eqref{b17} and with $\Delta$ the Laplacian on this metric we want to show that
\be 
\frac{\det(-\Delta)}{\det(-\Delta + 2/\sinh^2(r))}\,,\quad -\Delta = -\partial_r^2 + \partial_t^2\,,\label{b19}
\ee
goes to $1$ for $T\to\infty$, confirming that this gives no contributions in our double scaled regime of interest. Whilst we have not explicitly indicated the $\i \varepsilon$ regularization of \eqref{b17} at $r=\pm\infty$ in \eqref{b19}, it will play an important role.

This problem can be tackled in the most naive way, by simply computing the spectra $\lambda_1$ and $\l_2$ of respectively the differential operators $-\Delta$ and $-\Delta+2/\sinh^2(r)$ and explicitly computing
\begin{equation}
    \frac{\det(-\Delta)}{\det(-\Delta + 2/\sinh^2(r))}=\exp\bigg(\sum_i \log(\l_{1\,i})-\sum_j \log(\l_{2\,j})\bigg)\,.\label{b20}
\end{equation}
Both determinants are to be computed using Dirichlet boundary conditions because we fix the metric and dilaton fluctuations to vanish at the holographic boundaries $r=0$. We will show below that, due to the $\i\varepsilon$ regularization, the eigenfunctions of both operators of interest will exponentially decay towards the would-be horizon $r=\pm\infty$. This means that on this flat double cone \eqref{b17}, there are \emph{independent} modes in the left-and right wedges, there is no additional boundary condition or matching condition to be enforced at the would-be horizons. The determinants then factorize in the product of determinants in each wedges. In each wedge we can then compute the ratio first in Euclidean, and then analytically continue respectively as $\beta=\pm \i T$, with the sign being determined by that in \eqref{b17} \cite{Witten:2021nzp}.

To solve for the eigenvalues $\lambda_{1,j}$ and $\lambda_{2,j}$, note that the potential doesn't depends on $t$. So we can use separation of variables
\begin{equation}
    \phi_{n,k}(r,\t)=e^{\i \frac{2\pi}{\beta} n \tau} f_k(r)
\end{equation}
For the first determinant we have with $k$ non-negative (otherwise we have an over-complete basis)
\begin{equation}
    -\partial_r^2 f_k(r)=k^2 f_k(r)\,,\quad \lambda_{n,k}=k^2+\bigg(\frac{2\pi n}{\beta}\bigg)^2\,,
\end{equation}
imposing vanishing boundary condition at $r=0$, we have the orthonormal solutions
\begin{equation}
    f_k(r)=\sqrt{\frac{2}{\pi}}\sin(k r)\,.\label{mode1}
\end{equation}
Similarly for the second determinant, for the differential equation
\be 
(-\partial_r^2 + 2/\sinh^2(r)) f_k(r) = k^2 f_k(r)\,,\quad \lambda_{n,k}=k^2+\bigg(\frac{2\pi n}{\beta}\bigg)^2\,,
\ee
we have orthonormal solutions that satisfy Dirichlet boundary conditions at $r=0$ with $k$ non-negative 
\be 
f_k(r) = \sqrt{\frac{2}{\pi}} \frac{\coth (r) \sin(k r) - k \cos(k r)}{(1+k^2)^{1/2}}\,.\label{mode2}
\ee
These are related with Mehler functions as $f_k(r) = \sinh (r)^{1/2} P^{-3/2}_{-1/2 + \i k}(\cosh(r))$. Their orthonormality relation is an example of the Mehler-Fock transformation, a sort of similar transform as the Kontrovich-Lebedev transform for modified Bessel functions.

For every finite $r$ with $\varepsilon\to 0$ the differential equations are as above, and so are the wavefunctions. This fixes the spectrum to $\l_{n,k}=k^2+(2\pi n/\beta^2)$. Now we should look at the opposite regime $\abs{r}\gg \ln(1/\epsilon)$ and see if perhaps the behavior of the wavefunctions in this regime further restricts the spectrum, for instance by demanding a matching condition between both wedges. This turns out not to be the case. The potential in the bottom determinant can be ignored in this regime, and both differential equations reduce to
\begin{equation}
    (\l_{n,k}+\partial_r^2+\partial_r)\phi_{n,k}(r,\tau)=0\,,
\end{equation}
the solutions of which are exponentially decaying towards $\abs{r}=\infty$. So we require \emph{no} matching condition, or any extra boundary condition at the horizon. The implications of this were discussed below \eqref{b20}

Both modes \eqref{mode1} and \eqref{mode2} have continuous spectra, because of the infinite volume of spacetime. To compute their spectra reliably, as is conventional for quantum mechanical systems in infinite space, we should start with a finite space and then take the volume to $\infty$. The volume divergences will cancel in our ratio \eqref{b19}. Thus we consider a Dirichlet cutoff at $r=R\sim \log(1/\varepsilon)$ and eventually take $R\to\infty$. 

For the first set of modes \eqref{mode1} the quantization condition is
\begin{equation}
    \sin(k R)=0\quad \Rightarrow\quad  k_m=m \frac{\pi}{R}\,.
\end{equation}
The spectrum in the limit $R\to\infty$ follows from Newton's definition of integrals
\begin{equation}\label{append:dos}
    \sum_{m=0}^\infty \to \int_0^\infty \d k\,\frac{1}{\d k/\d m}\,,
\end{equation}
and we recover the known volume scaling of the density of states for quantum mechanics on a line
\begin{equation}
    \rho_1(k)=\frac{1}{\d k/\d m}=\frac{1}{k_{m+1}-k_m}=\frac{R}{\pi}\,.
\end{equation}
For the second set of modes \eqref{mode2} the quantization condition is at large $R$
\begin{equation}
    \cos(k R) k-\sin(k R)\coth(R)=0\quad \Rightarrow\quad \tan(k_m R)=k_m\,,
\end{equation}
The density of states is again determined via \eqref{append:dos}. Notice that we can write
\bea
\delta_k=k_{m+1}-k_m=\tan(k_{m+1} R) -\tan(k_m R)=\tan(\delta_k R)(1+k_m k_{m+1})\,,
\eea
in which we used the trigonometric identity for $\tan(\alpha-\beta)$. Now for large $R$, note that the solution is such that $\delta_k R\sim \pi$. So we have
\bea
\delta_k= (\delta_k R-\pi)(1+k(k+\delta_k)) \quad \Rightarrow\quad \rho_2(k)=\frac{R}{\pi}-\frac{1}{\pi} \frac{1}{k^2+1}
\eea
 
 We are now ready to compute the ratio of determinants \eqref{b17}, using \eqref{b20} (this is for just one wedge at the moment)
\begin{align}
    \frac{\det(-\Delta)}{\det(-\Delta + 2/\sinh^2(r))}&=\exp\bigg(\sum_{n=-\infty}^{+\infty}\int_0^\infty \d k (\rho_1(k)-\rho_2(k))\log\bigg(k^2+\bigg(\frac{2\pi n}{\beta}\bigg)^2\bigg)\bigg)\nonumber\\
    &=\exp\bigg(\frac{1}{\pi}\sum_{n=-\infty}^{+\infty}\int_0^\infty \d k \frac{1}{k^2+1}\log\bigg(k^2+\bigg(\frac{2\pi n}{\beta}\bigg)^2\bigg)\bigg)\\&=\exp\bigg(\frac{1}{\pi}\sum_{n=-\infty}^{+\infty}\log\bigg(1+\frac{2\pi\abs{n}}{\beta}\bigg)\bigg)\,.
\end{align}
Using Hurwitz Zeta function regularization we can compute the infinite product
\begin{equation}
    \prod_{n=1}^{\infty} (an+b)=a^{-1/2-b/a}\frac{\sqrt{2\pi}}{\Gamma(1+b/a)}\quad \Rightarrow\quad \prod_{n=1}^{\infty} \bigg(1+\frac{2\pi n}{\beta}\bigg)=\bigg(\frac{2\pi}{\beta}\bigg)^{-1/2-\beta/2\pi} \frac{\sqrt{2\pi}}{\Gamma(1+\beta/2\pi)}
\end{equation}
Using Stirling for large $\beta$ this reduces to $e^{\beta/2\pi}$.\footnote{One might be worried about Stokes phenomena, we can also first analytically continue $\beta=\pm\i T$, include the contribution from the other wedge and take $T\to\infty$ in the end. By doing that we get \bea \left(\frac{2 \pi}{i T}\right)^{-\frac{1}{2}-\frac{i T}{2 \pi}} \frac{\sqrt{2 \pi}}{\Gamma\left(1+\frac{i T}{2 \pi}\right)}\left(\frac{2 \pi}{e^{i \pi} i T}\right)^{-\frac{1}{2}-\frac{e^{i \pi} i T}{2 \pi}} \frac{\sqrt{2 \pi}}{\Gamma\left(1+\frac{e^{i \pi} i T}{2 \pi}\right)}\,. \eea
For large $T$ this behaves like $1-e^{-T}\to 1$.} Analytically continuing $\beta=\pm \i T$ and multiplying the ratio of determinants from both wedges, we confirm that the ratio exactly cancels in our double scaling limit where $T\to\infty$, completing the proof
\begin{equation}
    \frac{\det(-\Delta)}{\det(-\Delta + 2/\sinh^2(r))}\to e^{-\i T/\pi^2+\i T/\pi^2}=1\,.
\end{equation}

\subsubsection{Slit geometries}\label{sect:detsonslits}
To calculate the ratio of determinants on geometry with slits, we fist need to check the solution for the conformal factor $e^{2\omega_0}$. The metric is everywhere \eqref{b17}
\begin{equation}
    \d s^2=\frac{\d r^2+\d x^2}{\sinh(r)^2}\,,\quad x\sim x+\beta\,,
\end{equation}
where eventually one analytically continues $\beta=\i T$. The solution 
\begin{equation}
    e^{2\omega_0}=\frac{1}{\sinh(r)^2}\,.
\end{equation}
for the conformal factor holds on \emph{all} crotch geometries as we will now show.

As mentioned before, the light-cone diagrams we are interested have the property 
\begin{equation}
    \sqrt{\hat{g}}\hat{R}=-4\pi\sum_{\text{crotches}}\delta(x-x_i)\delta(r-r_i)\,,
\end{equation}
and by definition of course $g=e^{2\omega_0}\hat{g}$. Solutions to this are double covers of flat space, so if we write
\begin{equation}
    \d s^2=\d r^2+\d x^2=\d z\, \d \bar{z}\,,
\end{equation}
then a crotch at $x_c=0$ can be obtained by doing the coordinate transformation $w^2=\i (r_c-z)$ where $w=x+\i y$\,. Notice that when we travel around $w=0$ and $w$ picks up an argument $2\pi$ that $z$ picks up an argument $4\pi$, or more appropriately the $z$ plane covers only half of the $w$ plane, since turning $2\pi$ around the crotch in $z$ coordinates is only half a rotation around the origin in the complex $w$ plane, and we need to make a second rotation to came back to our starting point. The full geometry, obviously, is most accurately described using the uniformizing $w$ coordinate. If we use the $z$ coordinates, the geometry is flat locally everywhere, but there is a square root branchcut that we can lay from $r_c$ to $r_c+\i \infty$ and if we go through it we (smoothly) go onto the second sheet. To study the geometry in $w$ coordinates we should consider the regularized version
\begin{equation}
    \d s^2=4 (x^2+y^2+\gamma)(\d x^2+\d y^2)\,,\quad \gamma\to 0\,,
\end{equation}
which has $\text{Re}(\sqrt{\hat{g}})>0$ everywhere. Note that this metric is exactly flat (and in particular non-singular) at the origin. Using this metric one recovers indeed the delta function source in the full geometry
\begin{equation}
    \sqrt{\hat{g}}\hat{R}\,\d x\,\d y=-4\pi\,\frac{1}{\pi}\frac{\g}{(x^2+y^2+\g)^2}\,\d x\,\d y\to -4\pi\,\delta(x)\delta(y)\,\d x\,\d y\,.
\end{equation}
We have for this regularized metric
\begin{equation}
    \hat{\Delta}=\frac{\partial_w\partial_{\bar{w}}}{\abs{w}^2+\gamma}=\frac{\abs{z-z_c}}{\abs{z-z_c}+\gamma}\,4\partial_z\partial_{\bar{z}}\,,
\end{equation}
which vanishes when $z=z_c$ but equals the flat space Laplacian elsewhere. The JT action evaluated on the light-cone metrics is
\begin{equation}
   \frac{1}{2}\int \d^2 w \sqrt{g}\,\Phi(R+2)+2\pi\sum_\text{crotches}\Phi(w_c)=\int \d^2 w\sqrt{\hat{g}}\,\Phi(e^{2\omega_0}-\hat{\Delta}\omega_0)\,, 
\end{equation}
such that the dilaton path integral localizes to solutions which satisfy
\begin{equation}
    \sqrt{\hat{g}}\,(e^{2\omega_0}-\hat{\Delta} \omega_0)=0\,.
\end{equation}
Our proposed solution exactly satisfies $4\partial_z\partial_{\bar{z}}\omega_0=e^{2\omega_0}$ everywhere so this equation is satisfied if
\begin{equation}
    \sqrt{\hat{g}}\,\frac{\gamma}{\abs{w}^2+\gamma}=0\,,
\end{equation}
which is the case because $\sqrt{\hat{g}}=4(\abs{w}^2+\gamma)$ so indeed this goes to zero everywhere. So because $\sqrt{\hat{g}}$ vanishes on the crotches, our solution holds everywhere.

So we are being asked to compute a ratio of determinants with a ``simple'' and universal potential\footnote{For contrast, without a constrained instanton construction responsible for source terms in the action \cite{Usatyuk:2022afj}, the JT path integral localizes on Euclidean $\sqrt{g}(R+2)=0$ surfaces and this same equation would become
\begin{equation}
    \sqrt{\hat{g}}\,(e^{2\omega_0}-\hat{\Delta}\omega_0)+2\pi\sum_\text{crotches}\delta(x-x_c)\delta(y-y_c)=0\,.
\end{equation}
Locally near each crotch this admits the familiar (and singular) solution
\begin{equation}
    \omega_0=\frac{1}{2}\log \abs{w-w_c}^2\,.
\end{equation}
In this scenario the evaluation of the determinants looks utterly hopeless since the potential term would depend sensitively on the locations of the crotches via this $\omega_0$, and would be quite complicated.}
\begin{equation}
    \frac{\det(-\Delta)}{\det(-\Delta+2/\sinh(r)^2)}\,.
\end{equation}
In the remainder of this section we want to argue that this becomes independent of $T$ for $T\to\infty$ (our regime of interest in the main text). The argument is quite simple. Each determinant can be written as the exponential of a single particle path integral on the slit geometry. Time dependence comes from paths that wind around at least two crotches, those depend on the time differences between the crotches. However, for late times the length of those paths is proportional to $T$, and thus extremely large. The cluster decomposition principle says that the contribution of long paths in quantum mechanics decays to zero, if the length of the path goes to infinity. Thus those paths actually do not end up contributing to the determinants for late times. As a result, the ratio of determinants becomes $T$ independent. 

In the remainder we make this more concrete, by following these steps
 
\begin{enumerate}
    \item We can expand the ratio of determinants perturbatively in $V(r)$, the task is then to compute the free propagator on the slit geometries, which is a single particle path integral.
    \item Cluster decomposition still holds on the slit geometries, this means that contributions from long paths are suppressed. For late times, all paths that are time-dependent are long. So in the double scaling limit there is no time dependence.
    \item Off-shell, shorter paths can produce $r$ dependence. In the main text though, we consider on-shell geometries where the crotches sit at $r=\infty$. This means that all short paths are in a region where essentially $V(r)=0$. This also holds for the long paths. So on-shell the determinants cancel.
\end{enumerate}

For the first step we use the cluster expansion
\begin{align}
    \frac{\det(-\Delta)}{\det(-\Delta + V(r))}&=\exp\bigg(\Tr \log(-\Delta)-\Tr\log(-\Delta+V(r)) \bigg)\nonumber\\
    &=\exp\bigg(-\int_0^\infty \frac{\d t}{t}\Big(\Tr e^{\Delta t}-\Tr e^{(\Delta-V(r))t}\Big)\bigg)\nonumber\\
    &=\exp\bigg(\sum_{m=1}^\infty \frac{(-1)^m}{m}\int_0^\infty \d t_1\dots \int_0^\infty \d t_m \Tr e^{\Delta t_1}V\dots e^{\Delta t_m} V \bigg)\,.
\end{align}
Here the trace is the trace over the single particle Hilbert space\footnote{For the double cone without crotches one can also write very explicitly
\begin{equation}
    \Tr \mo = \sum_{n=-\infty}^{+\infty}\int_0^\infty \d k\, \rho_1(k)\, \bra{n,k}\mo \ket{n,k}=\prod_{i}O_{\lambda_i}\,.
\end{equation}
This reduces to \eqref{xint} after inserting a complete set of position states and using the fact that the wavefunctions $\braket{x}{n,k}=\phi_{n,k}(x)$ are complete.}
\begin{equation}
    \Tr \mo = \int \d x\,\bra{x} \mo \ket{x}\label{xint}
\end{equation}
Inserting complete sets of states we get
\begin{align}
    \Tr e^{\Delta t_1}V\dots e^{\Delta t_m} V =\int \d x_1\, V(x_1)\dots \int \d x_m\, V(x_m)\, K(t_1,x_1,x_2)\dots K(t_m,x_m,x_1)
\end{align}
This features the heat kernel on the slit geometry.
\begin{equation}
    K(t,x_i,x_f)=\bra{x_i}e^{\Delta t}\ket{x_f}
\end{equation}
We can do the $t_i$ integrals finally to get propagators
\begin{equation}
    \int_0^\infty \d t\,K(t,x_i,x_f)=G(x_i,x_f)=\bra{x_i}\frac{1}{-\Delta}\ket{x_f}\,.
\end{equation}
So we arrive at the following Feynman diagram decomposition of our ratio of determinants
\begin{equation}
    \frac{\det(-\Delta)}{\det(-\Delta + V(r))}=\exp\bigg(\sum_{m=1}^\infty \frac{(-1)^m}{m}\int \d x_1\dots \int \d x_m\, G(x_1,x_2)\, V(x_2)\dots G(x_m,x_1)\, V(x_1) \bigg)\,,\label{a72}
\end{equation}
so we have free propagation of particles on our geometries, but they can scatter off the potential, which gives rise to Feynman weights $V(x)$ for a scattering at position $x$.

Now we want to compute these propagators. As a warm up let's consider the double cone again
\begin{align}
    K(t,x_i,x_f)&=\sum_{n=-\infty}^{+\infty}\int_0^\infty \d k\,\rho_1(k)\, \phi_{n,k}(x_i)^*\, \phi_{n,k}(x_f) \exp\bigg( -k^2 t -\frac{4\pi^2n^2}{\beta^2} t \bigg)\nonumber\\&=\sum_{w=-\infty}^{+\infty}\frac{\beta}{4\pi t}\exp\bigg( -\frac{\ell_w(x_i,x_f)^2}{4 t}\bigg)+\text{images}\,,\label{sumimag}
\end{align}
where $\ell_w(x_i,x_f)^2=(r_i-r_f)^2+(x_i-x_f+w\beta)^2$ we used Poisson summation and did the integral over $k$ explicitly. The images are three other identical expressions where we choose all signs for $\pm r_i$ and $\pm r_f$ and add a sign prefactor $\pm \pm$. This corresponds with the image charges that implement the Dirichlet boundary conditions at $r=0$. From this we find
\begin{equation}
    G(x_i,x_f)=\frac{\beta}{2\pi}\sum_{w=-\infty}^{+\infty}\log(\ell_w(\bar{x_i},x_f)^2/\ell_w(x_i,x_f)^2)
\end{equation}
with $\ell_w(\bar{x_i},x_f)$ the length of the geodesic from an imagine charge at $\bar{x_i}$. We want to argue that the semiclassical generalization of this is
\begin{equation}
    G(x_i,x_f)\sim \sum_\gamma \log(\ell_\g(\bar{x_i},x_f)/\ell_\g(x_i,x_f))\,,\label{toshow}
\end{equation}
with $\gamma$ all topologically inequivalent trajectories, and $\ell_\g(x_i,x_f)$ the length of the shortest path of a certain topology. Before demonstrating \eqref{toshow}, let us explain why this would be useful. For that, we would like to show that the contribution of a certain class of paths $\gamma$ vanishes when $\ell_\gamma(x_i,x_f)\to\infty$ regardless of the specific dependence of $\ell_\gamma(x_i,x_f)$ on the coordinates $x_i$ and $x_f$ and the moduli of the path. By definition
\begin{equation}
    \ell_\g(\bar{x_i},x_f)\leq \ell_\g(x_i,x_f)+\ell(x_i,\bar{x_i})\,,\quad \ell(x_i,\bar{x_i})=2r_i\,,
\end{equation}
such that indeed for $\ell_\gamma(x_i,x_f)\to\infty$ for fixed $x_i$
\begin{equation}
    \log(\ell_\g(\bar{x_i},x_f)/\ell_\g(x_i,x_f))\leq \log(1+2r_i/\ell_\g(x_i,x_f)))\to 0\,.
\end{equation}
This is essentially the cluster decomposition principle: contributions from long paths are suppressed. Thus if we show that \eqref{toshow} holds true, we have shown essentially that in the double scaling limit the ratio of determinants is time-independent. The reason is that time dependence can only come from paths which wind around at least two crotches, this would depend on the time difference between the crotches. But in the double scaling limit $T\to \infty$ essentially all configurations have all crotches separated by order one fractions of $T$, thus paths winding around multiple crotches have lengths of order $T$. In other words, all those paths are extremely long, and thus their contributions to the propagator can be neglected. This means that there is essentially no $T$ dependence in the ratio of determinants.

On more complicated geometries it is no longer practical to compute \eqref{sumimag} exactly, however what we can do is find a semiclassical approximation, which looks a lot like \eqref{sumimag}. For this we can use the Gutzwiller trace formula \cite{Haake:1315494} (when the spacetime is a hyperbolic Riemann surface, this is the Selberg trace formula), which comes down to approximating a quantum mechanical path integral by (a sum of all) classical saddles, with a one-loop determinant.

For a massive particle the path integral for the propagator reads
\begin{equation}
    G(x_i,x_f)=\int_{x_i}^{x_f}\mathcal{D}x(\sigma)\exp\bigg(-m\int_0^1\d \sigma \bigg( g_{\mu\nu} \frac{\d x^\mu}{\d\sigma}\frac{\d x^\nu}{\d\sigma}\bigg)^{1/2}\bigg)\sim  \sum_{\gamma} e^{-m \ell_\gamma(x_i,x_f)}\label{b51}\,,
\end{equation}
where the second piece is the Gutzwiller trace approximation and $\ell_\gamma(x_i,x_f)$ is the length of a geodesic $\gamma$ connecting $x_i$ and $x_f$. Baring focal points in the geometry, we should think of $\gamma$ in Euclidean signature as labeling topologically distinct classes of paths, like winding $w$. This organization of the path integral into topologically distinct paths can be done before computing anything.

Consider now our slit geometries. In the single particle path integral \eqref{b51} there will for instance be contributions from paths $x(\sigma)$ that wind around crotches several times, go through slits etcetera. The path integral decomposes into a sum over $\gamma$ labeling topologically distinct classes of paths (paths in one class are homologous). A novelty of slit geometries is that many classes do not contain a geodesic. Geodesics in flat space are straight lines, thus geodesics do not wind around crotches for instance. An example of such a class $\gamma$ of paths $x(\sigma)$ is
\begin{equation}
    \begin{tikzpicture}[baseline={([yshift=-.5ex]current bounding box.center)}, scale=0.7]
 \pgftext{\includegraphics[scale=1]{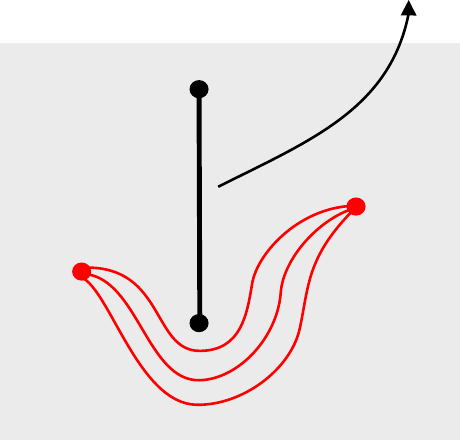}} at (0,0);
    \draw (0, -3) node {class of paths $x(\sigma)\subset\gamma $ without geodesic};
    \draw (1.9, 2.75) node {slit};
    \draw (-2, -0.5) node {\color{red}$x_i$};
    \draw (1.85, 0.05) node {\color{red}$x_f$};
    \draw (1.3, -1.5) node {\color{red}$x(\sigma)$};
  \end{tikzpicture}\label{app1}
\end{equation}
Nevertheless, the path integral \eqref{b51} obviously gets contributions from all classes. Even for classes $\gamma$ without a true geodesic, the path integral \eqref{b51} restricted to paths $x(\sigma)\subset \gamma$ would still be dominated by the paths with the shortest length, because the action $m \ell(x(\sigma))$ strongly favors short paths. Because of this the contribution from a class $\gamma$ to \eqref{b51} vanishes when the length of the shortest path in a class diverges $\ell_\gamma(x_i,x_f)\to\infty $.  

We now argue that this remains true for $m^2=0$, by arguing that a version of the Gutzwiller trace approximation on slit geometries gives \eqref{toshow}, which we've already showed vanishes for $\ell_\gamma(x_i,x_f)\to\infty $. For this we will use the equivalent Polyakov-type action\footnote{Integrating out $e(\s)$ reproduces the earlier Nambu-Goto type action, and the zero mode of $e(\sigma)$ is integrated only from $0$ to $\infty$ because we compute the propagator, not matrix elements of the WdW constraint \cite{Anous:2020lka,Casali:2021ewu}.}
\begin{align}
    G(x_i,x_f)&=\int_{x_i}^{x_f}\mathcal{D}x(\sigma)\int\mathcal{D}e(\sigma)\exp\bigg(-\int_0^1\d \sigma \bigg( \frac{1}{4 e} g_{\mu\nu} \frac{\d x^\mu}{\d\sigma}\frac{\d x^\nu}{\d\sigma}+ e m^2\bigg)\bigg)\nonumber\\
    &\sim \sum_\gamma \int_0^\infty \frac{\d e}{e}\, \exp\bigg( -\frac{\ell_\g(x_i,x_f)^2}{4 e}+e m^2\bigg) \sim  \sum_{\gamma} e^{-m \ell_\gamma(x_i,x_f)}\,.
\end{align}
The second line is true for geodesic trajectories, where the classical solution is $\d s/\d \sigma=\ell_\gamma(x_i,x_f) $ and we have gauge-fixed $e(\sigma)$ to its zero mode. The first equation on the second line remains true for $m^2=0$
\begin{equation}
    G(x_i,x_f)\sim \sum_\gamma \int_0^\infty \frac{\d e}{e}\, \exp\bigg( -\frac{\ell_\g(x_i,x_f)^2}{4 e}\bigg)\,,\label{b53}
\end{equation}
and essentially reproduces \eqref{sumimag}.

We want to prove that this remains true for any class of trajectories $\gamma$ in which there is no geodesic. Consider thereto trajectories of the following type, for instance
\begin{equation}
    \begin{tikzpicture}[baseline={([yshift=-.5ex]current bounding box.center)}, scale=0.7]
 \pgftext{\includegraphics[scale=1]{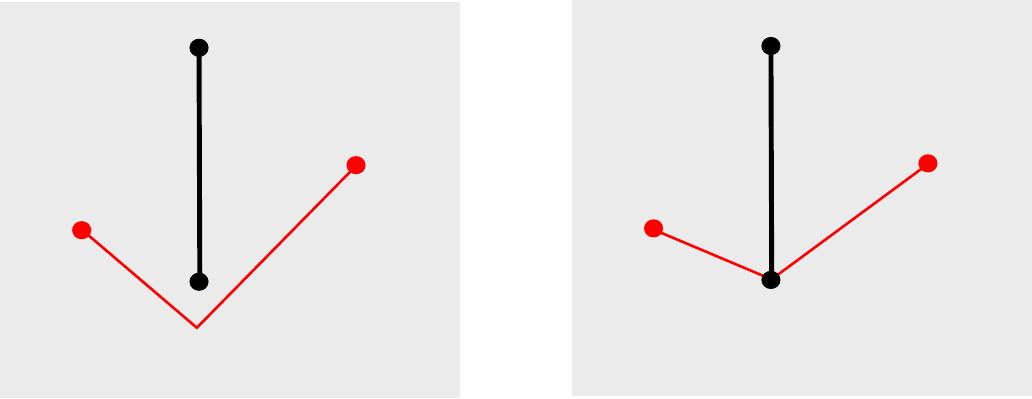}} at (0,0);
    \draw (-3.2, -2.6) node {almost everywhere geodesic};
    \draw (2.9, -2.6) node {dominant path};
    \draw (-3.1, -1.7) node {\color{red}$x_0$};
    \draw (-4.9, -0.25) node {\color{red}$x_i$};
    \draw (-1.05, 0.3) node {\color{red}$x_f$};
    \draw (0.95, -0.25) node {\color{red}$x_i$};
    \draw (4.8, 0.3) node {\color{red}$x_f$};
    \draw (-4.1, -1.1) node {\color{red}$\ell_1$};
    \draw (-2.4, -1.1) node {\color{red}$\ell_2$};
    \draw (1.9, -0.95) node {\color{red}$\ell_{\g\,1}$};
    \draw (3.85, -0.55) node {\color{red}$\ell_{\g\,2}$};
  \end{tikzpicture}\label{app2}
\end{equation}
These trajectories are geodesic almost everywhere, except that at certain special points $\sigma_0$ the particle gets a kick. Such kicks remind us of the singularities we had to allow in the geometry itself in the main text. We can take them into account for particles in the same way. For this it is most convenient to use the phase space path integral formulation (and specialize to our flat metric) for the propagator, with action\footnote{Notice the WdW Hamiltonian and its relation with the Laplacian $\Delta$ of targetspace \cite{Anous:2020lka,Casali:2021ewu}. If we integrate out $r$ and $\tau$ the momenta localize to constants and we see indeed that the non-zero Fourier modes of $e(\sigma)$ are redundant.}
\begin{equation}
    \exp\bigg(-\int_0^1\d \sigma\,\Pi_r \partial_\sigma r+\Pi_\tau \partial_\sigma \tau - e(\Pi_\tau^2+\Pi_r^2-m^2)  \bigg)
\end{equation}
Using similar methods as in the main text we can introduce charges $Q_r$ and $Q_\tau$
\begin{equation}
    1=\int_0^1\d \sigma_0\int_0^\infty \d r_0 \frac{1}{2\pi\i}\int_{-\i\infty}^{+\i\infty}\d Q_r\, e^{-Q_r r(\sigma_0)+Q_r r_0}\,,
\end{equation}
which introduce kinks in the particle trajectory. Indeed, classically the momentum jumps
\begin{equation}
    \Pi_r(\sigma_0+\varepsilon)=Q_r+\Pi_r(\sigma_0-\varepsilon)\,,\quad \Pi_r=\frac{1}{2 e}\partial_\sigma r\,,
\end{equation}
We consider fixed $Q_r$ and will vary it only in the end. On-shell, $x_0=x(\sigma_0)$. We denote the length of a straight trajectory between $x_i$ and $x_0$ by $\ell_1$, and that of the straight trajectory between $x_0$ and $x_f$ by $\ell_2$, then with $\d s/\d\sigma=\ell_1+\ell_2$ we have $\sigma_0=\ell_1/(\ell_1+\ell_2)$. Those straight trajectories are
\begin{equation}
    x_1=x_i+\frac{\s}{\s_0}(x_0-x_i)\,,\quad x_2=x_0+\frac{\s-\s_0}{1-\s_0}(x_f-x_0)\,,
\end{equation}
Resulting in the momenta
\begin{equation}
    \Pi_{r\,1}=\frac{1}{2e\sigma_0}(r_0-r_1)\,,\quad \Pi_{r\,2}=\frac{1}{2e(1-\s_0)}(r_f-r_2)\,,
\end{equation}
such that
\begin{equation}
    \Pi_{r\,1}^2+\Pi_{\tau\,1}^2=\frac{1}{4e^2\s_0^2}\ell_1^2\,,\quad \Pi_{r\,2}^2+\Pi_{\tau\,2}^2=\frac{1}{4e^2(1-\s_0)^2}\ell_2^2\,.
\end{equation}
The total on-shell action then reduces (after inserting $\sigma_0=\ell_1/(\ell_1+\ell_2)$) to
\begin{equation}
    \exp\bigg(-\int_0^{\s_0}\d \sigma\, e(\Pi_{\tau\,1}^2+\Pi_{r\,1}^2) -\int_{\s_0}^1\d \sigma\, e(\Pi_{\tau\,2}^2+\Pi_{r\,2}^2) \bigg)=\exp\bigg(-\frac{(\ell_1+\ell_2)^2}{4e}\bigg)\,.
\end{equation}
Without any topological obstructions (such as crotches), the extremum is of course the case without a kink, since then the total length $\ell_1+\ell_2$ is minimal. With a constraint, such as in \eqref{app2}, the dominant path is the one with minimal total length $\ell_{\g\,1}(x_i)+\ell_{\g\,2}(x_f)=\ell_\g(x_i,x_f)$.

At any rate, we have confirmed that \eqref{b53} holds also as approximation in classes of paths $\gamma$ that have no real geodesics in them, and which require the particle to get kicks at the crotches. This suffices to demonstrate \eqref{toshow} (we omit the standard computation of the one-loop determinant $\sim 1/e$). Thus cluster decomposition holds, and the determinants become independent of $T$ in the double scaling limit, because all $T$ dependent paths are very long and end up not contributing to the propagator. Here is an example of a representative long path $x(\sigma)$ of some class $\gamma$ for which $\ell_\gamma(x_i,x_f)=2(\t_2-\t_1)+\ell_{\g\,1}(x_i)+\ell_{\g\,2}(x_f)\to \infty$, because upon double scaling $\tau_2-\tau_1\to\infty$.
\begin{equation}
    \begin{tikzpicture}[baseline={([yshift=-.5ex]current bounding box.center)}, scale=0.7]
 \pgftext{\includegraphics[scale=1]{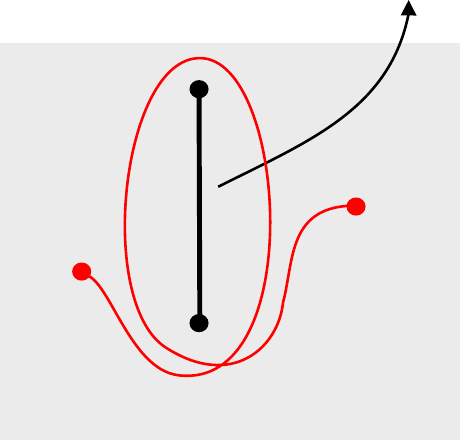}} at (0,0);
    \draw (0, -3) node {long paths};
    \draw (-0.7,-0.7) node {$\tau_1$};
    \draw (-1.3,1.3) node {$\tau_2$};
    \draw (-2, -0.5) node {\color{red}$x_i$};
    \draw (1.85, 0.05) node {\color{red}$x_f$};
    \draw (1.15, -1.3) node {\color{red}$x(\sigma)$};
    \draw (1.9, 2.75) node {slit};
    \draw (-4, -0.5) node {$G(x_i,x_f)\supset$};
  \end{tikzpicture}\label{app3}
\end{equation}
The contribution from this class $\gamma$ of paths $x(\sigma)$ to the propagator behaves as \eqref{toshow}, and thus indeed vanishes. The only finite contributions are from paths in \eqref{a72} that remain close to one crotch, and wind around it a few times. This gives a contribution that depends on the radial position of that crotch. In the classical configurations of the main text however, that radial coordinate is extremely large, such that $V(r)\to 0$. Thus for those configurations the ratio of determinants exactly cancels.
\section{Unwrapping the double cone}\label{app:orientation}
The purpose of this (colour book style) appendix is to clarify in more detail that the standard double cone picture \eqref{34} is equivalent to the unwrapped picture for the double cone that we use for instance in \eqref{genus2diag}. A second purpose is to explain the subtle differences between the identifications that one can make to built wormholes on the TFD, discussed in section \ref{sect:crotches}, versus those that we use on the double cone in section \ref{sect:2dplateau}. We start with the usual picture for the double cone, where one considers the TFD metric $\d s^2=\d\rho^2-4 E \sinh(\rho)^2\d t^2$ with an additional identification in Rindler time \eqref{doublecone} $t\sim t+T$ \cite{Saad:2018bqo}
\begin{equation}
    \begin{tikzpicture}[baseline={([yshift=-.5ex]current bounding box.center)}, scale=0.7]
 \pgftext{\includegraphics[scale=1]{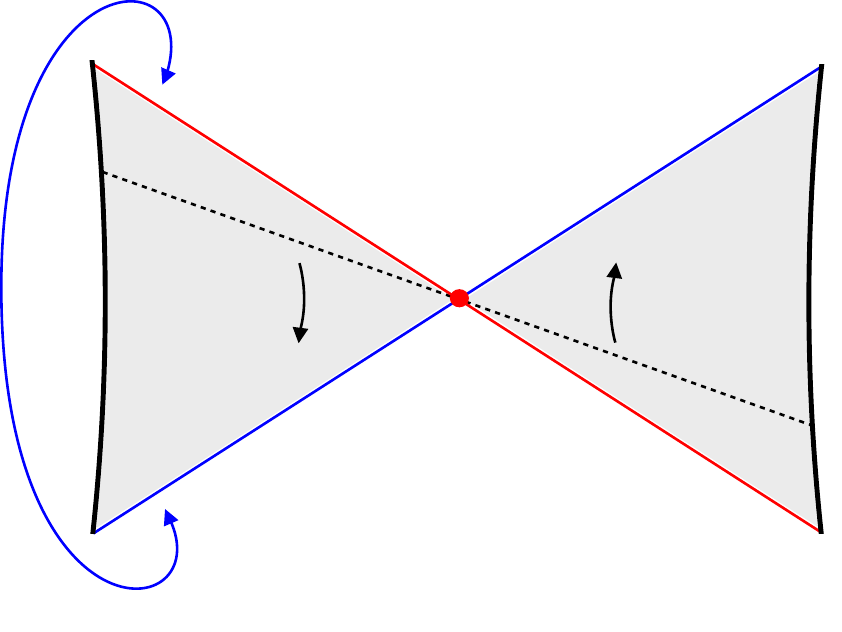}} at (0,0);
    \draw (3, 0.15) node {$t$ flow};
    \draw (-5.5, 0.15) node {\color{blue}identify};
  \end{tikzpicture}
\end{equation}
If we wrap this up we get a visualization of two cones, with Rindler time flowing in the same direction (from red to blue) on both sides (see also Fig. 2 in \cite{Mahajan:2021maz})
\begin{equation}
    \begin{tikzpicture}[baseline={([yshift=-.5ex]current bounding box.center)}, scale=0.7]
 \pgftext{\includegraphics[scale=1]{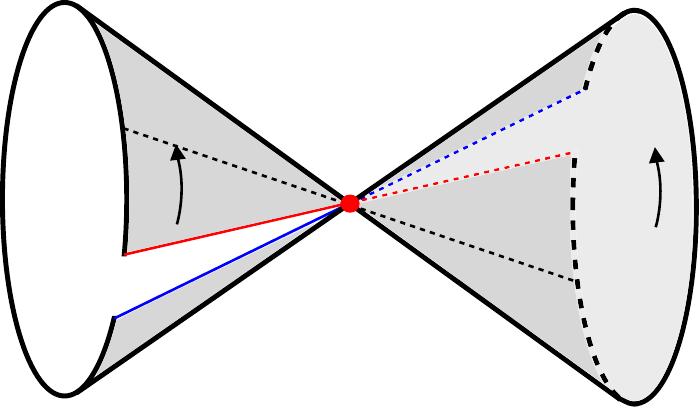}} at (0,0);
  \end{tikzpicture}
\end{equation}
Let us twist the right cone by $\pi$, this does not change the topology (twists never do)
\begin{equation}
    \begin{tikzpicture}[baseline={([yshift=-.5ex]current bounding box.center)}, scale=0.7]
 \pgftext{\includegraphics[scale=1]{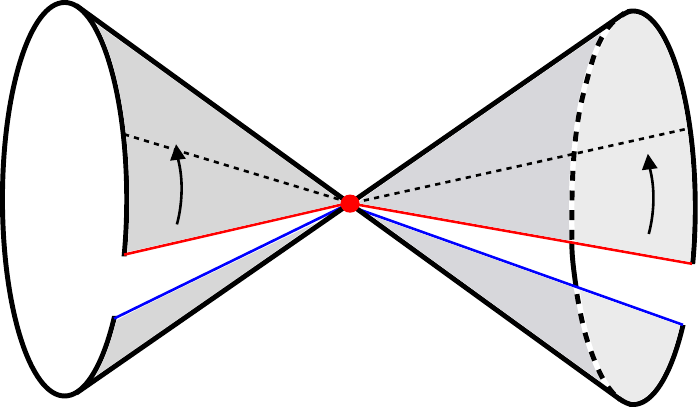}} at (0,0);
  \end{tikzpicture}
\end{equation}
Now we should remember that in the double cone metric the point $\rho=0$ is actually regularized as \cite{Saad:2018bqo} $\d s^2=\d\rho^2-4 E \sinh(\rho-\i\varepsilon)^2\d t^2$. This makes the metric Euclidean very close to $\rho=0$, and topologically has the effect of ``opening up'' the conical points (red dots), creating a tiny smooth wormhole connecting the two cones. So, a more accurate visualization of the double-cone topology is\footnote{The twist modulus is also very clear here, rotating both sides relative to one another.}
\begin{equation}
    \begin{tikzpicture}[baseline={([yshift=-.5ex]current bounding box.center)}, scale=0.7]
 \pgftext{\includegraphics[scale=1]{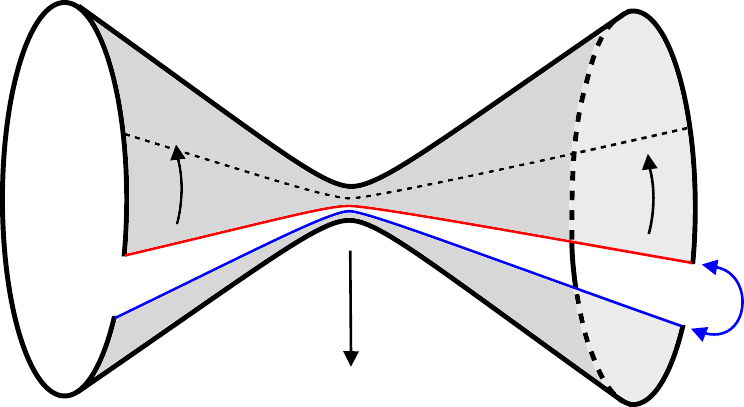}} at (0,0);
 \draw (5, -1) node {\color{blue}identify};
 \draw (0, -2) node {tiny wormhole};
  \end{tikzpicture}
\end{equation}
So, topologically this is just a $g=0$ wormhole connecting the two boundaries. One can now ``unwrap'' this last picture to obtain the representation of the double cone that we often used in appendix \ref{app:lightcone}
\begin{equation}
    \begin{tikzpicture}[baseline={([yshift=-.5ex]current bounding box.center)}, scale=0.7]
 \pgftext{\includegraphics[scale=1]{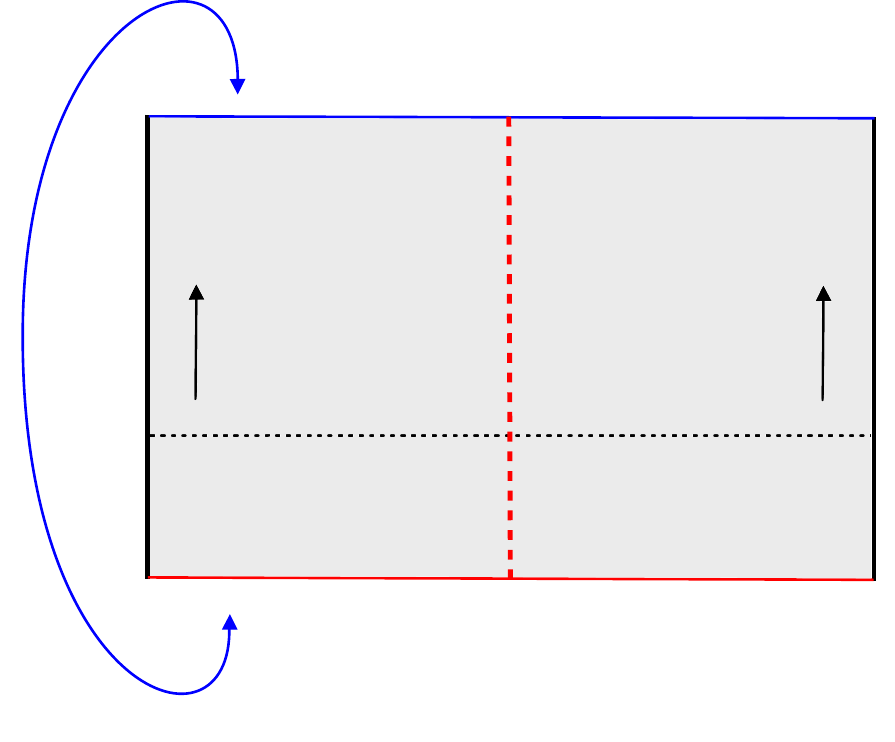}} at (0,0);
    \draw (-5.5, 0) node {\color{blue}identify};
  \end{tikzpicture}\label{c5}
\end{equation}

One reason why this last type of picture can be useful, is because it is visually easy to check that a certain identification results in an orientable surface. In particular we see that the following type of identification on the double cone (considered in section \ref{sect:2dplateau}) results in an orientable spacetime
\begin{equation}
    \begin{tikzpicture}[baseline={([yshift=-.5ex]current bounding box.center)}, scale=0.7]
 \pgftext{\includegraphics[scale=1]{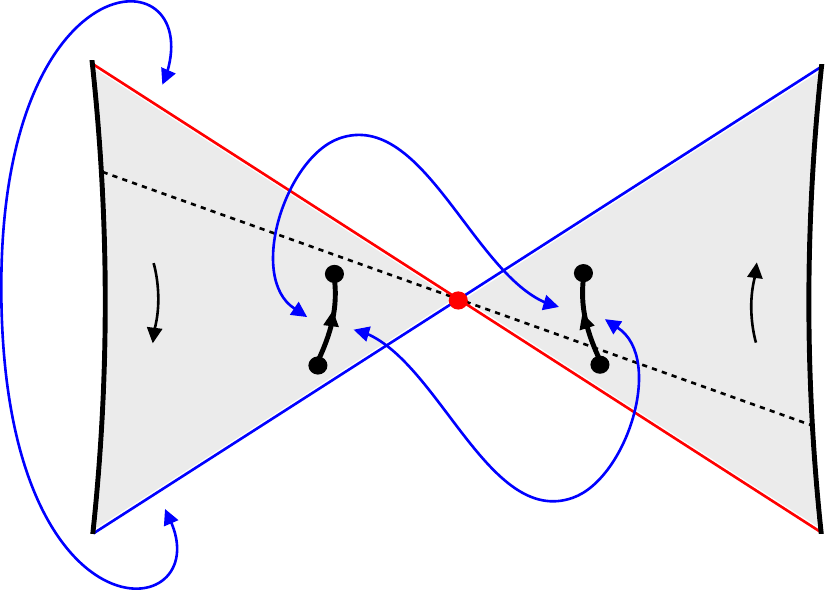}} at (0,0);
  \end{tikzpicture}
\end{equation}
This is easy to see because it maps to a $g=1$ version of \eqref{genus2diag}, which is clearly orientable. If one would flip the orientation (black arrow) on one of the slits, the identification would \emph{not} result in an orientable spacetime, so that does not occur in (for instance) JT gravity.

We would like to contrast this to the identifications one can make on the TFD geometry. The TFD geometry has essentially the same metric $\d s^2=\d\rho^2-4 E \sinh(\rho)^2\d t^2$ in its Rindler wedges, but it does not have the same $\i\varepsilon$ regularization of the point $\rho=0$, nor does it have an identification $t\sim t+T$. As a consequence, the orientable identifications on the TFD are as follows (see also section \ref{sect:crotches})
\begin{equation}
    \begin{tikzpicture}[baseline={([yshift=-.5ex]current bounding box.center)}, scale=0.7]
 \pgftext{\includegraphics[scale=1]{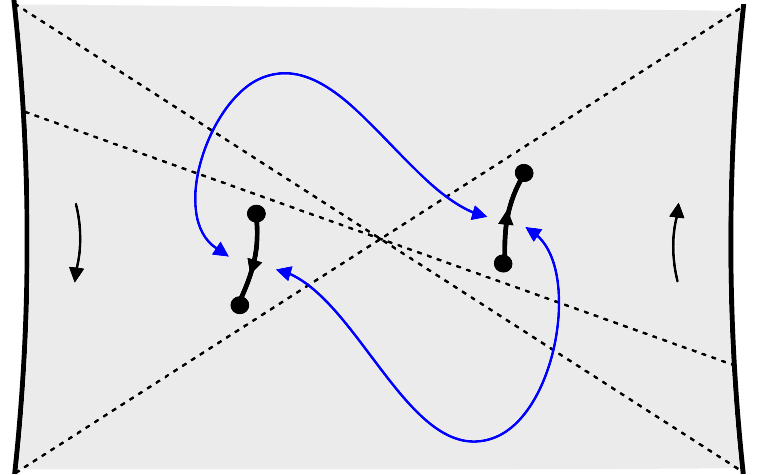}} at (0,0);
  \end{tikzpicture}\label{c7}
\end{equation}
Notice the orientation of the slits (black arrows), which is ``the oposite'' as on the double cone, despite the two having \emph{almost} the same metric. The point is that the two are topologically different very close to the horizon. In the double-cone case there is actually a tiny wormhole, in the TFD case there is not. Topologically, it is easiest to contrast \eqref{c7} for the TFD with \eqref{c5} for the double cone.

\bibliographystyle{ourbst}
\bibliography{Refs}
\end{document}